\documentclass[useAMS,usenatbib]{mn2e}
\usepackage{epsfig}
\usepackage{times}
\usepackage{color}

\def\cm3{cm$^{-3}$}
\def\kms{km~s$^{-1}$}

\def\msun{M$_{\odot}$}

\let\ts=\thinspace

\def\one{\ts {\,\sc i}}
\def\two{\ts {\,\sc ii}}

\def\beq{\begin{equation}}
\def\eeq{\end{equation}}

\def\lesssim{\mathrel{\hbox{\rlap{\hbox{\lower4pt\hbox{$\sim$}}}\hbox{$<$}}}}
\def\gtrsim{\mathrel{\hbox{\rlap{\hbox{\lower4pt\hbox{$\sim$}}}\hbox{$>$}}}}

\def\ip{\rho}

\def\aj{AJ}
\def\pasp{PASP}
\def\apj{ApJ}
\def\apjs{ApJS}
\def\apjl{ApJL}
\def\aap{A\&A}
\def\araa{ARA\&A}

\def\mnras{MNRAS}
\def\nat{Nature}

\def\jcp{J. Chem. Phys.}
\title[Modelling of Type II SN Polarisation]{Synthetic line and continuum linear-polarisation
signatures of axisymmetric type II supernova ejecta}

\author[Luc Dessart and D. John Hillier]
{
Luc Dessart$^{1}$\thanks{E-mail: Luc.Dessart@oamp.fr},
D. John Hillier$^{2}$
\\
$^{1}$ Laboratoire d'Astrophysique de Marseille, Universit\'e de Provence,
CNRS, 38 rue Fr\'ed\'eric Joliot-Curie, F-13388 Marseille Cedex 13, France \\
$^{2}$ Department of Physics and Astronomy, University of Pittsburgh, USA
}

\begin{document}

\date{Accepted . Received }

\pagerange{\pageref{firstpage}--\pageref{lastpage}} \pubyear{2010}

\maketitle

\label{firstpage}

\begin{abstract}
We present synthetic single-line and continuum linear-polarisation signatures due to electron scattering in
axially-symmetric Type II supernovae (SNe) which we calculate using a Monte Carlo and a long-characteristic
radiative-transfer code.
Aspherical ejecta are produced by prescribing a latitudinal scaling or stretching
of SN ejecta inputs obtained from recent 1-D non-LTE time-dependent calculations.
We study polarisation signatures as a function of inclination, shape factor, wavelength, line identity, and 
post-explosion time. 
At early times, cancellation and optical-depth effects make the polarisation intrinsically low, even causing
complicated sign reversals with inclination or continuum wavelength, and across line profiles.
While the line polarisation is positive (negative) for an oblate (prolate)  morphology at the peak and in the red wing,
the continuum polarisation may be of any sign.
These complex polarisation variations are produced not just by the asymmetric distribution of scatterers but also of the flux.
Our early-time signatures are in contradiction with predictions for a centrally illuminated
aspherical nebula, although this becomes a better approximation at nebular times.
For a {\it fixed} asymmetry, our synthetic continuum polarisation is generally low, may evolve 
non-monotonically during the plateau phase, but it systematically rises as the ejecta become optically
thin and turn nebular.
Thus changes in polarization over time do not necessarily imply a change in the asymmetry of the ejecta.
The SN structure (e.g., density and ionization) critically influences the level of polarisation.
Importantly, a low polarisation ($<$\,0.5\%)
at early times does not necessarily imply a low degree of asymmetry as usually assumed.
Asphericity influences line-profile morphology and the luminosity, and thus may compromise
the accuracy of SN characteristics inferred from these.
\end{abstract}

\begin{keywords}
polarization -- radiative transfer -- scattering -- supernovae: general -- supernovae: individuals: 1987A, 2007aa
\end{keywords}

\section{Introduction}

Supernova (SN) explosions are understood to stem from the
collapse  of the degenerate core of a massive star (see \citealt{woosley_janka_05} for
a recent review) or from the thermonuclear runaway of a Chandrasekhar-mass white dwarf
(see \citealt{HN00_Ia_rev} for a recent review).
Despite the very different  explosion mechanisms, radiation-hydrodynamic simulations of these events
yield ejecta that depart from sphericity both on large scale and small scale.
In core collapse SNe, large scale asymmetry is associated with the morphology of the revived shock
\citep{scheck_etal_06,burrows_etal_07b,burrows_etal_07a,marek_janka_09},
while small scale structure is expected from $^{56}$Ni fingers poking through the shock,
the associated mixing it triggers, and from Rayleigh-Taylor instabilities arising at the interface between
the hydrogen and helium shells after shock passage \citep{arnett_etal_89,fryxell_etal_91,mueller_etal_91,
kifonidis_etal_03,kifonidis_etal_06,joggerst_etal_09,hammer_etal_10}.
Such inhomogeneities may appear as density perturbations, composition variations, or a combination of both.
In SNe Ia, both small and large scale inhomogeneities seem to be related to the mechanism of
explosion, be it a deflagration, a detonation, or a combination of both \citep{gamezo_etal_05}.
Compared to an angle-averaged (1D) ejecta, such morphological variations may thus introduce spatial variations in the
properties of the gas that may alter the diffusion of heat in the ejecta, the excitation/ionisation of the gas etc.

Photometry and spectroscopy are the standard means to infer the properties of SN explosions through
the analysis of their radiation, allowing inferences on ejecta composition, temperature, density, expansion rate etc., to be made.
However, apart from observations performed at nebular times, such data usually convey little unambiguous information on the
multi-dimensional nature of SN explosions, thereby missing critical characteristics of such events.
However, linear-polarisation measurements can complement the information from spectra and light curves
and help to assess the level, and characterise the nature, of the asymmetry in a SN explosion \citep{SS_82}.
In scattering-dominated SN ejecta, a residual polarisation of the integrated continuum light received from the SN
(after a proper subtraction of any contribution from interstellar-polarisation; hereafter ISP) is understood
to stem from a non-spherical distribution of free electrons, either on the large scale (e.g.,
oblateness or prolateness), the small scale (e.g., clumps), or both.

Observationally, there is unambiguous detection of polarisation in numerous SNe, although it is generally
of a small magnitude, a function of the time since the explosion, and dependent on the SN type \citep{wang_wheeler_08}.
In Type II core-collapse SNe, polarisation tends to increase with time, being generally small at early times, and stronger as the photosphere recedes to regions close to the progenitor core \citep{jeffery_91a,jeffery_91b,
leonard_etal_01,leonard_filippenko_01,kawabata_etal_02,kawabata_etal_03,leonard_etal_06,chornock_etal_10}.
In Type Ib/Ic SNe, significant polarisation have been observed even at early times \citep{maund_etal_07,tanaka_etal_08,
tanaka_etal_09a,tanaka_etal_09b}. Their ejecta are believed to result from the explosion of the bare core of a massive star (a Wolf-Rayet
star), and in this context, reflect what is observed at late times in Type II SNe.
The observed polarisation signatures have  been interpreted as resulting primarily from the aspherical nature of the core-collapse
SN explosion mechanism \citep{hoeflich_etal_96a,hoeflich_etal_99,wang_etal_03a,leonard_etal_06},
rather than the asphericity of the progenitor star itself.
Although quite different in nature, SNe Ia also exhibit, albeit more modestly, polarisation variations in the optical
\citep{howell_etal_01,wang_etal_97,leonard_etal_05}, sometimes with
large intrinsic polarisations associated with high-velocity features in Ca\two\
\citep{wang_etal_03b,kasen_etal_03,mazzali_etal_05a}.
Independent evidence for departures from spherical symmetry appears in the morphology
of line profiles. Although this in principle applies at any time, it has been best emphasised
at nebular times for SN 1987A \citep{hanuschik_etal_88,spyromilio_93},
SN 1990I \citep{elmhamdi_etal_04}, SN 1993J \citep{spyromilio_94,matheson_etal_00}, or
SN 2004dj \citep{chugai_etal_05}.

Overall, the association of an ejecta asphericity with a polarisation measure is conditioned
by the details of the interaction of light and matter in the aspherical ejecta. Factors that influence
the observed polarisation level include the importance
of line versus continuum opacity sources, the ratio of scattering to absorption opacity,
the composition, the ionisation, and the viewing angle.
The interpretation of a polarisation measure thus requires detailed polarised-radiative-transfer
modelling.
Since the work of \citet{SS_82}, numerous studies have tackled this problem in the SN context.
\citet{hoeflich_91} discussed the importance of optical depth and occultation effects for continuum polarisation
and \citet{steinmetz_hoeflich_92} applied such a technique to 2D axisymmetric (prolate/oblate) ejecta produced from
rotating massive-star progenitors.
\citet{hoeflich_etal_96a} introduced the effect of lines, imposing Local-Thermodynamic Equilibrium (LTE) and
the line absorption probability, and modelled the observations of SN 1993J.
Using simplistic models,  \citet{chugai_92} provided insights into the polarisation
associated with an asymmetric distribution of $^{56}$Ni, either in clumps for
SN 1987A, or on a large-scale for SN 2004dj \citep{chugai_etal_05}.
More recently, a closer proximity to realistic SNe has been permitted with the use of 3D Monte Carlo
simulations that capture with better fidelity the 3D nature of the ejecta \citep{kasen_etal_03,kasen_etal_06}.
The various elements influencing polarisation are thus well identified today. These include
the absorption or the scattering nature of lines, the distribution of free electrons and the properties of the radiation field.
A necessity for accurate inferences based on the observed polarization is the determination of the ISP.

In this work, we study what determines the linear polarisation of Type II SN spectra. Type II SNe
arise from  the ``generic" mechanism of explosion following the collapse of the degenerate core
of a massive star, which may arise from a red-supergiant (RSG) star, or more rarely
from a blue supergiant (BSG) as in the case of SN 1987A  \citep{woosley_janka_05}.
Our polarisation modelling is based on detailed 1D non-LTE time-dependent simulations of SN 1987A \citep{DH10a}
and SNe II-Plateau (II-P) \citep{DH11}, which reproduce well the key spectroscopic and photometric
properties of each type. We employ the results from these 1D simulations at a few days after
explosion up to nebular times.

Our detailed simulations constitute a reliable physical basis for the computation
of polarisation signatures, which we perform with two different polarised radiative-transfer codes:
A Monte Carlo approach \citep{hillier_91} and a long-characteristic method \citep{hillier_94,hillier_96}.
At present, both solvers assume an axial symmetry, adopt but do not require a top/bottom symmetry,
and work for arbitrary viewing angles. 
In this first exploration, we focus on the polarisation of {\it individual lines} and delay to a forthcoming study the 
treatment of line overlap and line blanketing (Dessart \& Hillier 2011, in prep.). While simplistic, this facilitates
the interpretation of the complex line-polarisation signatures we obtain.
The ejecta asphericity is established
by enforcing a density scaling or a radial stretching as a function of latitude, leading to
an oblate or prolate configuration. The main merit of this approach is consistency.
For example, in former studies of SNe II-P, we have identified the importance of time-dependent effects
on the electron density \citep{DH08_time}, which conditions light polarisation through electron scattering.
Moreover, through line-profile modelling of SNe II-P spectra, we have found that line emission
and absorption stem from a confined region, within the SN ``photodisk'', and which
overlaps with the continuum photosphere \citep{DH05_epm, DH05_qs_SN}.
Working from such physical inputs is superior to prescribing how the line or the continuum
should form in the ejecta, or imposing a core-halo approximation
(e.g., an optically-thin scattering nebula or clumps of arbitrary optically thickness illuminated
by a central source).

  This paper is structured as follows.
   In the next section, we present the polarised radiative-transfer codes we employ
   to generate polarisation signatures of axisymmetric Type II SNe. A brief discussion
   of line formation in SN ejecta is presented in Section~\ref{sec_line_form}.
    We briefly present the 1D non-LTE time-dependent simulations we employ
   as a basis for such polarised radiative-transfer computations, and how we generate aspherical but
   axisymmetric ejecta from such 1D inputs (Section~\ref{sect_setup}).
   In the subsequent sections, we document the numerous results from our simulations.
   To lay down the key features that control line and continuum polarisation, we
   discuss our results for a radially-distorted oblate SN II-P ejecta
   halfway through the plateau phase (Section~\ref{sect_s15TO}).
   We then discuss in Section~\ref{sect_dependencies} the various dependencies of the polarisation
   signatures on inclination (Section~\ref{sect_incl}), shape factor (e.g., oblate versus prolate; Section~\ref{sect_shape_factor}),
   wavelength and line identity (Section~\ref{sect_lambda}), and time since explosion (Section~\ref{sect_time}).
   As an aside we discuss the important implications of ejecta asphericity on line-profile
   morphology (Section~\ref{sect_line_profile}) and the SN bolometric luminosity  (Section~\ref{sect_lbol}).
   Our conclusions are presented in Section~\ref{sect_concl}.

\section{Polarisation modelling and Numerical Codes}
\label{sect_codes}

In this work we assume that the polarisation is produced by electron scattering. The scattering of electromagnetic
radiation by electrons is described by the dipole or Rayleigh scattering phase matrix. To describe the ``observed"
model polarisation we adopt the Stokes parameters  $I$, $Q$, $U$, and $V$ \citep{Cha60_rad_trans}. Since we are dealing
with electron scattering, the polarisation is linear and the $V$ Stokes parameter is identically zero. For clarity we will
use the notation $I_Q$ and $I_U$ when discussing  the polarisation of the specific intensity, and $F_Q$ and $F_U$
when discussing the polarisation of the observed  flux.

For consistency with the  earlier work of \cite{hillier_94, hillier_96} we choose a right-handed set of unit vectors
$(\zeta_X,  \zeta_Y, \zeta_W)$. Without loss of generality the axisymmetric
source is centred at the origin of the coordinate system with its symmetry axis lying along $\zeta_W$, $\zeta_Y$ is in the plane of the sky,
and the observer is located in the XW plane.

We take $F_Q$ to be positive when the polarisation is parallel to the symmetry axis (or more correctly parallel
to the projection of the symmetry axis on the sky),
and negative when it is perpendicular to it. With our choice of co-ordinate system, and since the
SN ejecta is left-right symmetric about the XW plane,  $F_U$ is zero by construction. This must be the case since
symmetry requires that the polarisation can only be parallel, or perpendicular to, the axis of symmetry.
For a spherical source, $F_Q$ is also identically zero.

At times it will also be necessary to discuss $I(\ip,\delta)$, $I_Q(\ip,\delta)$ and $I_U(\ip,\delta)$, the observed
intensities on the plane of the sky. As usual, $I_Q$ is positive when the polarisation is parallel
to the radius vector, and negative when it is perpendicular. In the plane  of the sky we define a set of
polar coordinates ($\ip,\delta$)  with the angle $\delta$  measured anti-clockwise from $\zeta_Y$.
The polar coordinate, $\ip$, can also be thought of as the impact parameter
of an observer's ray. We also use the axes defined by the polar coordinate system to describe the polarisation.
$F_I$ is obtained from $I(\ip,\delta)$ using

\begin{equation}
F_I= {2 \over d^2}  \int_0^{\ip_{\rm max}} \int_{-\pi/2}^{\pi/2}  \, I(\ip,\delta) dA \, ,
\end{equation}

\noindent
where $dA=\ip d\delta d\ip$. Since $\zeta_\ip$ is rotated by an angle $\delta$ anticlockwise from $\zeta_Y$,
$F_Q$ is given by

\begin{equation}
F_Q = {-2 \over d^2} \int_0^{\ip_{\rm max}}  \int_{-\pi/2}^{\pi/2}
\left[ I_Q(\ip,\delta)  \cos 2\delta +  I_U(\ip,\delta)   \sin 2\delta \right] \, dA \,.
\end{equation}

\noindent
In a spherical system, $I_Q$ is independent of $\delta$, and $I_U$ is identically zero.

\noindent
The level of polarisation, p($\ip,\delta$), is given by

\begin{equation}
p(\ip,\delta) = \sqrt{I_Q(\ip,\delta)^2+I_U(\ip,\delta)^2}/I(\ip,\delta) \,,
\end{equation}

\noindent
while the integrated polarisation is

\begin{equation}
F_p = \sqrt{F_Q^2+F_U^2} \bigg/ F_I \, = \, |F_Q| \, \big / F_I \,.
\end{equation}

To summarize, given this geometrical setup, axisymmetric configurations can only produce
a residual polarisation $F_Q$,  which is positive (negative) when the electric vector
is parallel (perpendicular) to the symmetry axis.
Going from negative to positive polarization thus indicates a 90$^\circ$ shift in the direction of the
polarisation vector in the plane of the sky.
The positive definite polarisation $F_P$ is simply the absolute value of $F_Q$ in this context.
Our results can be directly compared to observations of axisymmetric SN ejecta, provided one rotates
the observed Stokes parameters to yield $F_U=0$. 

Before proceeding it is worth discussing how polarisation is achieved, and what determines
the level of observed polarisation. From the above it is apparent that for an unresolved source
the observed polarisation is determined by the following:

\begin{enumerate}
\item
The polarised flux produced at each location.
\item
The cancellation of the polarised flux arising from the integration over the plane of the sky.
\end{enumerate}

The first of these is controlled by the electron scattering opacity, the extent of the atmosphere,
the ratio of scattering to absorption opacity, and the source function. For example, an increase in the electron
density (at least in the optically thin limit) increases the amount of scattering, and hence polarised flux.
The same is true for the atmospheric extent --- maximum polarisation is generally achieved for a point source.

The cancellation effect is particularly interesting. For an axisymmetric optically thin nebula illuminated by
a point source the cancellation is related to the departure of the source from spherical symmetry, and
the inclination angle of the observer. For this model the intensity through the nebula is constant.
However, if we increase $\tau$ to $\sim$1 or more, this will no longer be the case. In such cases
it is possible to fundamentally alter the strength and sign of the polarisation --- the polarisation will not
necessarily be determined by regions with the highest electron density, but rather by the regions receiving,
and hence scattering, the highest flux.

To facilitate later discussions we also define two additional quantities, $\bar I_Q(\ip)$ and $dF_Q(\ip$)
by the following relations:

\begin{equation}
\bar I_Q(\ip) = {1 \over \pi} \int_{-\pi/2}^{\pi/2} \ip I_Q(\ip,\delta) \, d\delta
\end{equation}

\begin{equation}
dF_Q(\ip) = - \int_{-\pi/2}^{\pi/2} \left[ I_Q(\ip,\delta)  \cos 2\delta +  I_U(\ip,\delta)   \sin 2\delta \right] \, \ip d\delta \, .
\end{equation}

The first of these can be thought of as the angle-averaged polarised flux at a given polar coordinate, while
the second gives the contribution to the polarised flux at a given polar coordinate. A comparison between
the two helps to illustrate where flux cancellation is important, while the second also illustrates what
impact parameters contribute most to the observed polarisation flux.

 An important analytic estimate of the polarisation was obtained by \cite{BL77}  \cite[with a factor of 2 correction
 given by][]{BME78_pol}. Consider an axisymmetric aspherical nebula illuminated by a central point source.
 In such a case the observed polarisation, with the observer located at an inclination $i$ to the symmetry axis, is

 \begin{equation}
 F_p = 0.375 \tau (1-3\gamma) sin^2i \,,
 \end{equation}

 \noindent
 where  the angle-averaged electron-scattering optical depth, $\tau$, is defined by

 \begin{equation}
  \tau=0.5 \sigma_e \int_{-1}^1 \int_0^{\hbox{$r_{max}$}} N_{\rm e}(r,\mu) dr d\mu \,,
 \end{equation}

 \noindent
 and $\gamma$ is a shape factor defined by

  \begin{equation}
  \gamma = {\int_{-1}^1 \int_0^{r_{\rm max}}   \mu^2 N_{\rm e}(r,\mu)drd\mu \over  \int_{-1}^1 \int_0^{\hbox{$r_{max}$}}
                                                                                N_{\rm e}(r,\mu)drd\mu}.
 \end{equation}

 \noindent
Here, $N_{\rm e}(r,\mu)$ is the 2D electron-density distribution,  $\sigma_e$ is the Thomson cross section,
and we employ spherical polar coordinates $(r,\theta,\phi)$, with $\mu=\cos\theta$.

  Oblate configurations correspond to $\gamma < 1/3$ ($\gamma=0$ for a thin disk) and give rise to a
  positive polarisation (electric vector parallel to the symmetry axis)  while prolate configurations
  have $\gamma > 1/3$ ($\gamma=1$ for a pole line) and give rise to a negative
  polarisation (electric vector perpendicular to the symmetry axis).

The analytical result obtained by \citet{BL77} reveals the fundamental parameters that influence the observed
polarisation, and  a fundamental degeneracy --- the observed level of polarisation at a single frequency is related
to 3 quantities ---  the electron scattering optical depth, a shape correction factor $(1-3\gamma)$, and the viewing
angle, and that without additional constraints it is impossible to separate these three effects. In the context of the earlier discussion, $\tau$ is setting the polarisation levels, while the shape factor, and $\sin i$ determine the cancellation in the polarisation caused by integrating over the entire image.

  The formula also set an upper limit to the possible polarisation. For an equator-on view, this limit is 0.375$\tau$ for an
  infinitely-thin disk and $-2 \times 0.375\tau$ for a pole-line/blob (not symmetric about equatorial plane). Since $\tau$ is an average optical depth,
  these expressions are better written as

    \begin{equation}
       F_p=-0.375\tau_{\rm blob} (1.0 - \cos \theta) \sin^2 i \,,
    \end{equation}

\noindent
  where $\tau_{\rm blob}$ is the average radial optical depth of the blob, and $\theta$ is the half-opening angle of the blob; and

    \begin{equation}
       F_p= 0.375\tau_{\rm disk} \sin \theta' \sin^2 i \,,
  \end{equation}

  \noindent
  where $\tau_{\rm disk}$ is average radial optical depth in the disk and $\theta'$ is the half-opening angle of the disk.

  Many important assumptions were made by \citet{BL77} in the derivation of their polarisation result including the
  point source assumption, the neglect of  absorption, and  that $\tau$ is small (so that multiple scattering and
  changes in the source intensity can be ignored). In practice, the inclusion of these effects will tend to reduce
  the observed level of polarisation, and, as noted earlier, can even alter the sign of the polarisation.

  Polarization has been a useful tool for studying many asymmetric sources. It has been used to study, for example,
  Be stars \citep[e.g.,][]{WDB10_Be_pol,QBB97_pol_inter}, Wolf-Rayet stars \citep[e.g.,][]{HHH98_WR_pol},
  AGN \citep[e.g.,][]{L02_AGN_pol, M07_AGN_pol}
  and Symbiotic stars \citep[e.g.,][]{SS94_Symb_pol}. Many different effects have been found, which are also
  relevant to the study of SNe. These effects, for example, include the depolarisation in emission lines. Since emission
  lines generally form further out than the continuum, line photons tend to undergo less scattering, and hence
  show less polarisation \citep[see, e.g., review by][]{SMH92_rev}. The presence of a line-effect is proof that a
  source has intrinsic polarisation --- interstellar polarisation introduces only a slow wavelength dependence
  \citep{Ser73_pol} so that over a line the polarising effect of the interstellar medium is constant.

  Another important effect is that of continuum opacity.  For the wavelength range of interest electron scattering is
  gray, and hence, in the absence of other opacity sources, it will yield a polarisation which is wavelength independent.
  However, the presence of an absorptive opacity removes this property since the absorptive opacity is usually a
  function of wavelength. An obvious manifestation of the influence of absorptive opacity is the large change
  in polarisation across the Balmer jump  in Be stars  (e.g., \citealt{WBW96_abs_pol, WBB97_be_pol, QBB97_pol_inter}).

  An important assumption in this work is that the lines are intrinsically unpolarised \citep{hoeflich_etal_96a}, 
  and that any resulting polarisation
  is due to electron scattering. For H$\alpha$, which has a large recombination contribution and optical depth,
  this is likely to be a reasonable approximation, and is likely to be even more valid for higher Balmer lines. In support of this
  conclusion we note the study of  \citet{jeffery_91a} who adapted the Sobolev approximation \citep{Jef89_sob_pol} to
  handle polarisation. From his study of SN 1987A, \citet{jeffery_91a} concluded that both the continuum and line
  polarisation were mostly due to electron scattering.

    We perform our polarisation calculations using either a Monte-Carlo code (Sect.~\ref{sect_mc_code}),
    or a more conventional radiative transfer code (Sect.~\ref{sect_lc_code}). Most polarisation studies of SNe
    have been based on Monte-Carlo simulations (e.g.,  \citealt{hoeflich_91, KNW03_SN2001el, kasen_etal_03}).
    The main advantage is the relative ease with which complicated geometries and structures (e.g., clumping), and
    physics, can be handled.  A disadvantage is that, for a given problem, they are generally less efficient than
    more conventional transfer codes. However, with the advent of parallel processing the latter issue is less of
    a problem. Another issue, especially relevant to polarisation studies, is the necessity of collecting many photons
   per resolution element -- $10^4$ photons per resolution element will give an error of the same size as the typical
   polarisation signal (which is of order 1\%). Since we are essentially dealing with photon noise, a reduction in noise of
   a factor of 2 requires 4 times the computational effort. Although often ignored, and like more conventional transfer
   codes, Monte-Carlo codes also suffer from gridding and interpolation errors.

\subsection{The Monte-Carlo Code}
\label{sect_mc_code}

The Monte-Carlo code has been previously described by \citet{hillier_91, hillier_94}, although some important changes
have been made since then. The code can run in 1D, 2D, or 3D. It employs only a single processor, but due to the
nature of the simulations, independent simulations can be readily combined. The most significant change from earlier
work is that the choice of scattering angle depends on the Stokes parameters of the incident beam --- this is
necessary when there is a large number of scatterings (see below). To choose the two scattering angles we
adopt a procedure similar to that outlined by \cite{FHY94_mc_pol}.

 At present, the code is designed to handle a single line, and its adjacent continuum.  All emitted photons
 (or more correctly photon beams) carry a weight -- this weight is adjusted to account for the bias selection
 of photons and ensures that the correct fluxes are obtained at the end of the simulation. Biased selection
 is done to control and maximize the signal to noise. For example, when computing an observed spectrum
 it is pointless to follow photons from large optical depth since they will be destroyed before reaching the observer.

Opacities and emissivities for the code must be supplied. For opacities input from a 1D code we adopt simple
scaling laws for the opacity and emissivity. For the case where we scale the density by a factor $s(r,\theta)$
we generally scale the electron scattering opacity by $s(r,\theta)$, while other opacities and emissivities
(both line and continuum) are scaled by $s(r,\theta)^2$.

The later scaling arises under the assumption that the dominant process determining the level populations
are proportional to the square of the density (e.g., recombination). An alternative possibility, used in
the 2D code of \cite{BH05_2D}, is to assume the opacities are functions of the electron density only.
Both sets of assumptions are reasonable when we consider small departures from spherical
symmetry, but become more problematic as the departures from spherical symmetry increase.
When we distort the envelope, we assume that the opacities and emissivities on the distorted grid
are unchanged (see Section~\ref{sect_setup}).

To improve the signal to noise we treat all continuum photons across the line at the same time. These photons are
emitted from the same location and in the same direction, and experience the same scattering/continuum
absorption processes. As some of the continuum photons can be absorbed in the line, we associate a line absorption
weight with each photon.

 Line transfer is treated using the Sobolev approximation \citep{castor_70}. One consequence of the Sobolev
 approximation is that we ignore scattering of photons by electrons in the Sobolev resonance zone. For most lines
 this effect is relatively unimportant, but for lines with large optical depths, such as H$\alpha$, scattering in
 the resonance zone tends to increase the line flux and the strength of the electron scattering wings
 (for a fixed source function). For consistency,  the profile and line source function should be computed
 using the same Doppler width. Another consequence of the Sobolev approximation is that it does not produce
 for optically thick lines a redshift of order one Doppler width in the location of the line
 \citep[see, e.g.,][]{CPK84_shift,Hil00_prof} -- for SNe this effect will be unimportant.

In order to deduce the signal to noise of the results, we typically split the simulation into 10 sub-runs. The mean
and variance of these runs can then be computed using standard techniques. Because we treat all continuum
photons across the line at the same time, the errors are correlated, especially in regions where the line makes
only weak contribution to the flux. A practical effect of this correlation is that there will be a systematic offset between
profiles computed using an independent run or an alternative technique (e.g., with the transfer code).

Scattering of photons can be treated via either of three methods. First, we can scatter the photons isotropically.
Second, we can simply scatter the photons according to the dipole scattering function, and correct the photon weight
for the bias selection of scattering angles. Third, we scatter the photons according to the correct
scattering probabilities which are influenced by the polarisation.
As for the second method we correct the photon weight for the bias selection of scattering. For all methods the
change in intensity of the beam is accounted for when we apply the Rayleigh scattering phase matrix to the Stokes parameters
of the beam being followed. In principal all three methods should give the same answer although the statistical errors will be
different. In particular, the third method should give the lowest error since it scatters the photon (beam) according
to the actual probabilities.

In previous simulations we found that all three Monte-Carlo methods gave similar answers. However, for some of the SN simulations
the first two methods gave answers which were inconsistent with both the third method, and the long characteristic code.
The third method gave answers consistent with the long characteristic code.
An explanation for the observed behavior is as follows, and applies
to both methods  in which we do not choose the scattering
directions according to the scattering probabilities (but although to a lesser extent to the second method).
On each photon (beam) scattering we apply the
Rayleigh Scattering Matrix to the Stokes parameters. With the stokes parameters defined by the scattering
plane, $I$ and $Q$ transform according to
$$
\begin{array}{lll}
I' &=& 0.75(\cos^2 \theta +1)I + 0.75(\cos^2 \theta -1)Q  \\
Q' &=&  0.75(\cos^2 \theta -1)I + 0.75(cos^2 \theta +1)Q
\end{array}
$$
where $\theta$ is the angle between the incident and scattered beam.
Depending on the polarisation state of the beam, the scattering
intensity will be scaled by a factor that varies between 0 and 1.5. (e.g.,
when $Q=I$ , $I' = 1.5 \cos^2 \theta I$  ). For a few scatterings this procedure works well,
but in some of the SN models the effective number of scatterings
undergone by the observed photons was large -- in some cases over 40.

In the case of isotropic scattering the beam intensities ranged over seven orders of magnitude. The photons (beams)
contributing to the observed signal are basically those with intensities of order unity, or larger. Thus most of the
photons don't contribute to the observed signal. Crudely the distribution of beam intensities was log-normal. This is expected ---
the product of a series of random variables is approximately log-normal \citep[see, e.g.,][]{Limpert_etal_01}. 
Another issue arises from how we compute the signal
and error --- with a log-normal distribution the simple arithmetic mean is not the best estimate of the distribution mean.
Using the third method the intensities still have a log-normal distribution, but its breadth is considerably
smaller. Second, and more importantly, the photon weight needs to be allowed for --- the quantity that determines the intensity
is the beam weight times the Stokes parameters.

Another issue is the redistribution of photons in frequency induced by the scattering. One of two methods
is used to take into account the frequency shift. We can either choose the electron velocity components from a Maxwellian
velocity distribution, or we can use the angle averaged redistribution function. The later is done to allow comparison with
the transfer code which uses the angle-averaged redistribution function. The differences in predicted profiles, particularly for
the SN calculations, are small. The Compton shift is ignored. At present aberration and advection terms are not taken into
account.

An advantage of the Monte-Carlo code over the long-characteristic transfer code is the ability to investigate statistical
properties of the radiation field. For example, it is very easy to determine from where photons originate, where they are
last scattered, and the average number of scatterings undergone by an observed photon. A disadvantage of the current
implementation is that interpolations, when following the emission and scattering, are linear.
As a consequence the error only scales linearly with the grid size.

\subsection{The Long Characteristic Transfer Code}
\label{sect_lc_code}

The long-characteristic transfer code has been previously described by \cite{hillier_94, hillier_96}. Although the transfer and
Monte Carlo codes were written by the same author there is little overlap between the routines (such overlap is mainly related
to input/output and some interpolation routines), and since the two codes treat the electron scattering by two very different means,
the results from the codes are highly independent. Thus a comparison between the two codes provides an excellent means of
testing the accuracy and validity of the solutions. As for the Monte-Carlo code, aberration and advection terms are not taken into
account.

While small differences between the results are seen, these differences do not affect any of the conclusions drawn in this paper.
Importantly, the two codes give the same quantitative behavior for very different models -- models in which the level of polarisation
and its variation across line profiles and/or with the inclination are very different (we discuss and illustrate this agreement further below
in Section~\ref{sect_s15TO}).

The polarisation computation proceeds in a series of steps:

\begin{enumerate}

\item
Compute the moments describing the continuum polarisation. The moments are computed using an approximate lambda operator with
NG acceleration \citep{ng_74}. To start the calculation we use either moments from a previous calculation, or set them to zero.
An alternative starting procedure, although not  implemented, would be to use non-polarised estimates from a spherical or
non-spherical calculation.

\item
Compute the frequency dependent moments describing the continuum and associated line polarisation. This is done using a long-characteristic code, where the radiation at each location on the grid is determined using its own set of rays. While inefficient, its advantage is its accuracy ---  no interpolation of the intensities is required, and only the opacities, emissivities and the velocity field need to be interpolated.
Three options can be used to provide starting moments: (a) Use continuum values at all frequencies; (b) Use moments
from a previous calculation, perhaps done at lower resolution; (c) Use non-polarised moments supplied by a 2D-short characteristic calculation \citep{ZHG06_SC_meth}.  In the present work we utilized the first two methods.
The computational performance has been significantly improved with the use of Open Multi-Processing (OMP) instructions.
Typically we use 4 to 8 processors.

\item
Using the computed polarisation moments, we undertake a formal solution to compute the observed spectrum. Since we have moments  as a function of $(r,\theta)$ we compute the spectrum for multiple viewing directions. If necessary, the moments can be interpolated in $(r,\theta)$ to make a finer grid that can then be used to compute the spectrum with higher accuracy.  An option to store $I(\ip,\delta,\nu,i)$ is available --- this information is used by an IDL procedure to create 2D images.
\end{enumerate}
\noindent

\section{Comments on line formation in SN\lowercase{e}}
\label{sec_line_form}

Before proceeding to the discussion of continuum and line polarisation it is worth
emphasizing several important results that apply to the formation of an isolated line in SNe.

Due to the rapid expansion of the SN envelope, line photons emitted from a given location
in the envelope can only interact with the same line in a small zone of narrow spatial extent.
The extent of this zone is $\sim r (v_{\rm dop}/v)$ which  is $\ll r$ (since $v_{\rm dop} \lesssim 10$\,\kms,
although somewhat larger if we include micro-turbulence). As a consequence
line photons at a given observed frequency (velocity) originate from an iso-velocity surface, which,
for a  homologous expansion, is a plane perpendicular to the observer's sight line (see, e.g., Fig.~10
in \citealt{DH05_epm}).
Photons at line center originate from the plane passing through the center of the SN,
while photons near the blue (red)  edge originate far out in the SN ejecta, in a plane on the near
(opposite) side  of the observer.

During the nebular phase, the above approximation is excellent, but during the plateau phase the situation
is somewhat more complicated. If the line is optically thin, continuum emission simply adds to the line emission.
However, if the line is thick this is only true on the red side ---  continuum photons on the blue portion
of the line will have been processed by the line.

\begin{figure}
\epsfig{file=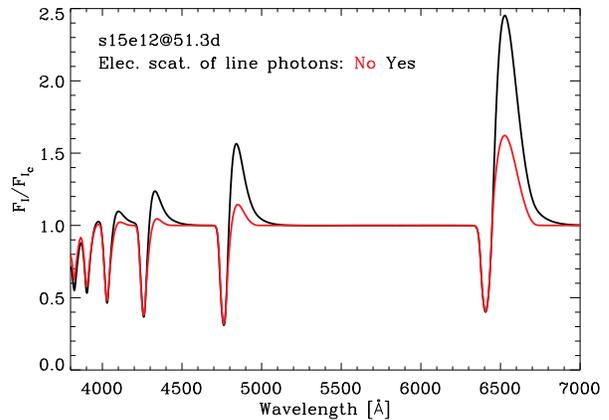,width=8.5cm}
\caption{
Synthetic spectrum for the model s15e12 of  \citet{DH11} at 51.3\,d after explosion
when allowance for bound-bound transitions of H\one\ are accounted for (black).
In red, we show the corresponding result when electron-scattered line photons
are neglected. The stark contrast implies that half the line photons or more undergo
at least one scattering with free electrons before escaping.
Such scattering can be a source of polarisation within the line, even though H\one\
Balmer lines form through recombination.
\label{fig_jcont}
}
\end{figure}

An additional complicating factor is electron scattering, which destroys the simple
correlation between velocity and the emission location presented above. The scattering
of optical photons off an electron is coherent in the frame of the electron (if we
ignore the small Compton shift). Because we have an homologous expansion
(with every point in the atmosphere seeing material moving away from it), photons
in the observer's frame will preferentially scatter to the red, as can be seen in
Figure~\ref{fig_jcont}. Electron thermal motions only have a relatively minor effect, since the
average electron thermal velocity ($\sim$\,550[$T/10^4$\,K]\,\kms) is much less than the
SN expansion velocities ($\sim$\,5000\,\kms).   The photons, emitted in a particular
line, but subsequently scattered by electrons, give rise to line polarisation, even
in a line formed by recombination or collisional processes. Depending
on the age and ejecta properties of the SN, 50\% or more of line photons may
experience an electron scattering (Fig.~\ref{fig_jcont}).

Another important issue affecting line formation in Type II SN is the slope of the
density distribution in the continuum/line forming region.
The density distribution in SNe is rather steep ($\rho \propto r^{-n}$ where $n > 8$)
during the plateau phase; this is much steeper than the density distribution in the line
formation region of, for example, Wolf-Rayet stars. This forces the continuum and line formation region
(for the Balmer lines) to overlap; alternatively, we can say that the average radii of
emission are rather similar. Larger $n$ gives rise to weaker Balmer emission, and
decreases the size of the spatial zone over which the line and continuum form. Since $n$ affects the
spatial extent of the envelope, $n$ will also affect the polarisation observed
for a SN of a given asphericity.

One final point worth stressing is that Type II SNe have photospheres that are
electron-scattering dominated. However, as is well known, the absorptive opacity,
and hence the thermalisation depth, play a crucial role in determining the flux. This
is a crucial point for polarisation studies --- the observed polarisation
is dependent on both the flux and how it is scattered. It is important that the
aspherical flux distribution be accurately modelled in order for the correct
polarisation to be predicted.

The importance of the flux distribution for understanding the polarisation is well known.
Indeed, it is possible to produce polarisation in a SN by having an inner aspherical
$^{56}$Ni source located in spherical ejecta (see, e.g, \citealt{chugai_92, HKW01_pol}).

   \section{Numerical setup}
\label{sect_setup}

\subsection{Presentation of the 1D reference models}

  The linear-polarisation signatures of axially-symmetric aspherical Type II SN ejecta that we present in this work
   are based on 1D non-LTE time-dependent radiative-transfer simulations of SN 1987A \citep{DH10a} and
   SNe II-P \citep{DH11}. These 1D simulations are performed in three steps.
   A progenitor star on the main-sequence  (here an 18\,\msun\ rotating star at the LMC metallicity
   for the SN 1987A model and a 15\,\msun\ non-rotating solar-metallicity star for the SN II-P model)
   is first evolved until iron-core collapse
   with {\sc kepler} \citep{weaver_etal_78}.
   This pre-SN model is then exploded by driving a piston outward from the surface of its degenerate core,
   yielding a SN ejecta with a prescribed asymptotic kinetic energy of 1.2\,B (model names lm18a7Ad and s15e12).
   Once the ejecta have reached homologous expansion (or are close to it), we remap the {\sc kepler} model
   onto the {\sc cmfgen} grid and evolve the associated gas and radiation. The time sequence was continued
   until 21\,d for the SN 1987A model and until $\gtrsim$1000\,d for the s15e12 model.
   These three modelling steps assume spherical symmetry and it is only in the fourth step
   that we enforce an asphericity (see next section).
   A full description of the {\sc cmfgen} simulations is provided in \citet{DH10a,DH11}.
   Here, we summarise the key assets of our approach and the results of relevance
   for the interpretation of the polarisation measures described in this paper.

     The strengths of our {\sc cmfgen} simulations are:
     \begin{enumerate}
    \item
    the treatment of non-LTE for {\it all} level populations
     (and thus all lines);
     \item
     the explicit treatment of line-blanketing (our computation includes
     $\sim$10$^5$ lines);
      \item
       the consideration of all time-dependent terms appearing in the statistical-equilibrium,
     energy, and radiative-transfer equations;
      \item
      the basis of the computation on physical inputs of the SN ejecta
     describing the density/temperature/radius/velocity distribution as well as the complex chemical stratification; and
     \item
      the modelling of the {\it entire} ejecta at all times.
 \end{enumerate}

     The main approximations of this {\sc cmfgen} modelling are spherical symmetry, homologous expansion, and the treatment
     of radioactive decay as a local heating source. At nebular times, the neglect of non-thermal excitation and ionisation
     leads to an underestimate of the free-electron density, the ionisation state of the gas, and the excitation of ions
     \citep{KF92,KF98a,KF98b}. This causes a disagreement between observations and our {\sc cmfgen}
     predictions for H$\alpha$ and the continuum flux at the end of the plateau phase in SNe II-P
     (Section~\ref{sect_time}; \citealt{DH11}).

     To set the context for our polarisation computations, let us summarise these time-dependent simulations.
     After shock breakout, Type II SN ejecta progressively cool, with a recession of the photosphere to
     deeper/slower-moving mass shells and a radial expansion of the photosphere.
     As hydrogen recombines at the photosphere, a plateau forms in the light curves.
     In the SN II-P model, this plateau persists as long as the  ejecta electron-scattering optical depth
     stays above unity, up to $\sim$130\,d here. Subsequently, the photosphere recedes through the
     progenitor helium core, thus probing the inner regions of the ejecta where the shock was launched.
     The post-plateau fast fading ceases when the luminosity
     becomes equal to the instantaneous energy-deposition rate from radioactive decay.
     In the SN 1987A model, a post-breakout plateau exists too but it is halted at $\sim$20\,d after explosion
     when the heat wave generated by radioactive decay at depth induces a brightening.
     Importantly, we find that the SN photosphere resides in the outer ejecta layers for days to weeks (10-20 days for
     SN 1987A but up to 50 days in the SN II-P model) after shock breakout. Hence, SN radiation
     probes the outer ejecta layers early on and the deep ejecta layers close to the explosion site at late times.

     As discussed in \citet{DH05_qs_SN}, line formation in SNe II-P is significantly affected by the generally
     steep density fall-off in the outer ejecta.
     This is constrained in observations through the sizable blueshift and the weakness of P-Cygni profile emission
     at early times \citep{DH05_epm,DH05_qs_SN}. Line formation is thus
     quite confined and overlaps significantly with the continuum formation zone. This property  dramatically affects
     the polarisation from such objects.

     For a power-law density distribution
     $d_{\rm 1D}(r) = d_0 (r_0/r)^{N_d}$ ($d_0$ is the density at a reference radius $r_0$), observations
     suggest a density exponent $N_d$ of $20$ to $50$ at very early times \citep{dessart_etal_08}, converging to
     a more modest value of 10 as time proceeds \citep{DH06_SN1999em}.
     In the SN II-P simulations used here, the density exponent is $15$ beyond 9000\,\kms,
     $\sim25$ between 7500--9000\,\kms, $\sim10$ between 5000-7500\,\kms, and
     flattening further at smaller velocities (the density is constant below 3000\,\kms\ and the edge of the helium core at 1500\,\kms).
     These values are somewhat larger than derived from observations during the plateau phase.
     For SN 1987A, a more gradual density fall off in our model yields good fits to observations, with $N_d$ of $\sim$8 \citep{DH10a}.

\begin{figure}
\epsfig{file=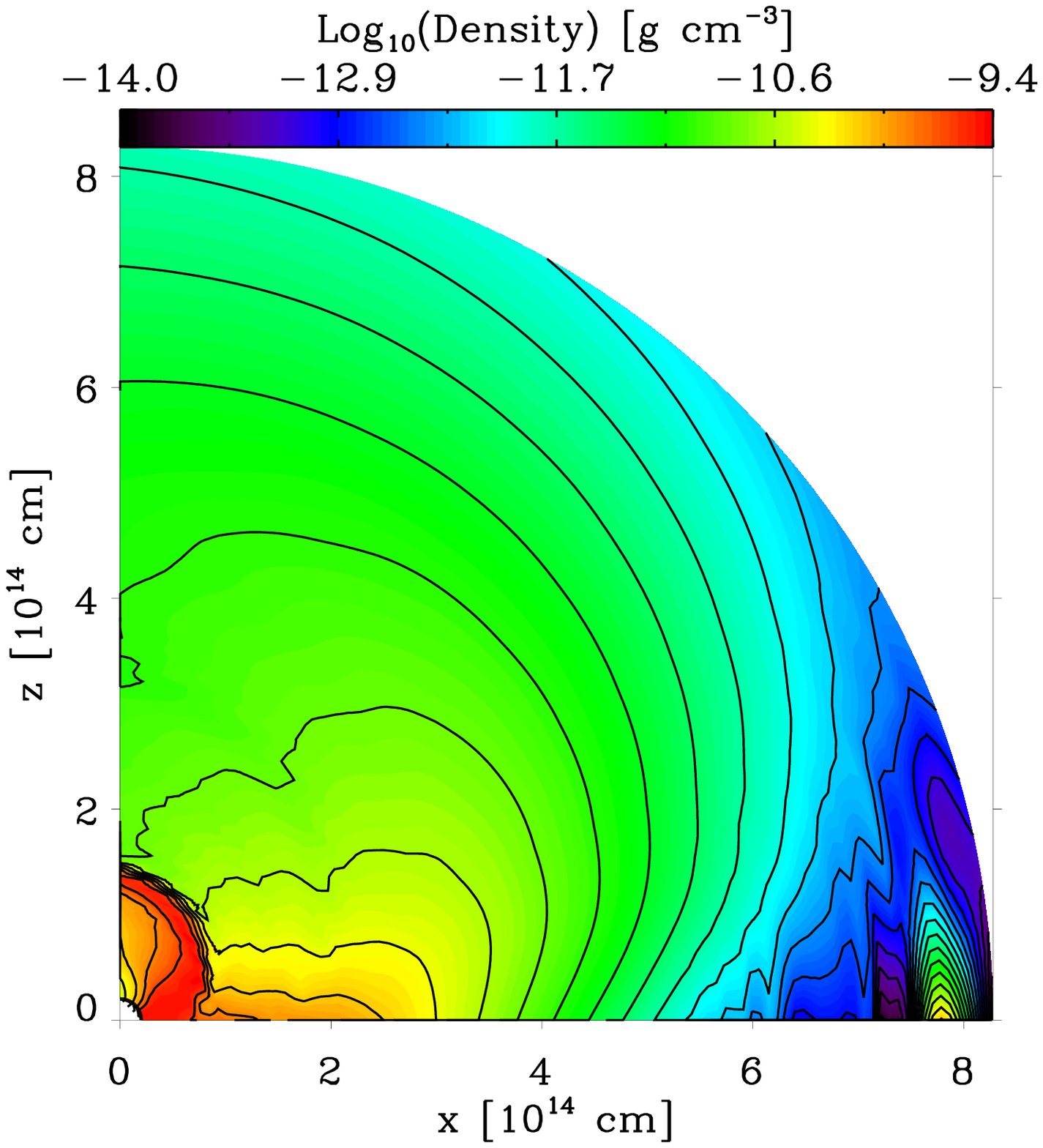,width=9cm}
\epsfig{file=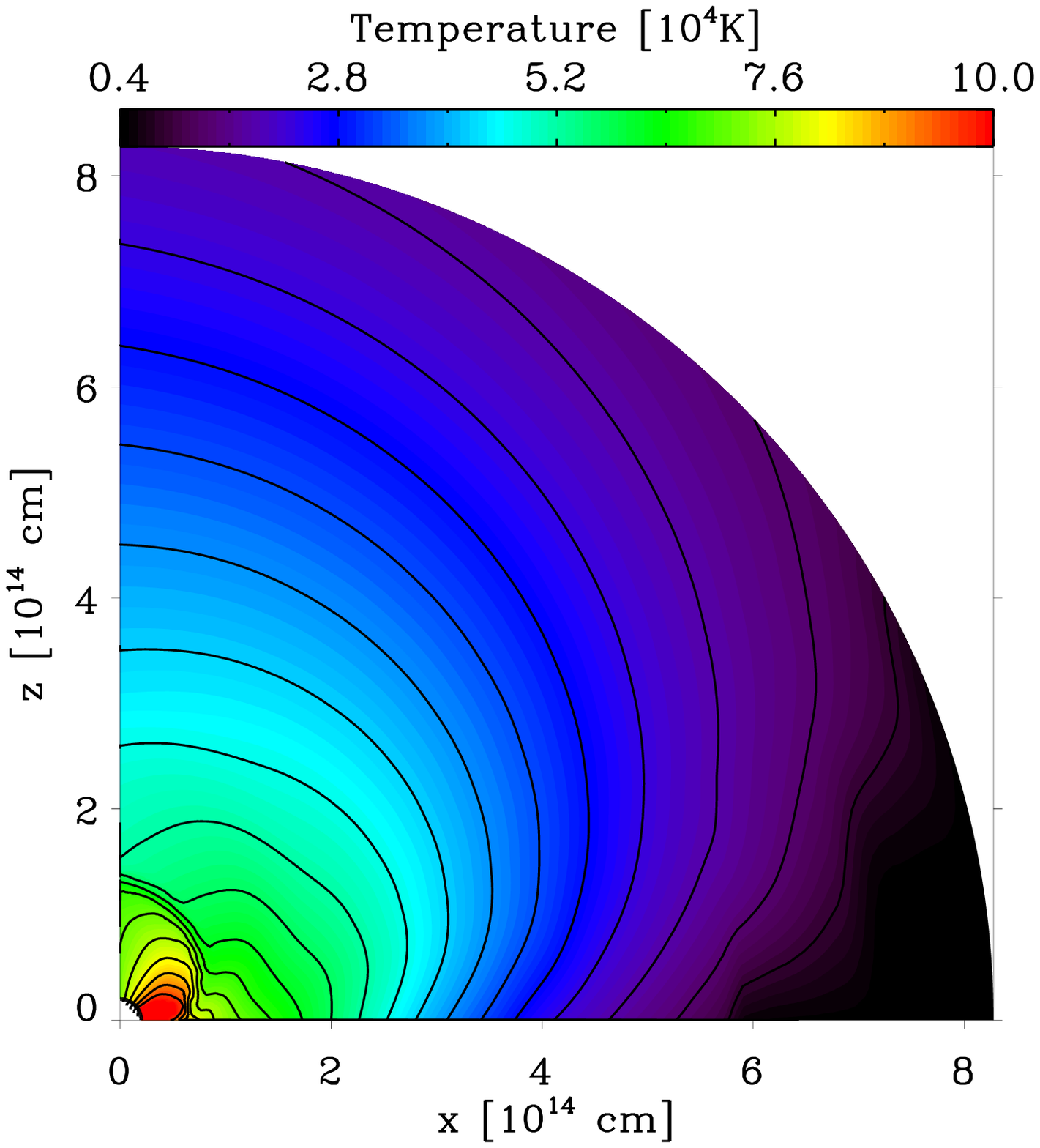,width=9cm}
\caption{Colour map of the 2D ejecta density (top) and temperature (bottom) in an axi-symmetric (and top-bottom
symmetric) explosion, which
is constructed by stacking in latitude the results from independent {\sc v1d}
radiation-hydrodynamics
simulations of piston-driven explosions with a range of explosion energies \citep{DLW10a}.
These simulations are based on the s15 pre-SN progenitor model of \citet{WHW02} and range
from 0.18 to 1\,B, in increments of $\sim$0.1\,B (explosion energy increases with latitude).
This montage illustrates the density/temperature distribution versus latitude and, in the present setup,
the morphological evolution from prolate at large radius/velocity to oblate at small radius/velocity
for the density and temperature contours.
\label{fig1_V1D_2D} }
\end{figure}

\begin{figure}
\epsfig{file=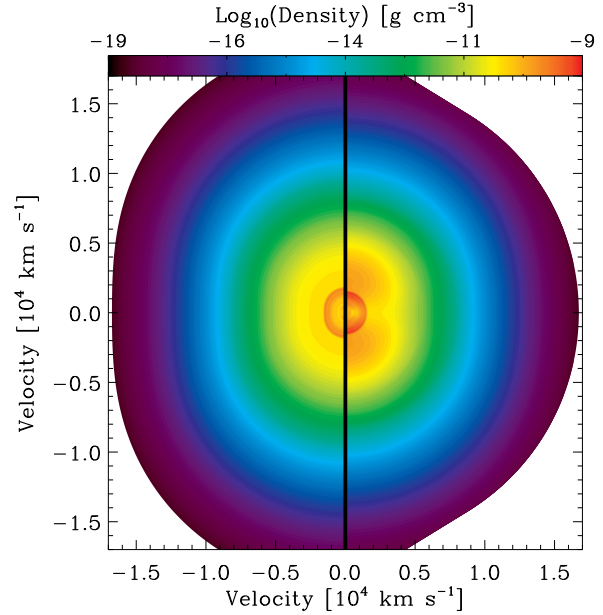,width=9cm}
\caption{Colour map of the density distribution in a 2D axisymmetric prolate ejecta produced by applying a radial distortion (left) or
a density scaling (right) to the 1D s15e12 model of \citet{DH11} at 17.9\,d after explosion. For the most part, this 1D ejecta is
well described by a power-law density distribution $d_{\rm 1D}(r)$ with $d\ln d_{\rm 1D}/d\ln r = -10$.
In this illustration, we use a radial stretching yielding $d(r,\mu) = d_{\rm 1D}( r/ (1 + A_2 \mu^2))$ with $A_2 = 0.175$,
or a latitudinally dependent density scaling yielding $d(r,\mu) = d_{\rm 1D}(r) \times (1 + A_1 \mu^2)$, with $A_1 = 4$.
\label{fig_sn2d}}
\end{figure}

  \subsection{Parametarisation of latitudinal variation in density with imposed axial symmetry}
\label{sect_sn2d}

    Starting from such 1D snapshots of the ejecta gas properties, we build 2D ejecta by
    prescribing a latitudinal variation of ejecta properties.
    We limit our investigation to (2D) axially-symmetric configurations, with reflection symmetry about
    the equatorial plane --- this is somewhat more simplistic than current polarisation studies based on 3D
    hydrodynamical inputs, but it allows a better treatment of the radiative transfer (see Section~\ref{sect_codes})
    and yields synthetic polarisation signatures that are more easily interpreted.
    Furthermore, although quite important, we do not explore the effect of small-scale inhomogeneities (i.e., $^{56}$Ni fingers, clumps;
    \citealt{fryxell_etal_91,kifonidis_etal_03}).

   The aspherical, but axially symmetric, configurations we study correspond to oblate or prolate morphologies, which may arise
   in Nature from bipolar explosions. We performed exploratory hydrodynamics simulations of such aspherical explosions
   to help understand the 2D ejecta properties. Lacking a 2D radiation-hydrodynamics code for this work,
   we build 2D ejecta using independent 1D ejecta characterised by different kinetic energies.
   These were computed with the 1D radiation-hydrodynamics code {\sc v1d} \citep{livne_93,DLW10a, DLW10b},
   starting from the s15 pre-SN model of \citet{WHW02}, and depositing an energy $E_{\rm dep}$
   at the surface of the degenerate iron core ($\sim$1.8\,\msun) over a timescale of 1\,s.
   We produced nine 1D ejecta with an asymptotic kinetic energy comprised between 0.18 and 1\,B, and
   separated by $\sim$0.1\,B.
   We then assigned the least energetic explosion model of 0.18\,B to the equatorial direction ($\theta=\pi/2$, where
   $\theta$ is the polar angle), and stacked more energetic models at constant $\theta$ increment up to the polar direction.
   At a given post-explosion time (after shock breakout and once homologous expansion is established), the
   iso-velocity curves are concentric shells about the star centre. This has to be since we take the same explosion time for all
   directions. The density and temperature distributions are however strongly latitude-dependent (Fig.~\ref{fig1_V1D_2D}).
   We find that the larger the energy, the larger the mass at large velocity. With our adopted arrangement, this produces a
   prolate outer ejecta but an oblate inner ejecta.
   Hence, the ``shape factor" (see Section~\ref{sect_codes}) reflecting the oblateness/prolateness of the ejecta
   \citep{BL77} switches with depth. The same behaviour is seen for the temperature.
   It is unclear if this property, to be expected on dynamical grounds, has been observed.

   This artificial and simplistic approach is instructive. An aspherical explosion, characterised by a given explosion time,
   has a velocity distribution that is given by $r/t$, irrespective of direction: Iso-velocity curves are concentric shells.
   Further, axially-symmetric configurations can be approximated by a radial stretching or by a latitudinal scaling of the 1D
   input from {\sc cmfgen}. In this work, we use both approaches. A radial stretching translates into a large pole-to-equator
   density contrast if the ejecta scale height is very small (e.g., as in a hydrostatic stellar atmosphere). However, in regions
   where the density profile is flat (e.g., the inner layers of a Type II SN ejecta), such a stretching yields no pole-to-equator
   density contrast and a latitudinal scaling is thus preferred.

   For a radial stretching, we distort the 1D ejecta distribution and leave all quantities unchanged on this distorted grid.
   For example, for the density $d(r,\mu)$ at radius $r$ and along $\mu=\cos\theta$,
   we have $ d(r,\mu) = d_{\rm 1D}\left( r /(1 + A_2 \mu^2)\right)$, where $A_2$ is a positive (negative) scalar
   for a prolate (oblate) ejecta morphology. All continuum and line opacities/emissivities are stretched accordingly.
   For a latitudinal scaling, the density is given by $d(r,\mu) = d_{\rm 1D}(r) \times (1 + A_1 \mu^2)$, where
   $A_1$ is a positive (negative) scalar for a prolate (oblate) ejecta morphology.\footnote{
   In practice the scaling function is actually $c(1+A_1\mu^2)$ where $c$ was chosen so that
   the integral over the scaling function is unity. Thus we scale, or stretch, about the 1D model, and
   for the simple functional dependence above  the 2D density structure at 54$^\circ$ (i.e.,  $\mu=1/\sqrt{3}$) is identical to the 1D model.
   Since the crucial factor is the pole to equatorial density contrast, we neglect the factor $c$ when quoting the scaling
   law.}
   In this case,
   opacities/emissivities have a latitudinal dependence that scales with the square of the density, except
   electron scattering whose opacity scales linearly with the density.

   Under certain ejecta conditions, stretching and scaling can yield a similar asphericity.
   Using the power-law density distribution above, we have  $(1 + A_2 \mu^2) =  (1 + A_1 \mu^2)^{1/N_d}$.
   For $N_d=10$, a density contrast of 5 between pole and equator corresponds to $A_1 = 4$ and $A_2 = 0.175$.
   In Fig.~\ref{fig_sn2d}, we illustrate the  corresponding latitudinal and radial variation of the density for each
   scaling option and using our s15e12 1D simulation at 17.9\,d \citep{DH11}.
   The two scalings produce similar deviations from spherical symmetry at large velocity (or radius)
   where the density fall-off is indeed steep.
   In the inner regions of the ejecta, where the density profile is flatter, the two scaling options yield significant differences.
   In practice, at early post-explosion times, when the photosphere resides in the outer ejecta, we use
   either scaling, but at late times, we use the latitudinal scaling only.

\begin{figure}
\epsfig{file=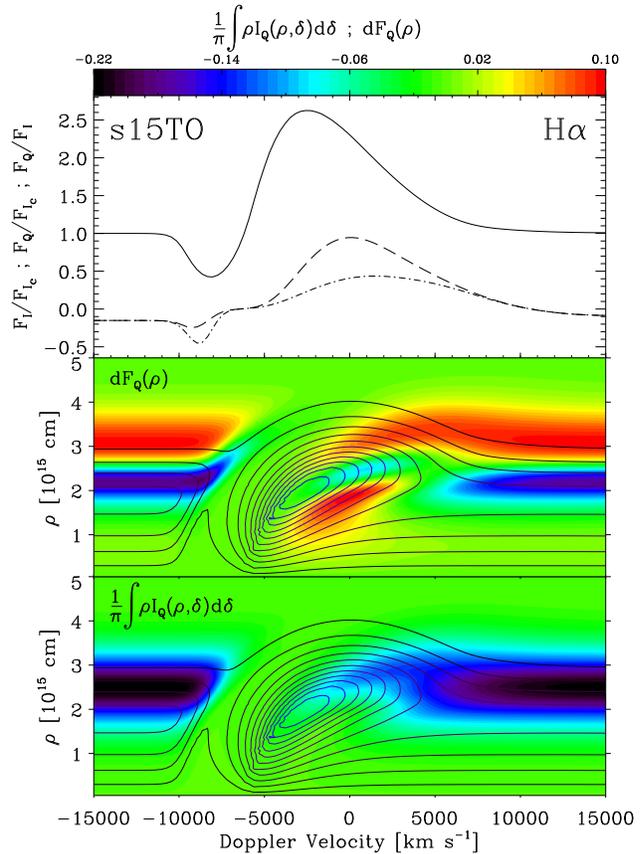,width=8.5cm}
\caption{Illustration of various radiative quantities over the H$\alpha$ region
(the x-axis shows the Doppler velocity centred on its rest wavelength)
obtained for model s15TO at 51.3\,d after
explosion and for an inclination of 90$^{\circ}$ (edge-on view of an oblate spheroid).
{\it Top:} Variation with Doppler velocity of the normalised flux $F_{I}/F_{Ic}$  (solid),
polarised flux normalised to the continuum flux $F_{Q}/F_{Ic}$ (dashed) or normalised to
the local flux $F_{Q}/F_{I}$ (dash-dotted).
{\it Middle:} Observed angle-integrated value of  $dF_Q(\ip)$.
In these lower and middle panels, we also overplot contours (shown in black) for the
angle-integrated quantity $\ip I(\ip)$, whose integral over $\ip$ yields the flux $F_I$ (at a given wavelength).
{\it Bottom:} Variation of $\bar I_Q(\ip)$ with Doppler velocity
and impact parameter $\ip$. This angle-averaged value of the local $I_Q(\ip)$ is systematically
negative and reaches a minimum at the limb of the photodisk ($\ip \sim$2.5$\times$10$^{15}$\,cm).
\label{fig_q_versus_p_freq_map}
}
\end{figure}

\begin{figure}
\epsfig{file=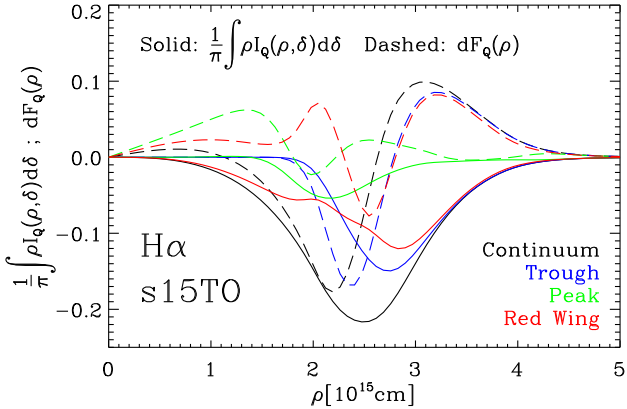,width=8.5cm}
\caption{Variation of $\bar I_Q(\ip)$ (solid) and d$F_Q(\ip)$ (dashed) with impact
parameter $\ip$ for selected locations within the H$\alpha$ line profile.
We show slices  at $-14,910$\,\kms\ (continuum; black), $-8005$\,\kms\ (trough; blue), $-2501$\,\kms\ (peak; green),
and $2001$\,\kms\ (red wing; red).
\label{fig_q_versus_p_freq_slice}
}
\end{figure}

   In the simulations presented below, we use the following nomenclature. Model s15TO is
based on model s15e12 of \citet{DH11}, sTretched radially to yield an Oblate morphology.
Similarly, model s15CP is based on model s15e12, sCaled latitudinally to yield a Prolate morphology.
We also include results for the SN 1987A model lm18a7Ad, which we call 87A (e.g., with suffix TO etc.).
For each simulation, we specify in the figures and in the text the post-explosion time and the magnitude of
imposed asphericity (controlled through $A_1$ or $A_2$).
All the results presented in this paper are converged --- increasing the resolution further
does not visibly alter the resulting Stokes parameters. Typically, we use 130 radial depth points,
150 impact parameters, 41 azimuthal angles (to cover from 0 to $2\pi$), 21 polar angles (to cover from 0 to $\pi/2$),
and 500 frequency points (to cover from $-$20,000\,\kms\ to 40,000\,\kms\
for 87A models or $-$15,000\,\kms\ to 30,000\,\kms\ for s15 models).
We compute polarisation signatures for the s15 model from 8.5\,d to 150\,d after explosion,
stepping through the sequence of \citet{DH11} at photospheric epochs, but show only
a few calculations at nebular times.

Most of the polarisation signatures we discuss in this paper were computed
with the long-characteristic code, as  it yields the Stokes parameters for a variety of inclinations
at no extra cost. However, the code currently requires an optically-thick inner boundary, preventing its
use at and beyond the end of the plateau phase. Although the Monte Carlo code operates
for a prescribed inclination, it can be used with an optically-thin inner boundary (i.e., transparent
hollow core). Hence, we use the Monte Carlo code to produce polarisation signatures
during the nebular phase - these are discussed near the end of the paper in Section~\ref{sect_time}.

\section{Polarisation signature for one reference model/case}
\label{sect_s15TO}

In this section, we discuss in detail the total and polarised flux for model s15TO at 51.3\,d
after explosion. We produce an oblate ejecta by means of a 25\% stretching ($A_2=-0.25$) in the
radial direction. To facilitate the interpretation  we adopt an inclination of 90$^{\circ}$ (i.e.,  we observe
the axisymmetric ejecta edge-on).
In the top panel of  Fig.~\ref{fig_q_versus_p_freq_map}, we present the normalised H$\alpha$ flux $F_I/F_{Ic}$ at this epoch
(solid), together with the polarised flux $F_Q/F_{Ic}$ (dashed) and $F_Q/F_I$ (dash-dotted).
The wavelength coverage includes the neighbouring continuum regions. Despite the sizeable
asphericity, the polarisation is below 1\% both in the continuum and within the line (irrespective
of the normalisation to $F_I$ or $F_{Ic}$).

The polarisation shows a complicated behaviour across the P-Cygni line profile.
$F_Q/F_{Ic}$ varies from $\sim$$-0.15$\% in the continuum, to $\sim$$-0.2$\% in the P-Cygni
trough, becomes zero at a Doppler velocity of $\sim$$-6000$\,\kms, peaks at 0.9\% in the red wing of the profile, before
gradually returning to the continuum value of $\sim$$-0.15$\% at a Doppler velocity of $\sim$15,000\,\kms.
In the $Q-U$ plane, this excursion would correspond to a flat loop along the line $U=0$. Observationally, it
would correspond to linear excursions in the $Q-U$ plane offset by interstellar-polarisation, and
at an angle set by the orientation of the ejecta on the sky.
The lower panels in Fig.~\ref{fig_q_versus_p_freq_map} help explain the origin of this low polarisation.

\begin{figure*}
\epsfig{file=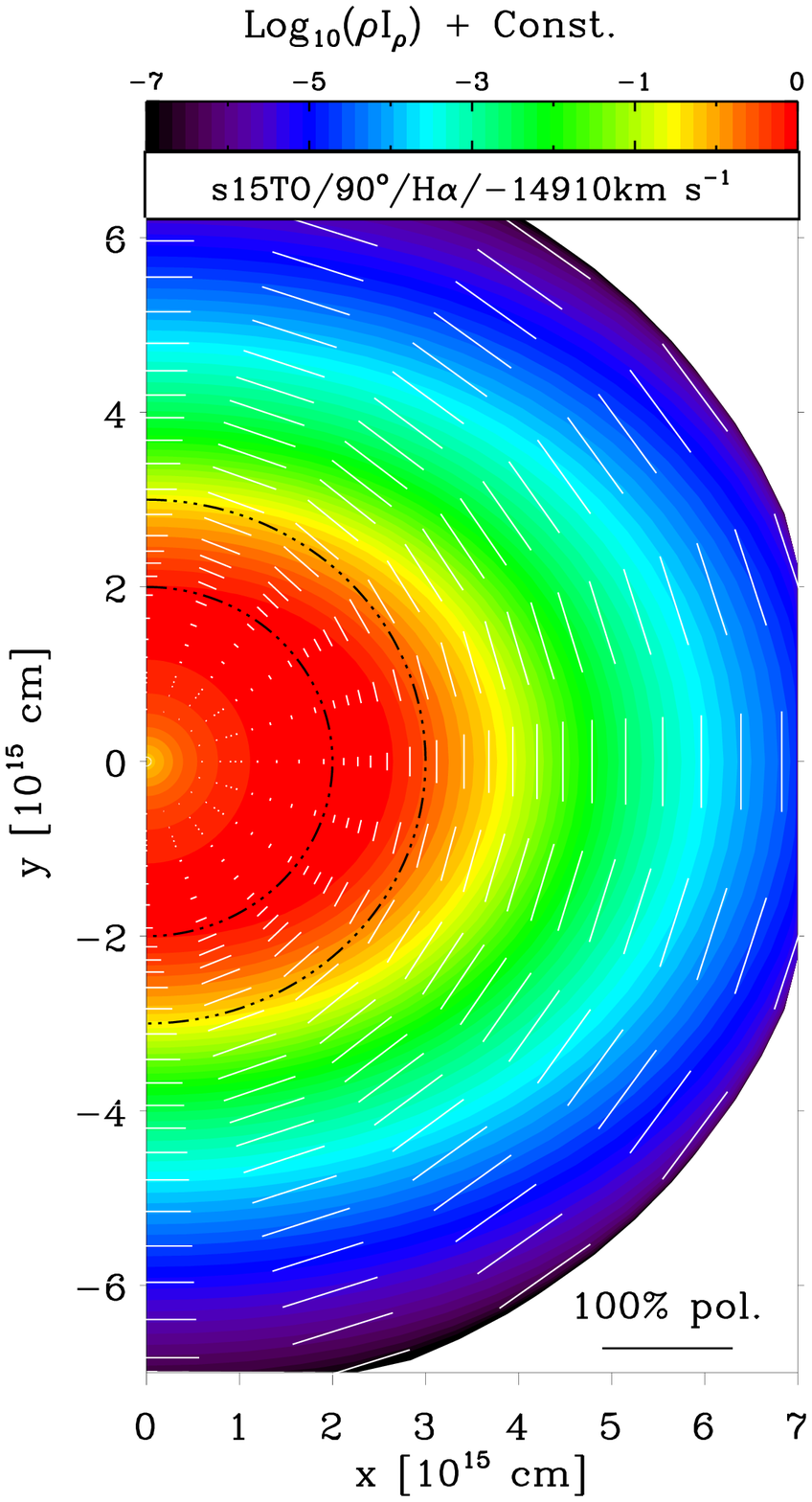,width=4cm}
\epsfig{file=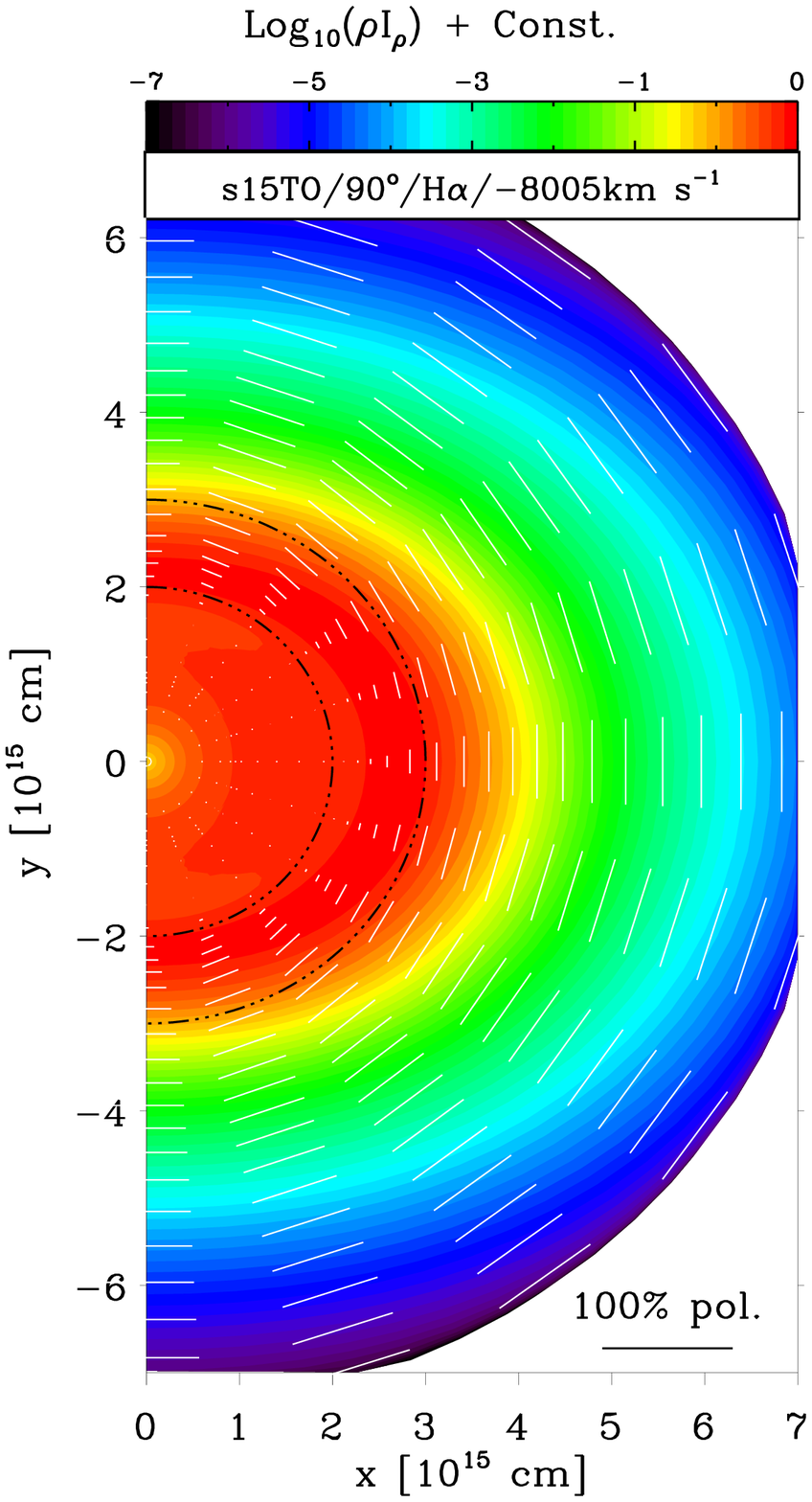,width=4cm}
\epsfig{file=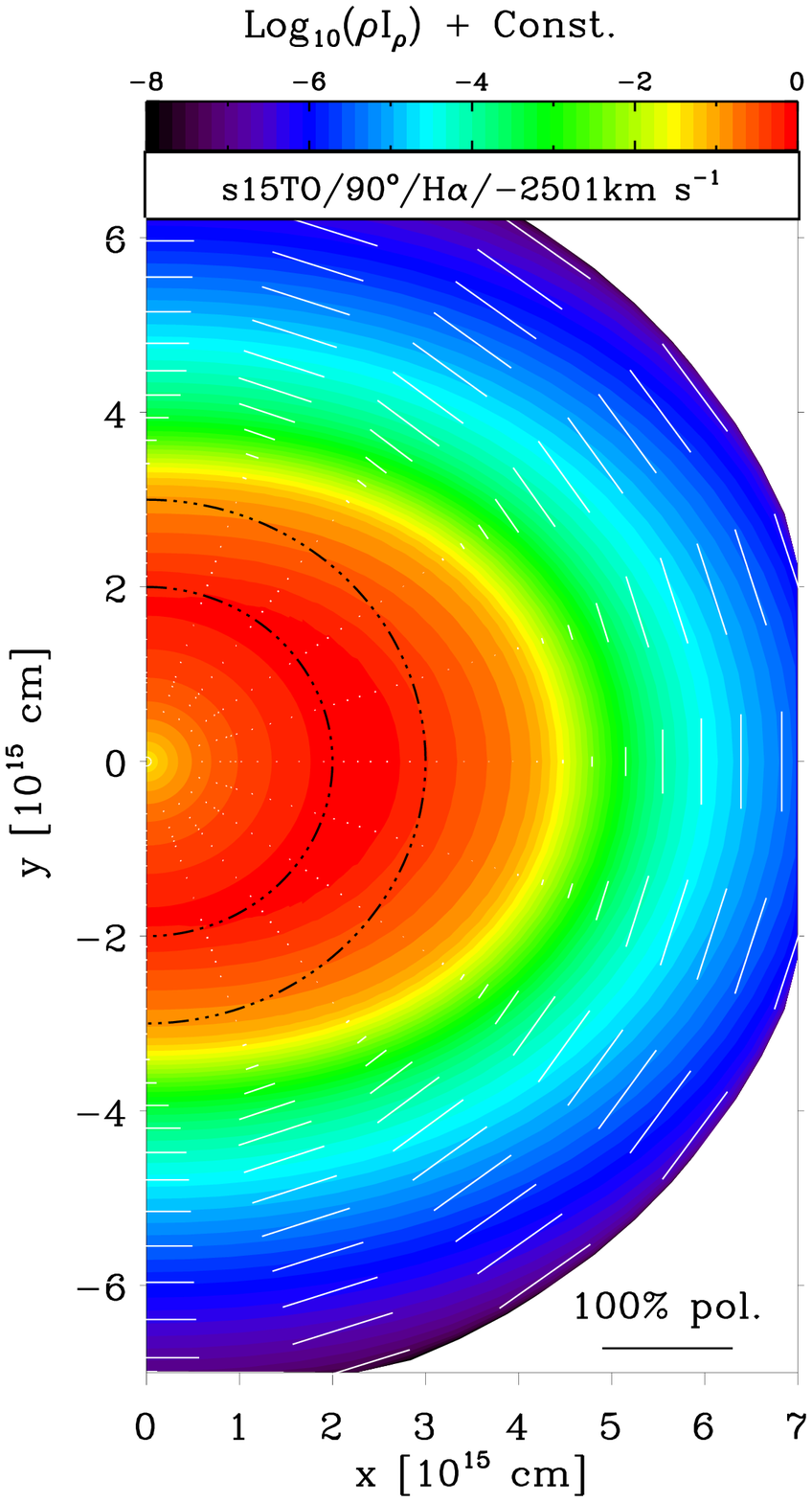,width=4cm}
\epsfig{file=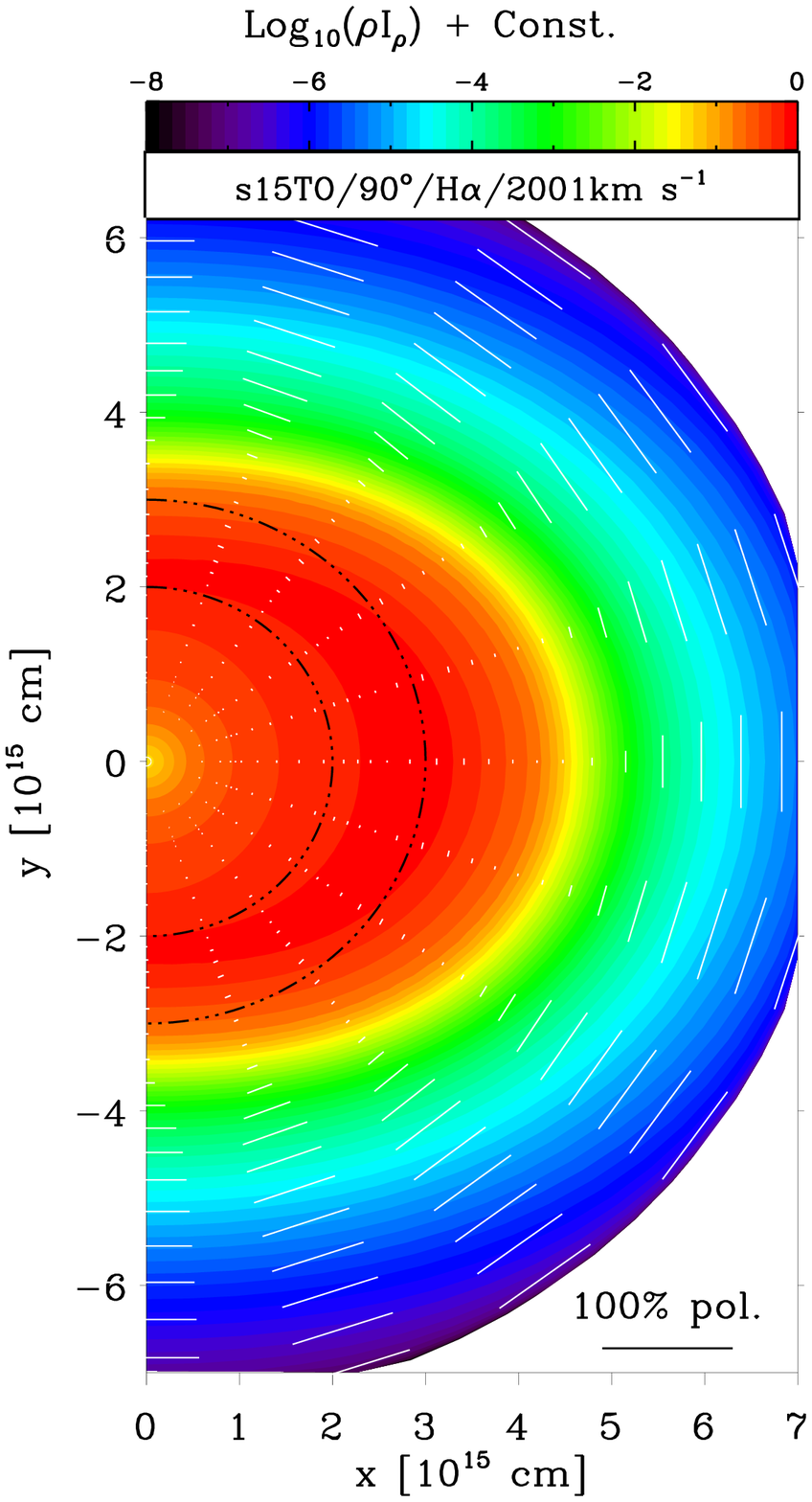,width=4cm}
\caption{Colour map of the specific intensity (scaled by the impact parameter $\ip$)
on the plane of the sky for the s15TO model on day 51.3 and seen at
a 90$^{\circ}$ inclination (i.e., edge on) and characterised by a 25\% radial stretching ($A_2=-0.25$).
We overplot segments (white bars) showing the local polarisation strength and orientation
(the length corresponding to 100\% polarisation is shown at the bottom right of each panel).
From left to right, we show the results across the H$\alpha$ line profile (as 
in Fig.~\ref{fig_q_versus_p_freq_slice}), i.e.,
the continuum blueward of the line at  $-14,910$\,\kms\ (from line center),
the H$\alpha$ trough at $-8005$\,\kms, the H$\alpha$ peak at $-2501$\,\kms,
and the red wing of H$\alpha$ at 2001\,\kms. The two semi-circular black contours in
each figure are provided to help illustrate the departure from circular symmetry.
\label{fig_2D_pol}}
\end{figure*}

The lower panel depicts
$\bar I_Q(\ip)$  which gives the variation with $\ip$ of the local value of $I_Q(\ip,\delta)$ across the profile.
As the polarisation is perpendicular to the radial direction, $I_Q(\ip,\delta)$ is negative, irrespective of
$\delta$ and wavelength.\footnote{In reality the polarisation is generally not perfectly perpendicular to
the radial vector (i.e.,  $I_U$ is non-zero) but in practice the departure is not large (see Fig.~\ref{fig_2D_pol})
and will not affect our qualitative discussions.}
 The magnitude of $\bar I_Q(\ip)$ varies considerably with $\ip$, peaking at $\sim$$-0.22$
at the edge of the photodisk ($\ip \sim$2.5$\times$10$^{15}$\,cm). The quantity $dF_Q$, which gives the observed
angle average of the polarisation at $\ip$ on the plane of the sky with respect to a fixed set of axes,
shows both negative and positive values (we show specific slices of $\bar I_Q(\ip)$ and $dF_Q$ in
Fig.~\ref{fig_q_versus_p_freq_slice}). When performing the integral over $\ip$, the strong
cancellation yields a small negative residual continuum polarisation. The line polarisation is
negative on the blue side, but becomes positive as we move towards the red wing.

SN polarisation is often discussed in the context of an optically thin scattering nebula illuminated by a central source.
For the same oblate geometry, allowance for scattering in the continuum, and
a line like H$\alpha$ forming over a large volume above that central source,
the resulting polarisation would be positive throughout the continuum and the line, merely
reduced across the profile due to flux dilution.
Instead we obtain a negative continuum polarisation and a polarisation sign reversal  across the profile
--- this behavior arises from  the specificities of line and continuum formation in SN ejecta.

In the bottom and middle panels of Fig.~\ref{fig_q_versus_p_freq_map}, we show contours
of the quantity $\ip I(\ip)$ which, when integrated over impact parameter $\ip$, gives the total flux at the
corresponding Doppler velocity (wavelength). As discussed in \citet{DH05_qs_SN,DH05_epm},
the steep ejecta density fall-off at the photosphere at that time causes the line and continuum flux
to arise from a similar spatial region. In fact, a significant fraction of the H$\alpha$ line flux arises primarily from regions
that are optically thick ($\tau \gtrsim 1$) in the electron scattering. The bulk of the SN radiation, whether line or continuum, comes primarily
from the photodisk --- an imaginary disk on the plane of the sky with a radius on the order of $R_{\rm phot}$.
There is little contribution from the side lobes, in contrast to P-Cygni profiles formed in mass-losing
stars like P-Cygni or Wolf-Rayet \citep{DH05_qs_SN}. This property tends to produce low polarisation
values because the observed flux is strongly {\it diluted} by a dominant contribution from forward-scattered
photons originating from the photo-disk and thus essentially unpolarised. It is the reduced contribution of such
unpolarised photons that causes the polarisation excess in P-Cygni troughs.

As noted in Section~\ref{sec_line_form}, there is a significant probability that H$\alpha$ line photons scatter with
electrons at least once before escaping. However the effective number of scatterings undergone by the
observed line photons is significantly less than that undergone by the observed continuum photons. This,
combined with the larger extension of the H$\alpha$ formation region results in the positive
polarisation (in the red portion of the line) that is expected for the adopted oblate morphology and edge-on view.

  \begin{figure}
   \epsfig{file=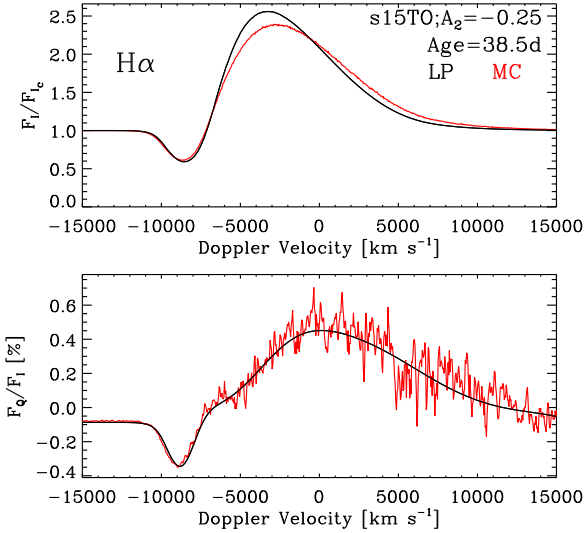,width=9cm,scale=1.15}
   \caption{Comparison between the normalised H$\alpha$ synthetic flux $F_I/F_{Ic}$ (top) and the corresponding normalised
   polarisation flux $F_Q/F_{I}$ (normalised to the local flux and given as a percentage; bottom),
   computed with the 2D long-characteristic solver (LP; black)
   and the Monte Carlo code (MC; red; 20 million photons were used; the results are rebinned on a 100\,\kms\ velocity grid
   to reduce the apparent noise).
   The model is s15TO at 38.5\,d after explosion,
   i.e., the s15 model stretched radially by 25\% ($A_2 = -0.25$) to yield an oblate spheroid.
   Ejecta properties are identical in both cases. For reasons unclear at present,
   the solution from the Monte Carlo code has to be globally shifted vertically, here by 0.02\%, to match the continuum polarisation
   level obtained with the 2D long-characteristic code. Apart from this global, but small shift, the agreement is very good.
   \label{fig_comp_mc_lp}}
   \end{figure}

To illustrate further how polarisation cancellation operates, we show intensity and polarisation
maps in Fig.~\ref{fig_2D_pol} for the four Doppler velocities selected in Fig.~\ref{fig_q_versus_p_freq_slice}.
For each panel, we show a colour map for the quantity $\ip I(\ip)$ at each of the four Doppler velocities,
from continuum, to the trough, the peak, and the red wing of the line profile.
In each panel, we normalise $\ip I(\ip)$ to the maximum within that map (the relative difference in
intensity between these panels can be inferred using the top panel of Fig.~\ref{fig_q_versus_p_freq_map}).
We overplot black circles which reveal the oblateness of the intensity contours.
We also draw polarisation bars, with a scale corresponding to 100\% given at the bottom right of
each panel.

In all panels, the highest polarisation is achieved for regions at large impact parameters (the maximum
is 89\% polarisation; at such impact parameters, 90$^{\circ}$ scattering of photons emitted from a central point source
would produce a 100\% polarisation). However, these external regions contribute very little flux
and thus do not influence the observed polarisation. Instead, the polarisation is primarily
controlled by the inner regions, which have a smaller intrinsic polarisation but contribute a much
larger fraction to the observed flux.
In the two panels on the left, we see that the maps and polarisation segments have similar distributions,
with the exception that in the P-Cygni trough (second panel from left), the amount of unpolarised
light forward scattered or directly escaping to the observer is reduced, yielding an enhanced
polarised flux. Here, the cancellation works the following way. For $\ip \sim 2 \times$10$^{15}$\,cm,
the intensity and the polarisation is larger for polar regions, biasing the observed polarisation towards negative values
(polarisation vector perpendicular to polar axis).
For $\ip \sim 3 \times$10$^{15}$\,cm, the polarisation is larger for polar regions, but the intensity is larger
along the equatorial regions, biasing the observed polarisation towards positive values
(polarisation vector parallel to polar axis).
At the flux peak (which is blueshifted by 2500\,\kms) and in the red wing of the line profile, the polarisation
tends to rise only at large impact parameters, with the largest flux contribution from the equatorial regions
which bias the polarisation towards positive $F_Q$. The intensity map is clearly oblate in those Doppler-velocity
slices, and the effect of electron-scattering of line photons is key to yield the positive residual polarisation.

From the analysis of this case, we find that the polarisation of such an oblate ejecta
is small for several reasons. The line and continuum formation regions  in such SNe II-P models
overlap, which causes strong optical-depth effects, significant scattering of line photons, little
emission from the side lobes (i.e.,  which would mimic a nebula). Multiple scattering also acts to depolarise the light.
Since there is no central radiating source the
simplistic view of a central source shining on an extended aspherical nebula is particularly unsuited to
describe the polarisation conditions in aspherical SN II ejecta.
At this time, this makes the intrinsic polarisation of SN II light small by design.
Added to this, strong cancellation over the plane of the sky reduce this polarisation further.
Given these conditions, even for significant departures from spherical symmetry, such SN-II ejecta
seem unfit to produce strong polarisation signatures. An important corollary is that a small
observed polarisation does not necessarily imply a small departure from spherical symmetry.

Before wrapping up this section, we emphasize that the results obtained with the long-characteristic
code are closely matched by the independent Monte Carlo code for the same ejecta conditions
(Fig.~\ref{fig_comp_mc_lp}).
This provides a check on each code and further grounding for our results.

\section{polarisation dependencies}
\label{sect_dependencies}

  Having laid out the basic features controlling the polarisation for one reference case, we
  now discuss its dependence on inclination, shape factor, wavelength and line identity,
  and time after explosion.

\begin{figure*}
 \epsfig{file=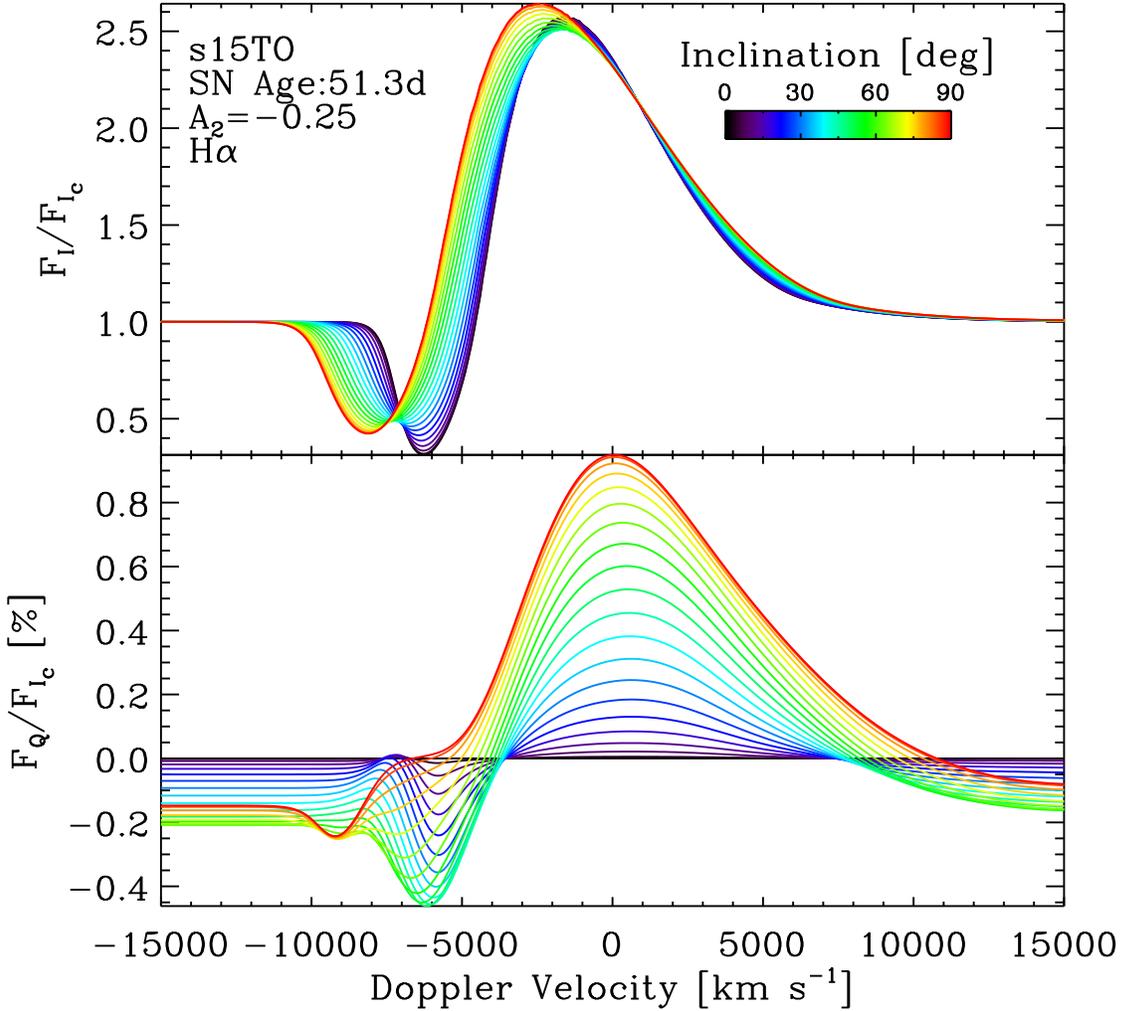,width=15cm}
 \caption{Inclination dependence (colour coded) of the flux $F_I/F_{Ic}$ (top) and polarised flux $F_Q/F_{Ic}$
 (bottom) for the model s15TO (age of 51.3\,d, 25\% radial stretching).
 [See Section~\ref{sect_incl} for discussion.]
  \label{fig_s15TO_incl} }
\end{figure*}

\subsection{Dependency of polarisation on inclination}
\label{sect_incl}

  The edge-on view of the oblate ejecta for model s15TO discussed in the previous
  section was selected for simplicity (e.g., apparent top-bottom symmetry).
  However, the long-characteristic code outputs
  the inclination-dependent intensities naturally since the radiation field is solved for along
  all 21 polar angles (each corresponding to a different inclination).

  In Fig.~\ref{fig_s15TO_incl}, using the same reference model s15TO as in Section~\ref{sect_s15TO},
  we show the variation of the total normalised flux $F_I/F_{Ic}$ (top panel) and the polarised flux
  $F_Q/F_{Ic}$  (bottom panel) versus inclination (indicated by a colour coding).
  The inclination dependence is strong for both quantities.
  While the peak flux $F_I/F_{Ic}$ does not vary much, the absorption and the emission parts of the
  P-Cygni profile shift monotonically to the blue for larger inclinations. The location of the peak
  flux varies by as much as 1000\,\kms\ and the location of the maximum absorption by as much as
  2000\,\kms, corresponding to a maximum change of $\sim$30\%. This has important
  implications for inferences of the expansion rate and the ejecta kinetic energy (Section~\ref{sect_line_profile}).
  Similarly, the total flux varies considerably with inclination (not shown), since for a pole-on
  view the oblate ejecta offers a much larger radiating surface than for an edge-on view.
  In the present case, the continuum flux is 50\% larger for a pole-on view than for an
  edge-on view. We discuss the implication for the inference of the SN luminosity in Section~\ref{sect_lbol}.

\begin{figure*}
\epsfig{file= 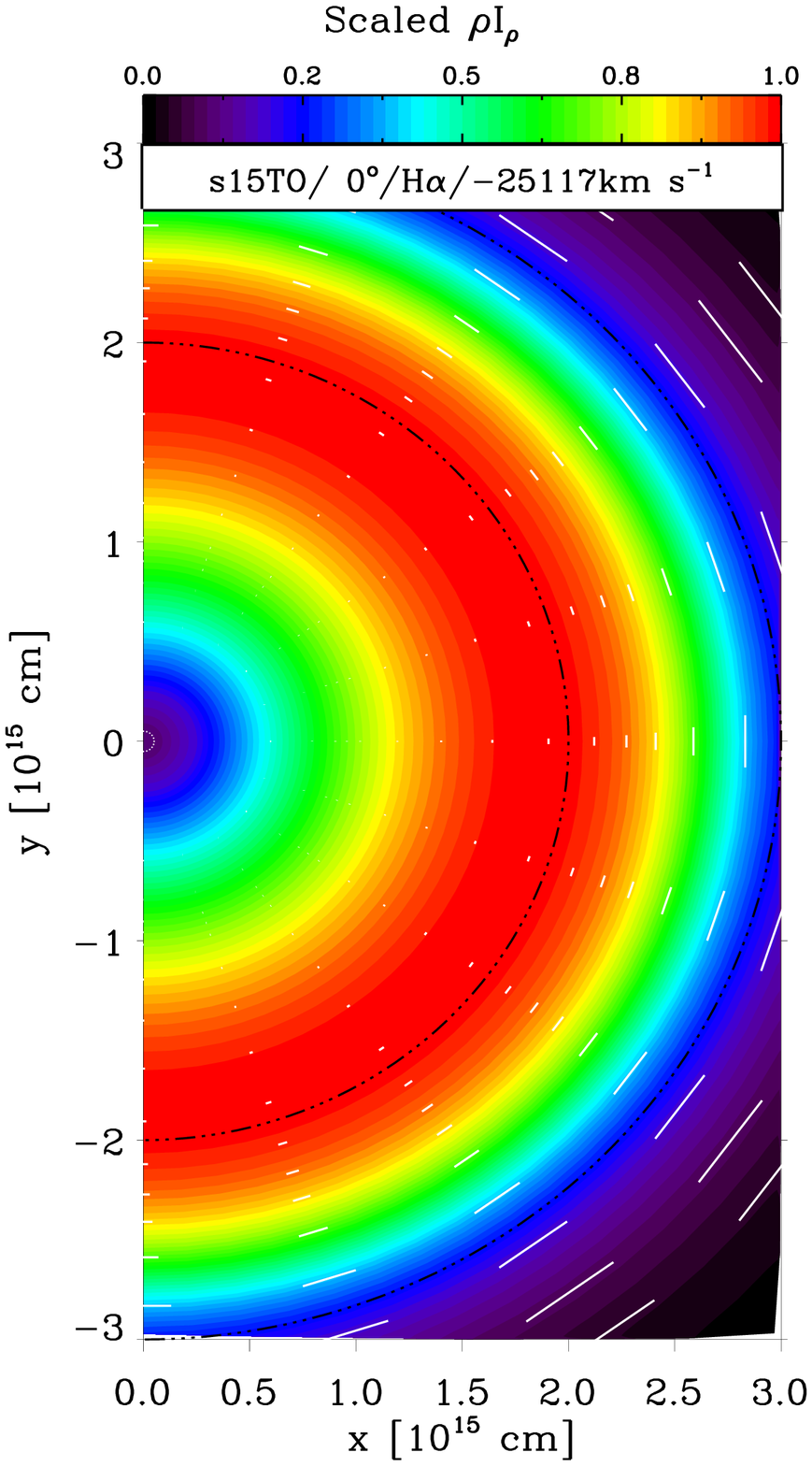,width=3.25cm}
\epsfig{file= 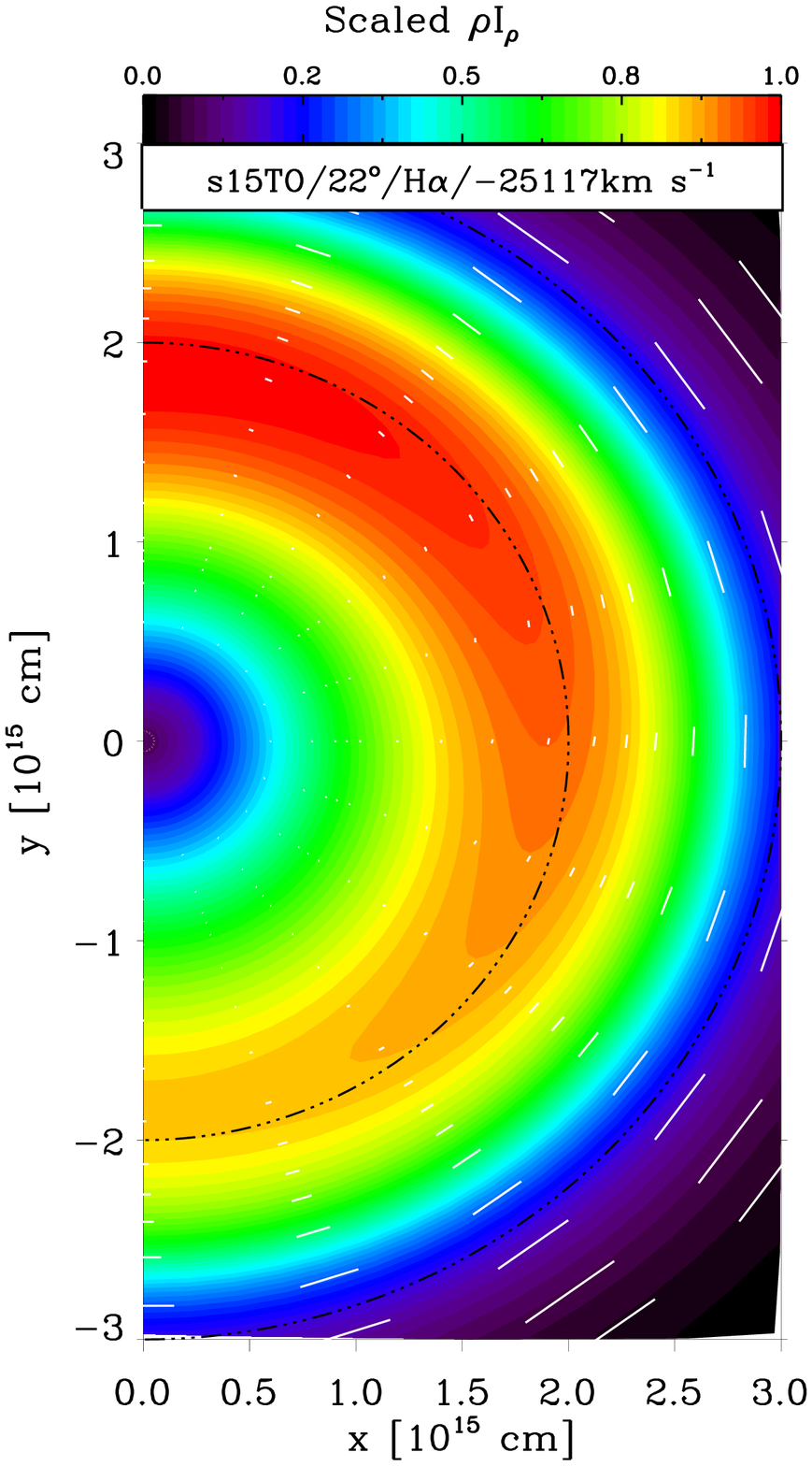,width=3.25cm}
\epsfig{file= 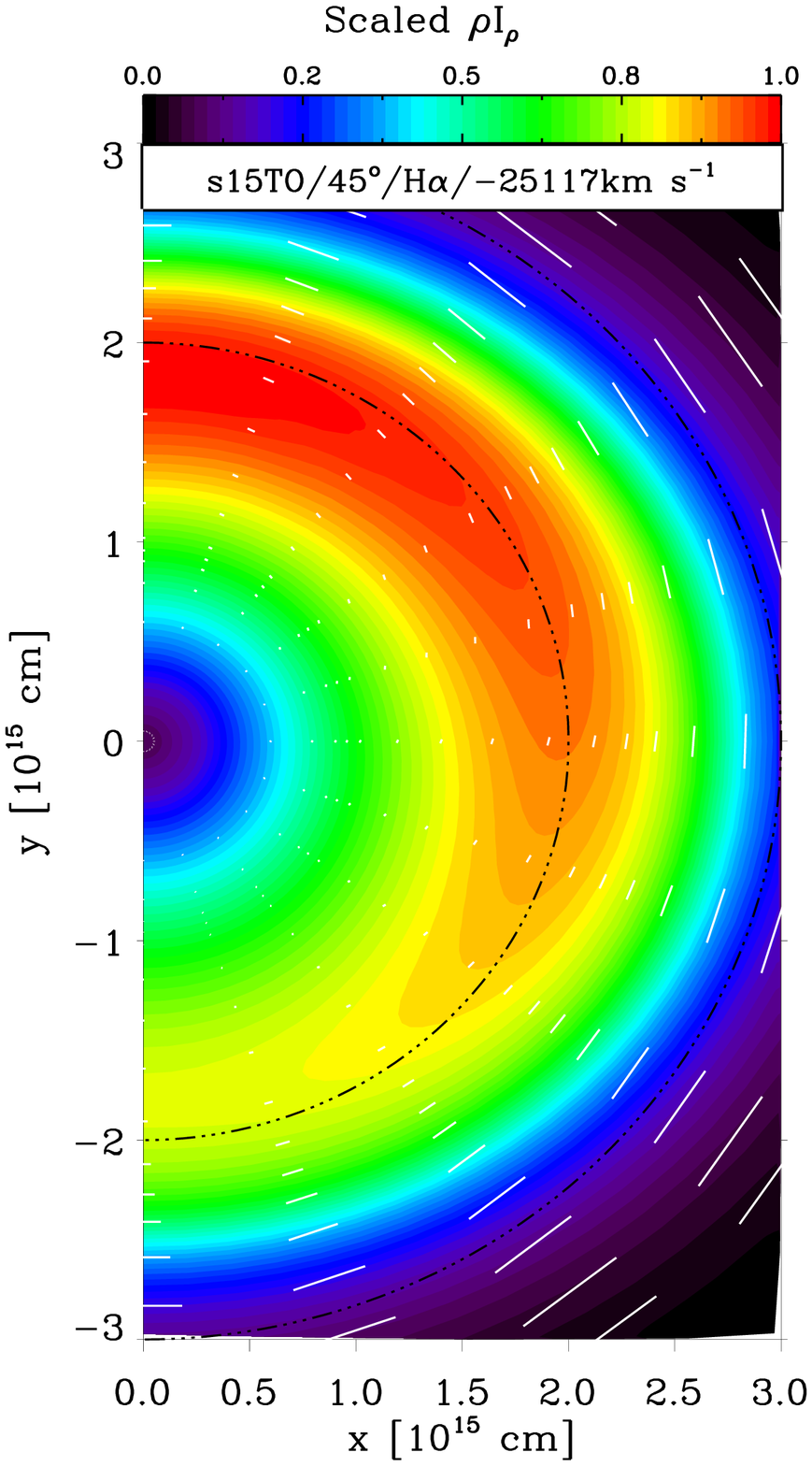,width=3.25cm}
\epsfig{file= 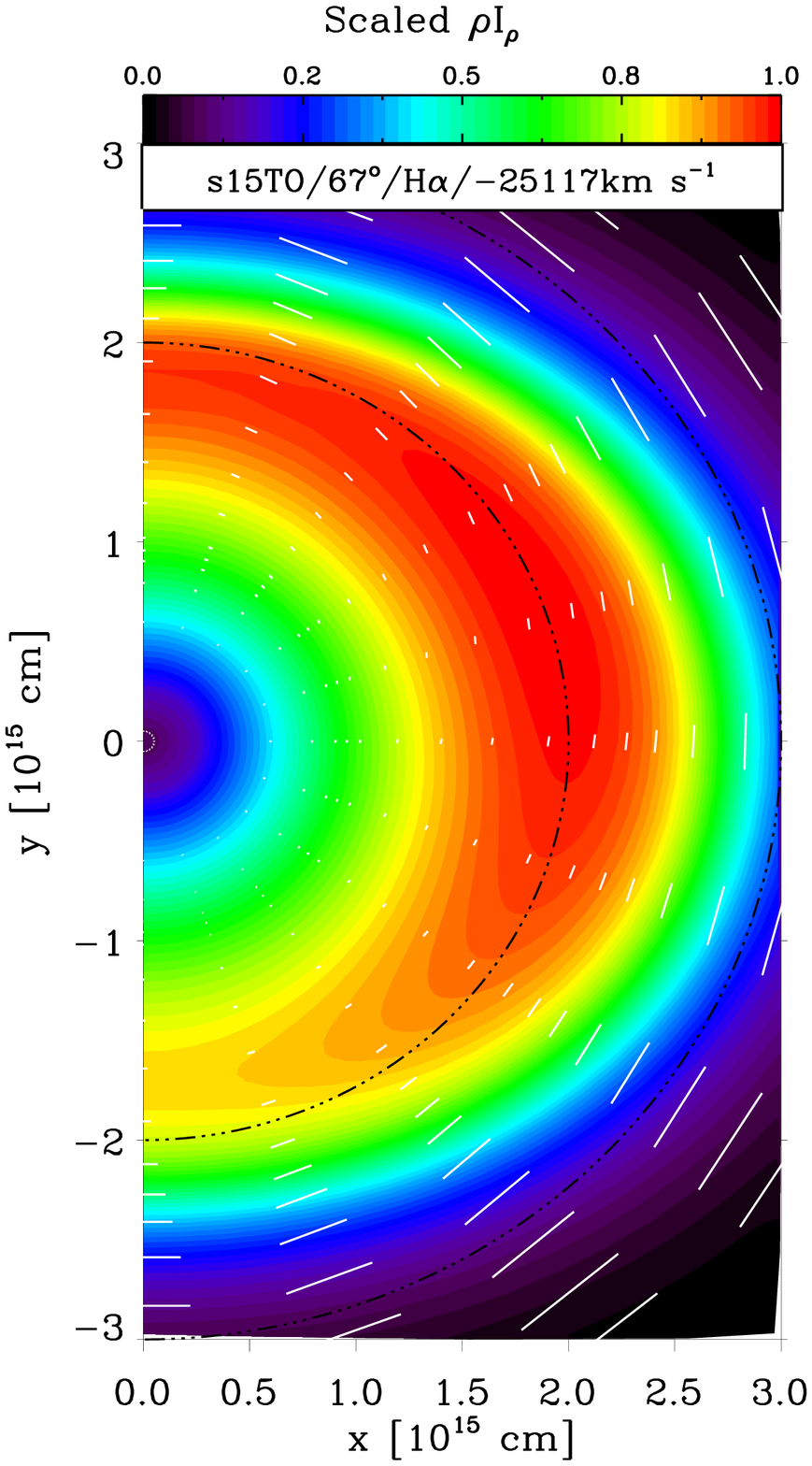,width=3.25cm}
\epsfig{file= 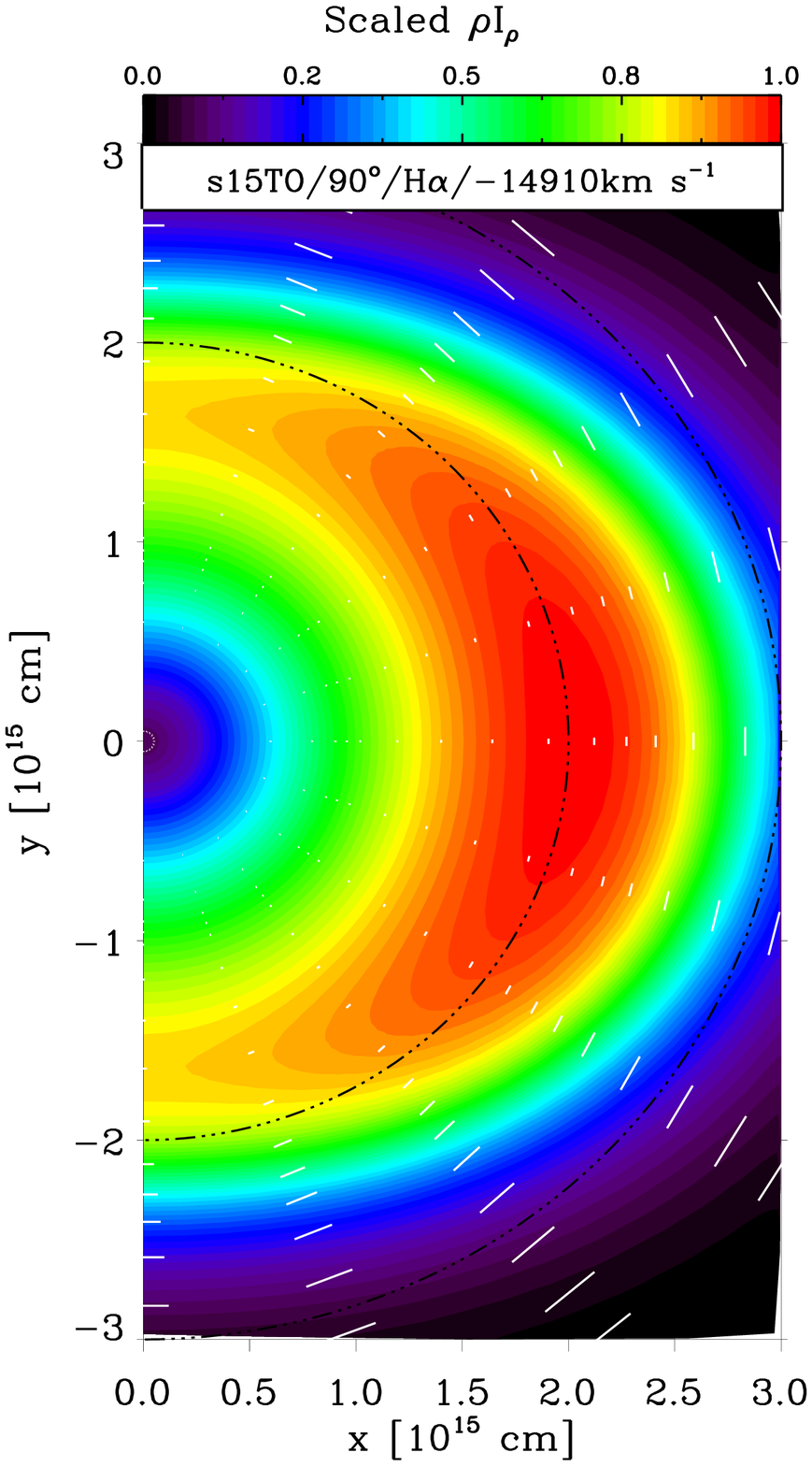,width=3.25cm}
\caption{Illustration of the {\it inclination dependence} of the continuum polarisation
(chosen at $\sim$$-$25,000\,\kms\ from the H$\alpha$ line center)
for model s15TO (oblate spheroid) at 51.3\,d after explosion.
In each panel, we show a colour map of the specific intensity (scaled by the impact parameter $\ip$
and normalised to a maximum value of one).
We overplot segments (white bars) showing the polarisation-intensity magnitude and orientation
(the maximum polarisation magnitude is $\sim$90\% at large $\ip$).
From left to right, the inclination is varied from 0$^{\circ}$, to 22.5$^{\circ}$, 45$^{\circ}$,
67.5$^{\circ}$, and 90$^{\circ}$.
\label{fig_s15TO_cont_incl}}
\end{figure*}

  The variation of the polarised flux $F_Q/F_{Ic}$ with inclination reflects in part those seen in the total flux. The location
  of maximum and minimum polarisation across the line reflects the wavelength shifts of the trough and peak.
  More importantly, the polarisation varies differently across the profile. Its magnitude varies
  non-monotonically with inclination, being an extremum in the continuum and in the trough for an inclination
  of 60$^{\circ}$.
  What causes the reversal are subtle variations in the cancellations with inclination, with different
  regions, primarily within or at the edge of the photodisk, contributing varying levels of polarisation
  (i.e., the magnitude of the normalised polarisation)
  as well as different intensities (i.e., how much flux comes from different regions on the plane of the sky).
  This supports the finding  from the previous section that the polarisation is not born in an
  optically-thin aspherical nebula, but instead from a medium that is optically thick and thus both
  emitting and scattering.

  At the profile peak and in the red wing, the variation is monotonic and reaches a maximum
  for a 90$^{\circ}$ inclination. Interestingly, in those profile regions, the variation of  $F_Q/F_{Ic}\sin^2 i$
  is essentially degenerate, as obtained for a scattering nebula surrounding a central source  \citep{BL77}. Here,
  the source of radiation is made of line and continuum photons emitted from interior regions.

  To illustrate this inclination variation in the continuum,
  we show in Fig.~\ref{fig_s15TO_cont_incl} maps of the specific intensity $I(\ip,\delta)$ (scaled by the impact parameter) and
  polarisation strength and orientation for inclinations of 0$^{\circ}$, 22.5$^{\circ}$, 45$^{\circ}$, 67.5$^{\circ}$, and
  90$^{\circ}$ (ordered from left to right).
  A pole-on view of an oblate spheroid yields zero polarisation due to full cancellation.
  As inclination is increased, the cancellation is incomplete and operates as described in the previous section.
  Compared to the edge-on view case, there is now a bias due to the varying flux across the plane of the sky,
  with a maximum in the north pole regions for small inclinations and shifting to the equatorial regions at large
  inclinations. This shift is in part what causes the non-monotonic behaviour of the continuum polarisation.

\subsection{Dependency of polarisation on shape factor}
\label{sect_shape_factor}

   A major goal of polarisation observations is to infer the level and nature of asymmetry
   of a given SN ejecta. For our axially-symmetric configuration, we can investigate
   oblate and prolate configurations, and in each case vary the magnitude of the asphericity
   by modulating the pole-to-equator contrast.

\begin{figure*}
 \epsfig{file=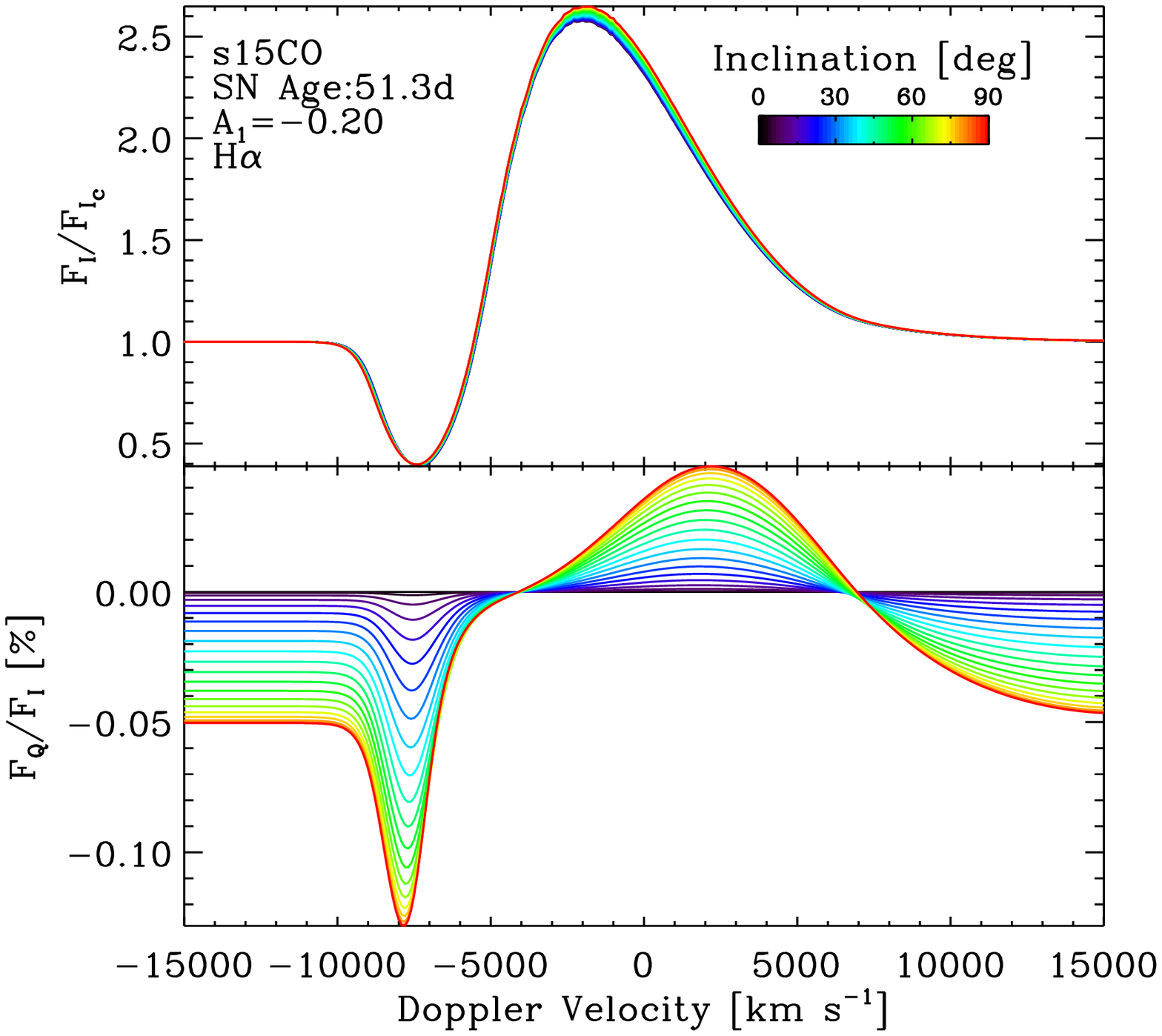,width=7.5cm}
 \epsfig{file=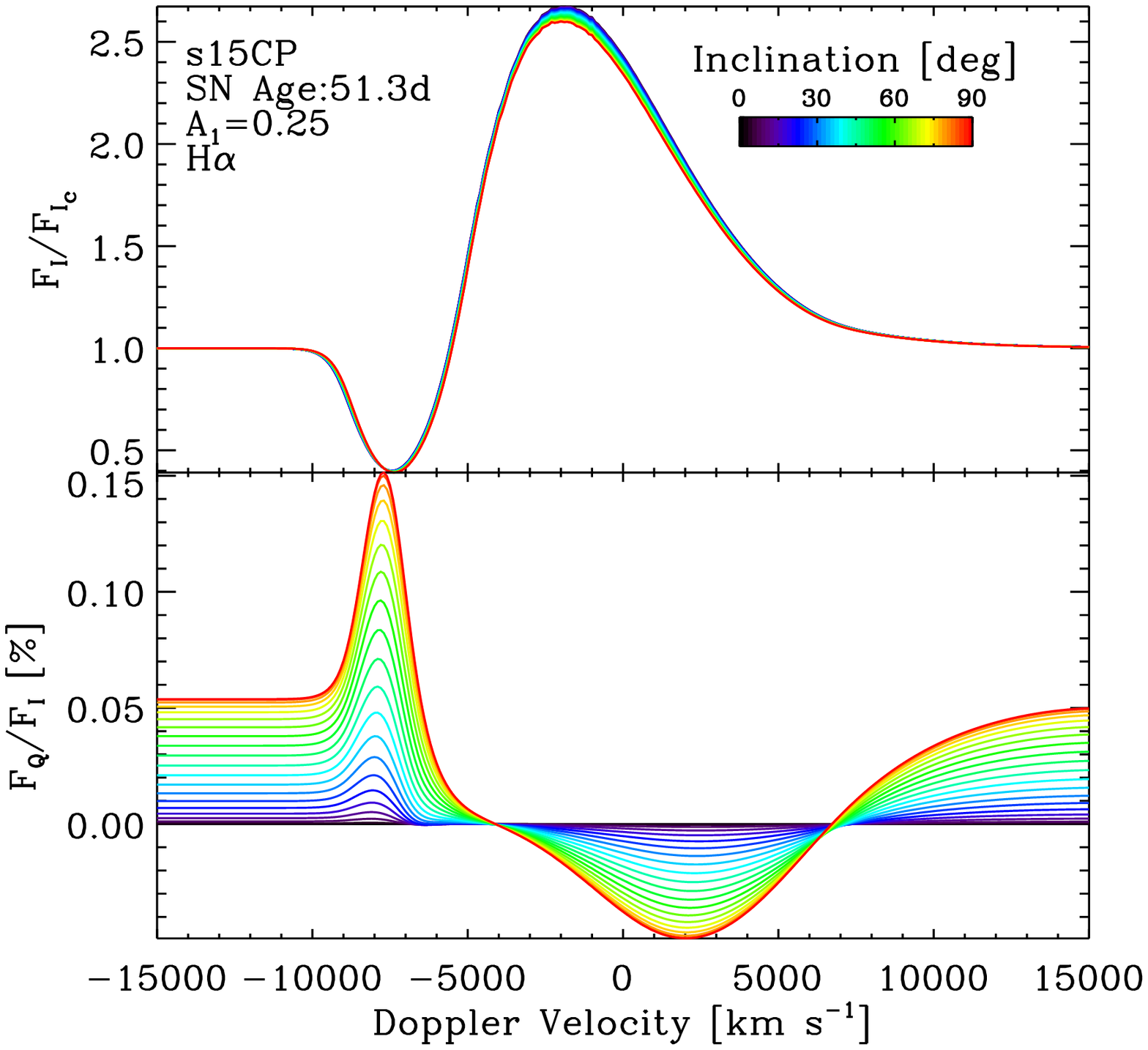,width=7.5cm}
 \epsfig{file=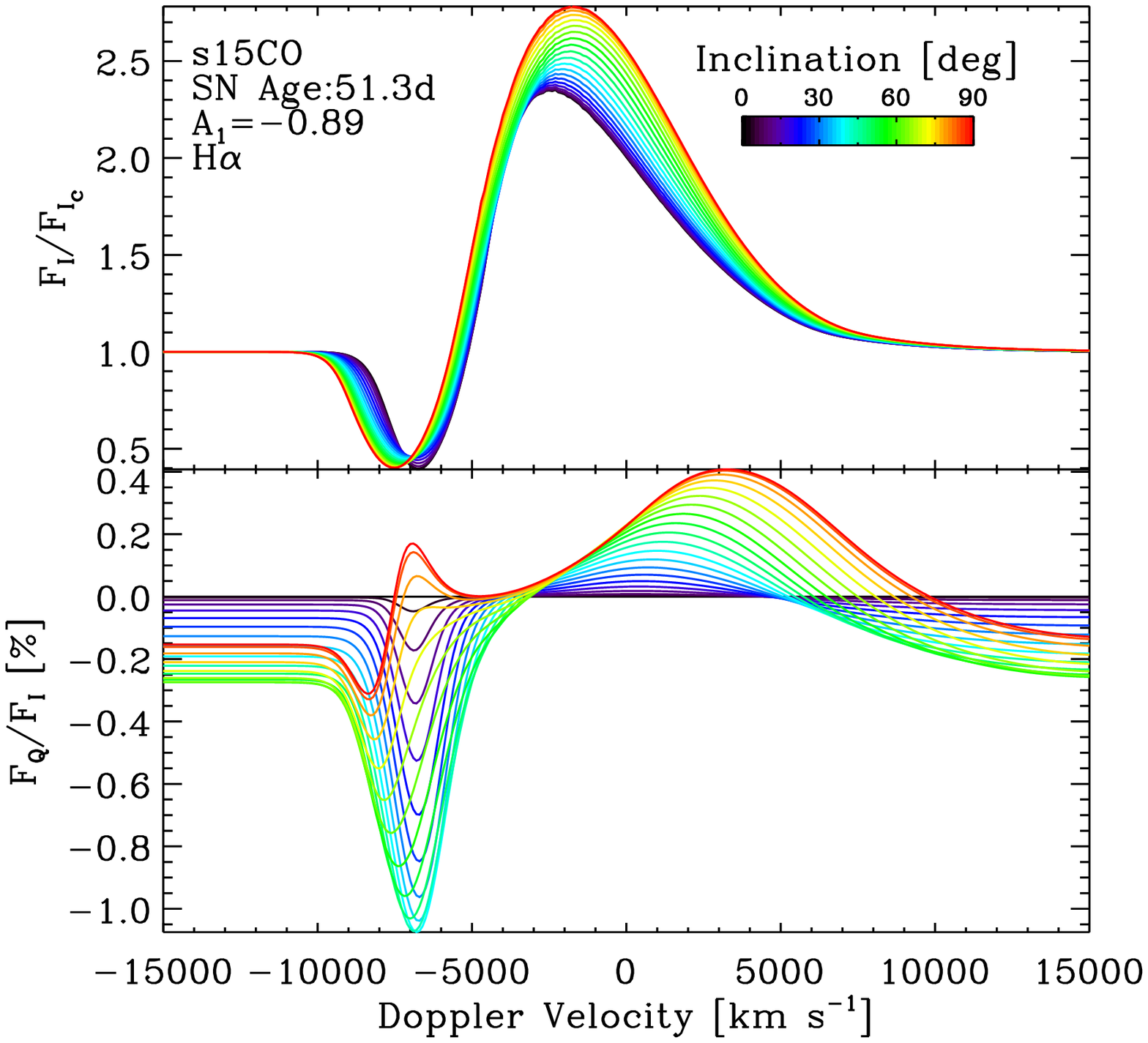,width=7.5cm}
 \epsfig{file=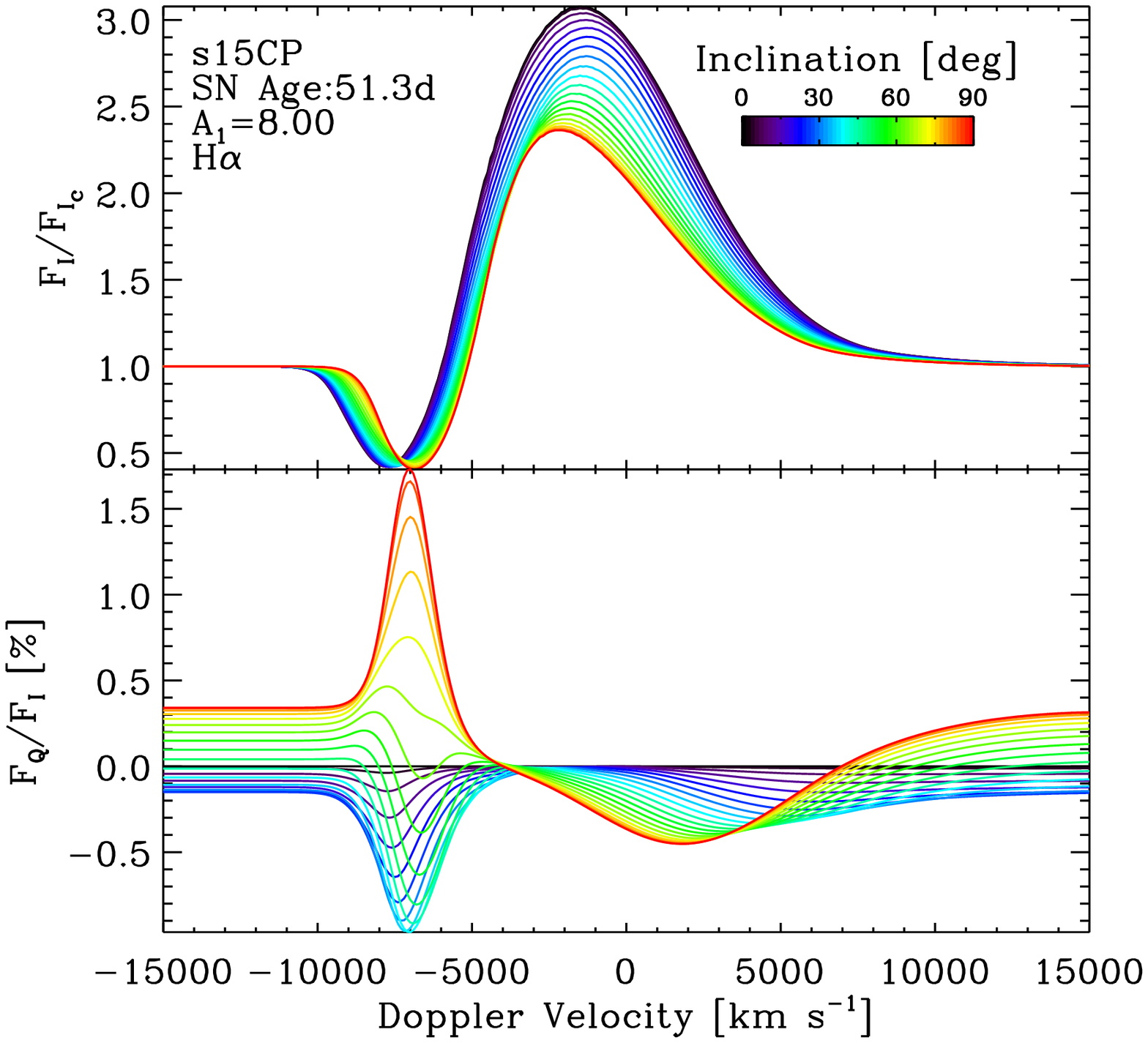,width=7.5cm}
\caption{{\it Left column:}
Synthetic H$\alpha$ normalised line flux (upper panels) and normalised linearly-polarised
 flux (lower panels) based on model s15e12 at 51.3\,d after explosion, and shown for
 an increasing oblateness corresponding to $A_1=-0.2$ (top) and -0.89 (bottom).
 {\it Right column:} Same as left column, but now in the right panel of each row, we show synthetic profiles
 for a model having the inverse of the pole-to-equator density ratio shown in the left panel.
 Notice how this switch from oblateness to prolateness  (with equivalent density contrast between pole and equator)
 has for main effect to flip the sign of the polarisation,
 with subtle differences occurring within the region of P-Cygni profile absorption.
 \label{fig_oblate_prolate}
}
\end{figure*}

   We explore this dependency with H$\alpha$ for models s15CO and s15CP,  adopting a density
   scaling with latitude to yield an oblate or a prolate spheroid. In Fig.~\ref{fig_oblate_prolate},
   we show simulations corresponding to a pole-to-equator ratio of 1.25 (top row) and 9 (bottom
   row).
   In the former case, the asymmetry is small and the variation of the resulting total flux
   is negligible. The polarised flux is very small, with a maximum magnitude of $\sim$0.1\%
   in the P-Cygni trough. Its variation is monotonic with inclination but we observe the
   same sign reversal across the line profile as in model s15TO (Section~\ref{sect_s15TO}).
   Switching here from an oblate to a prolate configuration merely flips the polarisation:
   the resulting curves are mirror images of each other.
   As before, outside of the peak and red wing region, the polarisation has a sign that is
   opposite to that obtained for an optically-thin nebula of the same shape, being negative (positive)
   for an oblate (prolate) configuration.

   In the bottom row of Fig.~\ref{fig_oblate_prolate}, we show the corresponding results when
   the adopted asymmetry yields a pole-to-equator density contrast of 9.
   The effect on the total flux is now no longer negligible.
   Furthermore, as discussed for the reference model in Section~\ref{sect_s15TO}, cancellation
   effects complicate the polarisation measure, yielding non-monotonic and even sign-reversal
   of the polarisation in the continuum and in the trough. The peak and red-wing
   regions are, however, better behaved and show a monotonic evolution of their polarisation with inclination.
   Maximum polarisations can be larger, in particular in the profile trough (model s15CP with $A_1 = 8$),
   but cancellation effects can also conspire to make it small  (model s15CO with $A_1 = -0.89$).

   Increasing/decreasing the pole-to-equator density ratio from 1.25 to 9 thus yields very ambiguous results
   for the continuum and P-Cygni absorption regions, with sign reversals and non-monotonic behaviour.
   Interestingly, the polarisation in the peak and red wing regions follows a gradual and monotonic trend,
   increasing by about a factor of ten (from $\pm$0.05 to $\pm$0.4\%).
   The polarisation values in those spectral regions are of the same magnitude, but opposite sign,
   between corresponding oblate/prolate configurations, and in agreement with the expectations
   for an optically-thin scattering nebulae. Furthermore, we find that the polarisation in the peak/read-wing
   region follows closely the scaling of  \citet{BL77}, as the scaling quantity $(1-3\gamma)$ varies
   from 0.06 to 0.34 for model s15CO and from -0.06 to -0.58 for model s15CP. In fact, adopting
   $\bar \tau \sim 2$ and these case-dependent shape factors gives the peak/red-wing polarisation
   magnitudes we obtain.

    \subsection{Polarisation dependency on continuum wavelength and line identity}
\label{sect_lambda}

   We have so far focused on the polarisation of the H$\alpha$-line flux and the neighbouring continuum.
However, spectropolarimetric observations usually cover the entire optical (and sometimes the near-IR) range,
thus probing a wide range of continuum wavelengths and spectral lines.
Hence, we now discuss the dependency of the polarisation with wavelength. First, we
present the variation with continuum wavelength (Section ~\ref{sect_cont}), followed by the polarisation signatures
of individual Balmer and Paschen lines (Section ~\ref{sect_line}), before turning to lines of species other than hydrogen, i.e.,
Fe\two, Na\one, and Ca\two\ (Section ~\ref{sect_other_lines}).

\subsubsection{Variation with continuum wavelength}
\label{sect_cont}

\begin{figure}
\epsfig{file=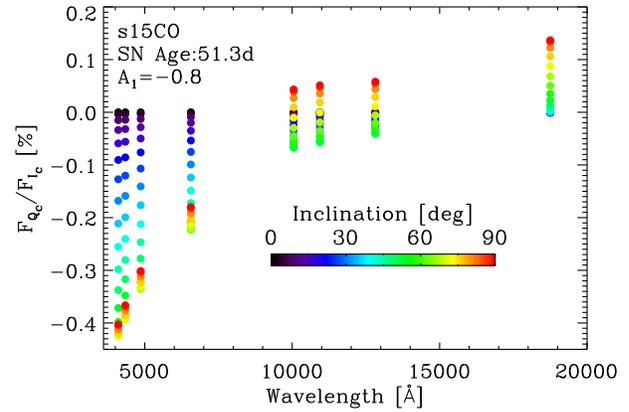,width=8.5cm}
\caption{Wavelength dependence of the normalised continuum polarisation flux $F_{Qc}/F_{Ic}$,
from the optical to the near-IR, for model s15CO at 51.3\,d and $A_1 = -0.8$ (pole-to-equator density
ratio of 0.2). Notice the large polarisation change, even associated at this epoch with a sign reversal.
\label{fig_var_cont}}

\end{figure}

   To illustrate this dependence with model s15CO at 51.3\,d after explosion ($A_1 = -0.8$),
   we compute the continuum polarisation in the spectral regions adjacent to the
   H\one\ Balmer and Paschen lines, thus covering from $\sim$4000\,\AA\ to 2\,$\mu$m. We show the
   results in Fig.~\ref{fig_var_cont}.
   The continuum polarisation varies enormously with wavelength, even showing a sign reversal
   between optical and near-IR wavelengths for a 90$^{\circ}$ inclination (variation from -0.4\% up to 0.1\%).
   Crossing the Balmer jump blueward (not shown), the continuum polarisation for a 90$^{\circ}$ inclination
   jumps from -0.5\% to -2\%, while it is essentially constant through the Paschen jump.
   Such a large polarisation in the blue part of the spectrum is difficult to observe,
   as it requires early-time observations when blanketing in the optical is weak, and an SN ejecta sufficiently
   asymmetric in its outer parts. These conditions seem to be rarely met in Nature.

\begin{figure}
   \epsfig{file= 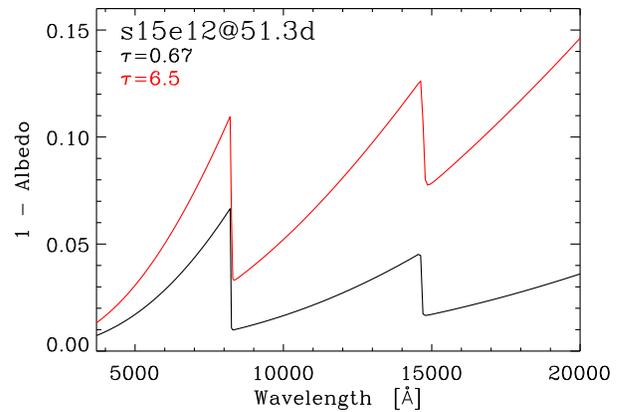,width=8.5cm}
   \caption{Wavelength dependence of the quantity ($1- {\rm Albedo}$), which is equal to
   $\chi_{\rm abs} / \chi_{\rm tot}$, for model s15e12 at 51.3\,d and shown at a Rossleand-mean optical
   depth of 0.67 (black) and 6.5 (red).
   \label{fig_albedo}}
   \end{figure}

\begin{figure}
   \epsfig{file=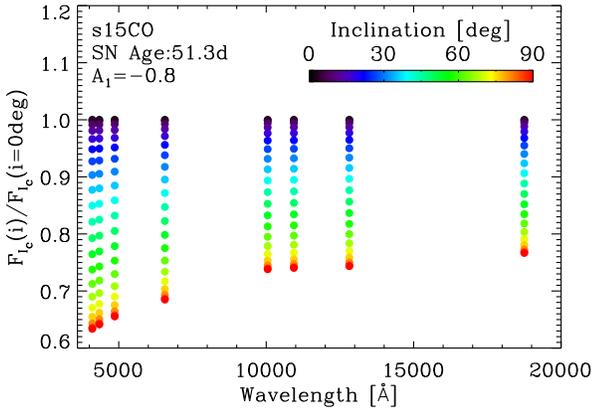, width=8.5cm}
   \caption{Illustration of the variation in the  continuum flux (normalised to the flux value
   for a pole-on view) as a function of inclination and wavelength for model s15CO at 51.3\,d.
   }
   \label{fig_ref_cont}
   \end{figure}

   We associate the variation in continuum polarisation with the variation in thermalisation
   depth with wavelength. This arises from both the variation in albedo (the ratio of scattering to total
   opacity; Fig~\ref{fig_albedo}) and continuum source function with wavelength and depth. While the
   electron-scattering opacity is grey, bound-free and free-free absorption vary with wavelength,
   approximately as $\lambda^3$. In addition, we would expect to see changes  in polarisation near the
   Balmer ($\lambda \sim$\,3650\,\AA) and Paschen ($\lambda \sim$\,8200\,\AA) jumps.
   Below 3650\,\AA, absorption by the $n=2$ state is allowed, while below 8200\,\AA, absorption from the $n=3$ state
   of H\one\ is allowed. A change in the albedo effects both the thermalisation depth and the photon escape.
   When the albedo is
   low, a photon can only undergo a few scatterings before it is destroyed by a continuum absorption. However,
   when the albedo is high (i.e.,  close to 1), the photon can undergo many scatterings before it is destroyed.
   A consequence of the influence of the albedo is that the angular flux distribution will be a function of
   wavelength.  This is shown in Fig.~\ref{fig_ref_cont} where we plot  the flux along inclination $i$,
   normalised to its value along $i=0^\circ$.
   This result emphasises that both the angular distribution of the electrons, and the
   angular distribution of the flux, affect the observed level of polarisation.

\begin{figure*}
\epsfig{file=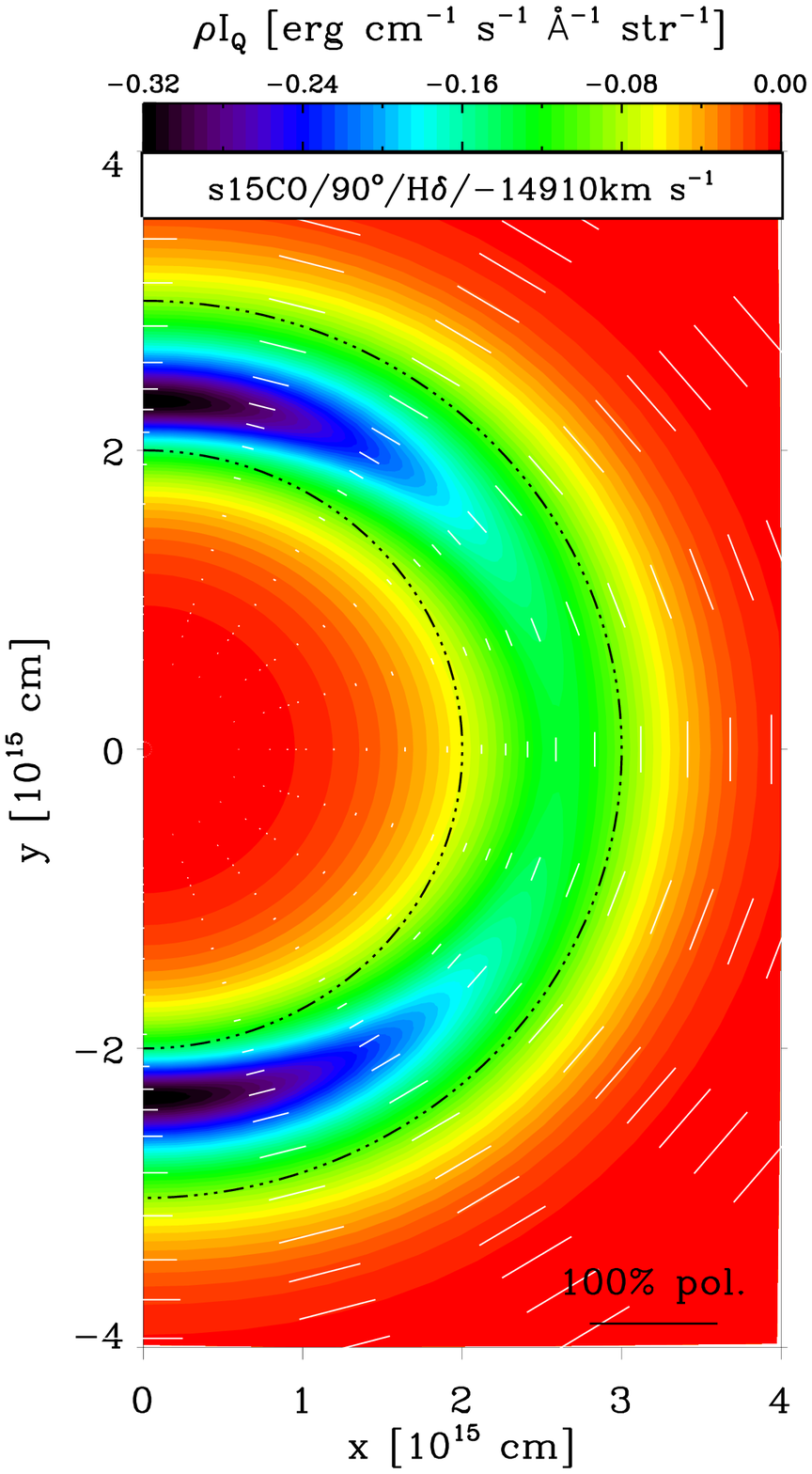,width=4cm}
\epsfig{file=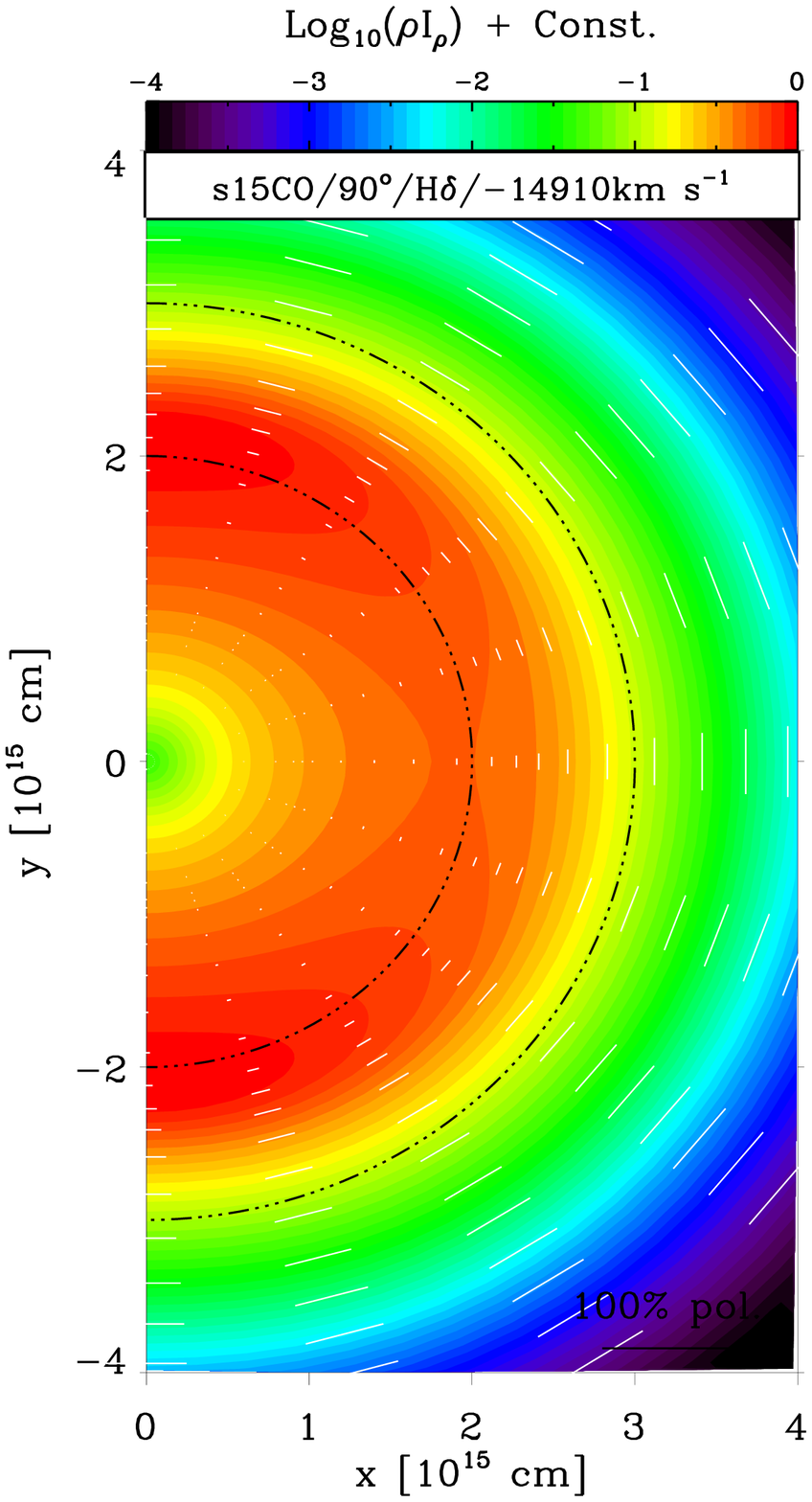,width=4cm}
\epsfig{file=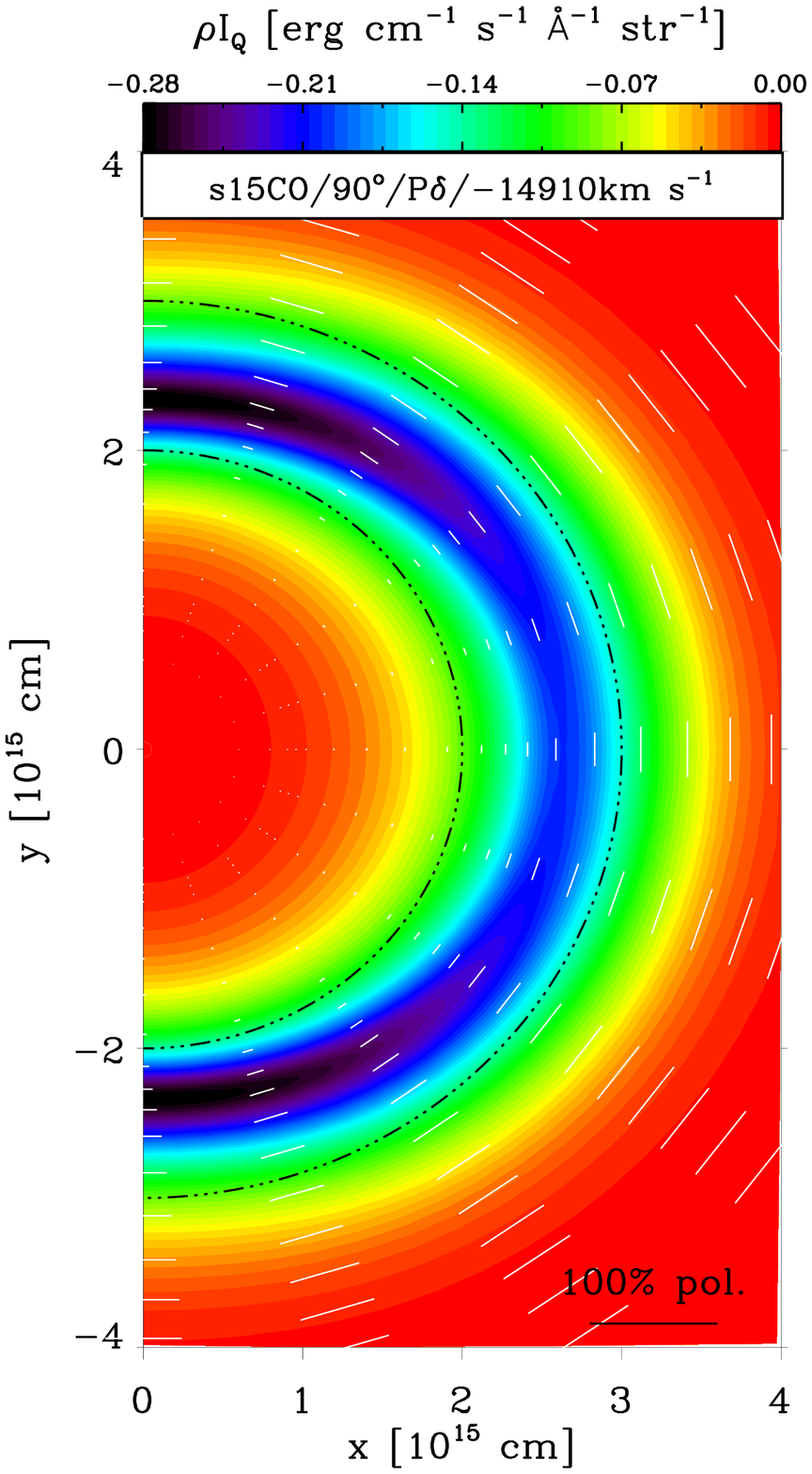,width=4cm}
\epsfig{file=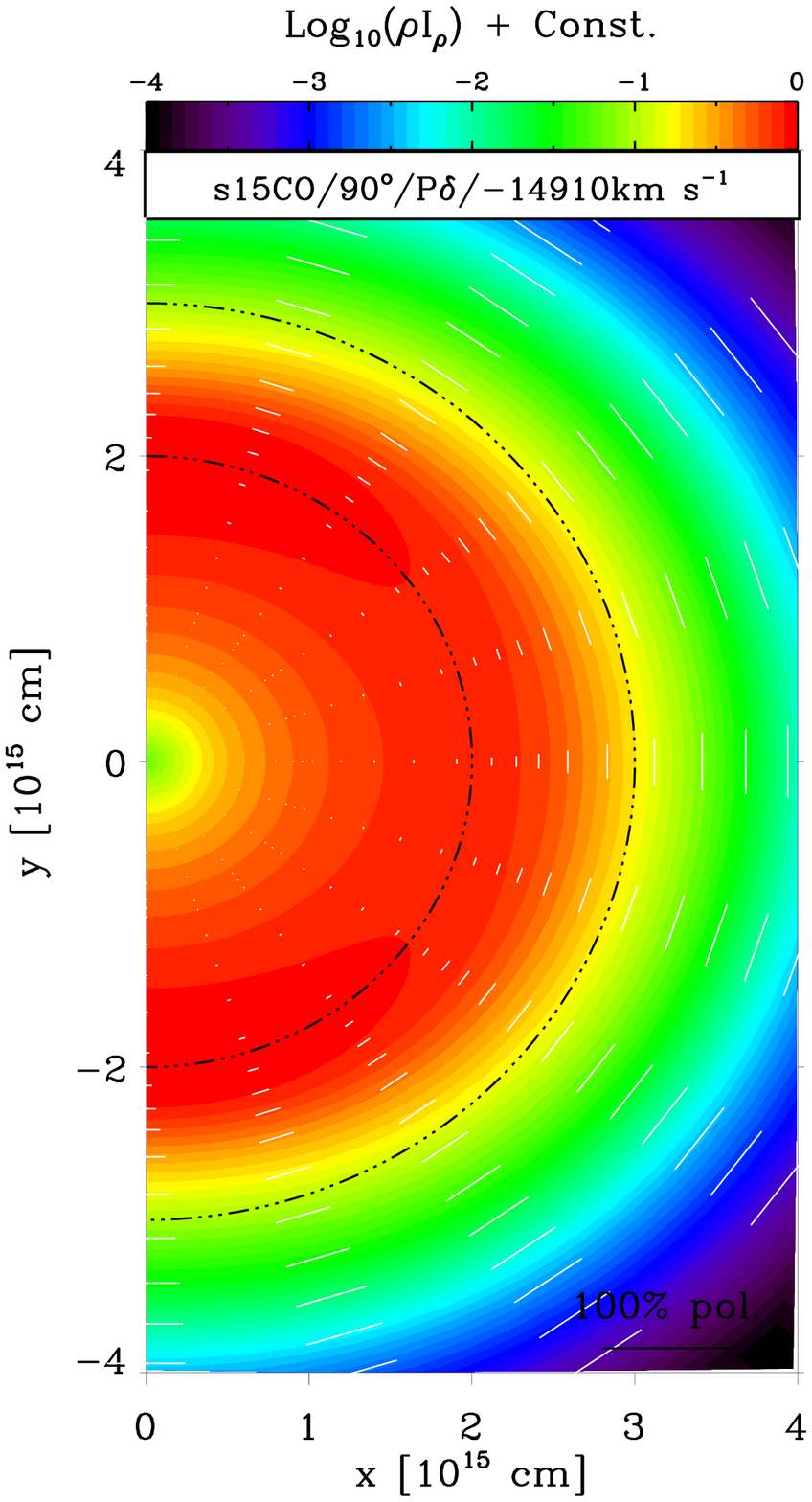,width=4cm}
\caption{
{\it Left two panels: } Colour map of the quantity $\ip I_Q$ (left) and $\ip I_{\ip}$ (right) in the continuum
adjacent to the H$\delta$ line for model s15CO at 51.3\,d after explosion and $A_1=-0.8$ (see also
Fig.~\ref{fig_var_cont}).
{\it Right two panels:} Same as left, but now for P$\delta$.
The stronger flux along the polar directions causes the residual polarisation to be negative
in the continuum adjacent to the H$\delta$ line. In contrast, the continuum flux in the P$\delta$
region is more uniformly distributed on the plane of the sky and contributes to the near cancellation
of the polarisation. This is caused by an optical-depth effect.
\label{fig_pd_hd_cont_map}
}
\end{figure*}

   In Fig.~\ref{fig_pd_hd_cont_map}, we show the distribution of $\ip I_Q$ and $\ip I_{\ip}$
   in the continuum adjacent to H$\delta$ (left columns) and P$\delta$ (right columns).
   The stronger flux along the polar directions causes the residual polarisation to be negative
   in the continuum adjacent to the H$\delta$ line. In contrast, the continuum flux in the P$\delta$
   region is more uniformly distributed on the plane of the sky and contributes to the near cancellation
   of the polarisation. This is an optical-depth effect: The equatorial regions have a higher density at
   a given radius and are more optically thick, so that more radiation escapes through the poles.
   The bias is enhanced at shorter wavelengths.

   Comparisons between stretched and scaled models confirm the wavelength dependent
   polarisation discussed above, but the precise quantitative behavior is a function of both the
   assumed asymmetry and the structure of the SN ejecta. In SNe, it is essentially the same medium
   that emits and scatters so that subtle, and rather unpredictable, cancellation effects take place.
   These are essentially impossible to predict without detailed radiative-transfer calculations.
   It also appears essential to start from physical inputs of the aspherical ejecta rather than use
   ad-hoc prescriptions for their morphology, as we presently do in this exploratory work.

   Observationally, the wavelength variation of continuum polarisation in the optical is generally
   associated with the differential magnitude of line blanketing. Its dominance shortward of 5000\,\AA,
   in particular because of a forest of Fe\two\ lines in cool SN spectra, is expected to drive the
   SN polarisation to zero in that spectral region.

   Our calculations suggest that SNe during the plateau phase should show, if asymmetric, a continuum polarisation
   which is a function of wavelength, and which may show sign reversals.  Such signatures are difficult to observe
   for two reasons. First, the polarisation level is low, even if there is
   a large asymmetry. Second, the observations need to be corrected for interstellar polarisation which is difficult
   to accurately estimate, and often it is constrained by a requirement that the polarisation variation with wavelength
   is ``simple.''. However the predicted continuum polarisations have been observed. In particular, \citet{chornock_etal_10}
   obtained observations of SN 2007aa during the plateau phase, 50 (i.e.,  $-50$\,d) and 22 days before the plateau phase ends.
   Observations at $-50$\,d show a continuum polarisation which increases from blue to red, which is negative below
   $\sim$5000A, and which is positive above that wavelength. The observation at $-22$\,d seems to be flatter, and shows
   no polarisation reversal. Complicated polarisation signatures across the lines are seen, although in the spectrum at $-50$\,d
   the polarisation appears to go to zero close to line center, rather than on the blue side of the emission as we predict.
   At $-22$\,d, H$\alpha$ shows a  P Cygni profile in polarisation, with a significantly enhanced polarisation in the red-wing.
   The polarisation observations of SN 2006ov near the end of the plateau phase by \citet{chornock_etal_10} also show a wavelength
   dependent polarisation, although in this case no polarisation reversal is seen.

\subsubsection{Variation with Balmer line/series}
\label{sect_line}

   The above wavelength dependence of the continuum polarisation across the optical
   and near-IR naturally affects the polarisation of overlapping lines. For example, the
   Balmer series starts at the 3660\,\AA\  edge and extends to H$\alpha$ at 6562\,\AA,
   covering here a range of continuum polarisation from  -0.4\% to 0.03\%.
   This matters because, in the P-Cygni trough, we have shown in the previous sections that one
   obtains a polarisation  similar to that in the continuum, only amplified due to the lack of forward-scattered,
   or intrinsically, unpolarised radiation.
   Line absorption and/or line emission over a large volume may wash out this continuum polarisation
   and leave a distinct polarisation. Indeed, in the previous sections, we obtained peak/red-wing polarisation
   of the opposite sign from that in the continuum.

\begin{figure}
\epsfig{file=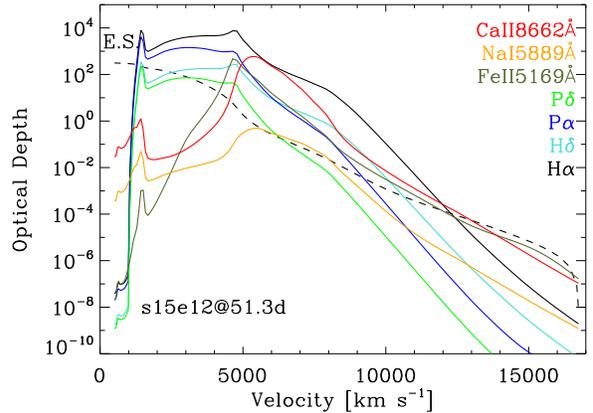,width=8.5cm}
\caption{Sobolev optical depth versus velocity for a variety of lines in model s15e12 at 51.3\,d
after explosion. Using a colour coding, we show H\one\ Balmer and Paschen lines,
the blue component of Na\one\,D, Fe\two\,5169\,\AA, and the red component at 8662\,\AA\ of the Ca\two\
triplet resonance transitions that appear as a broad feature at $\sim$\,8500\,\AA.
We also overplot the electron-scattering optical depth (labelled ``E.S.", dashed line).
\label{fig_tau_dist}}
\end{figure}

   The effect on polarisation of the line absorption/emission/scattering is conditioned by the location
   and extent of its formation region.
   Lines that are thinner tend to form at smaller radii, at a larger continuum optical depth, and produce
   P-Cygni profiles that have a weaker emission (e.g., H$\delta$ and P$\delta$). In contrast, lines that are
   thicker tend to form at larger radii, extending above the photosphere and over a larger volume,
   and produce strong lines  (e.g., H$\alpha$ and P$\alpha$; Fig.~\ref{fig_tau_dist}).
   In all cases, lines tend to form at a continuum optical depth that is not small, so that scattering
   with free electrons occurs (Fig.~\ref{fig_jcont}).

\begin{figure*}
\epsfig{file=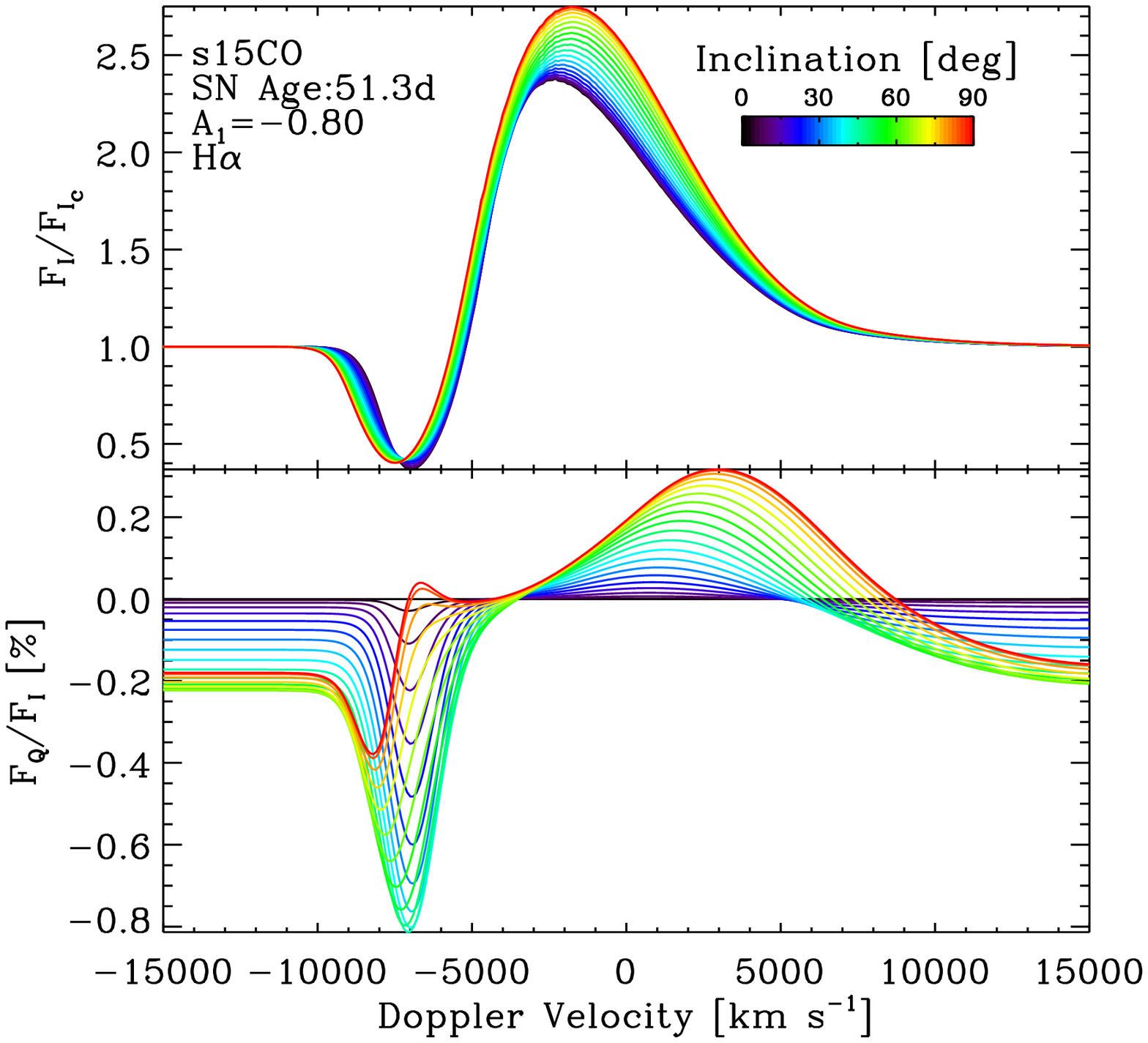,width=7.5cm}
\epsfig{file=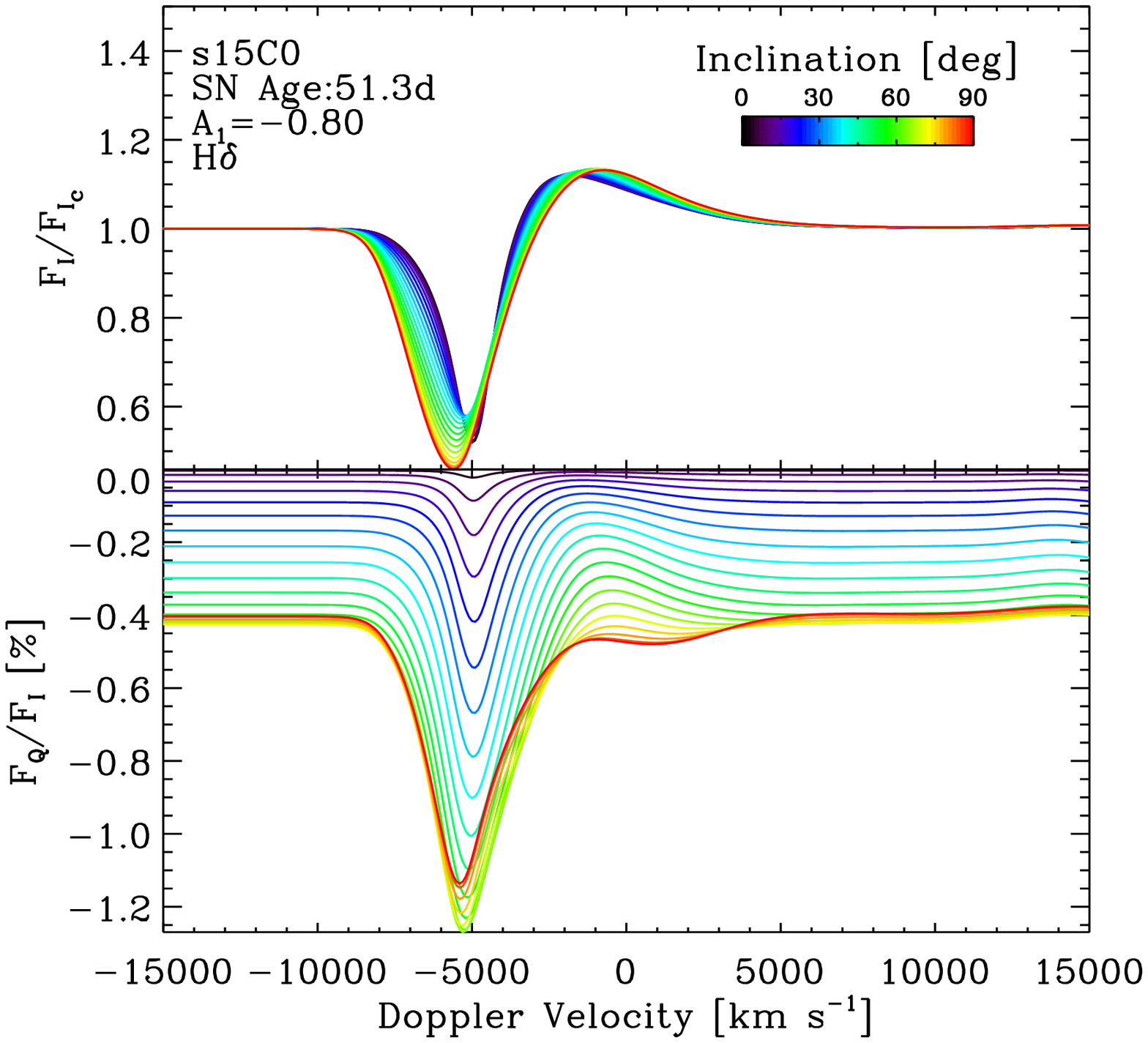,width=7.5cm}
\epsfig{file=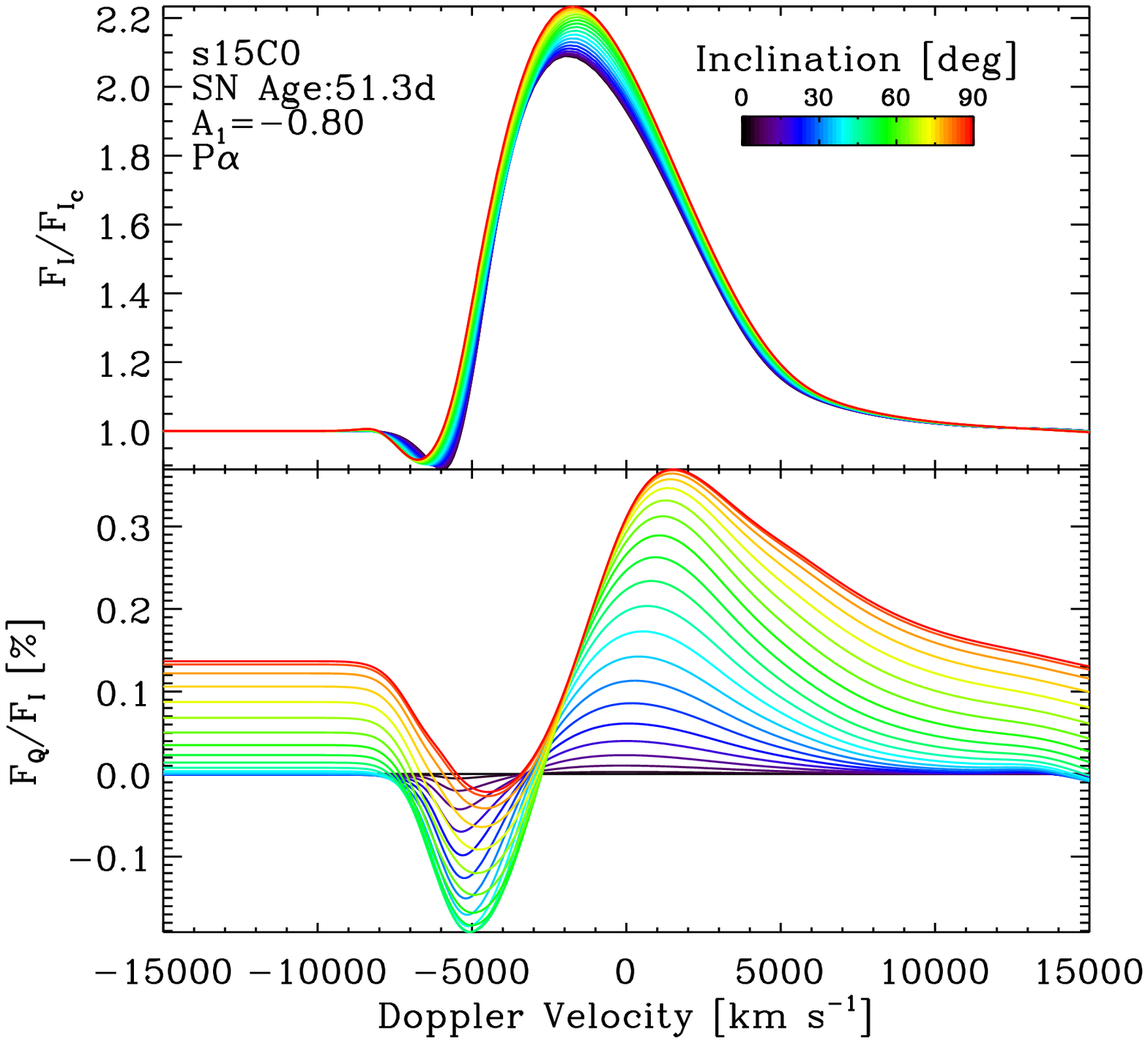,width=7.5cm}
\epsfig{file=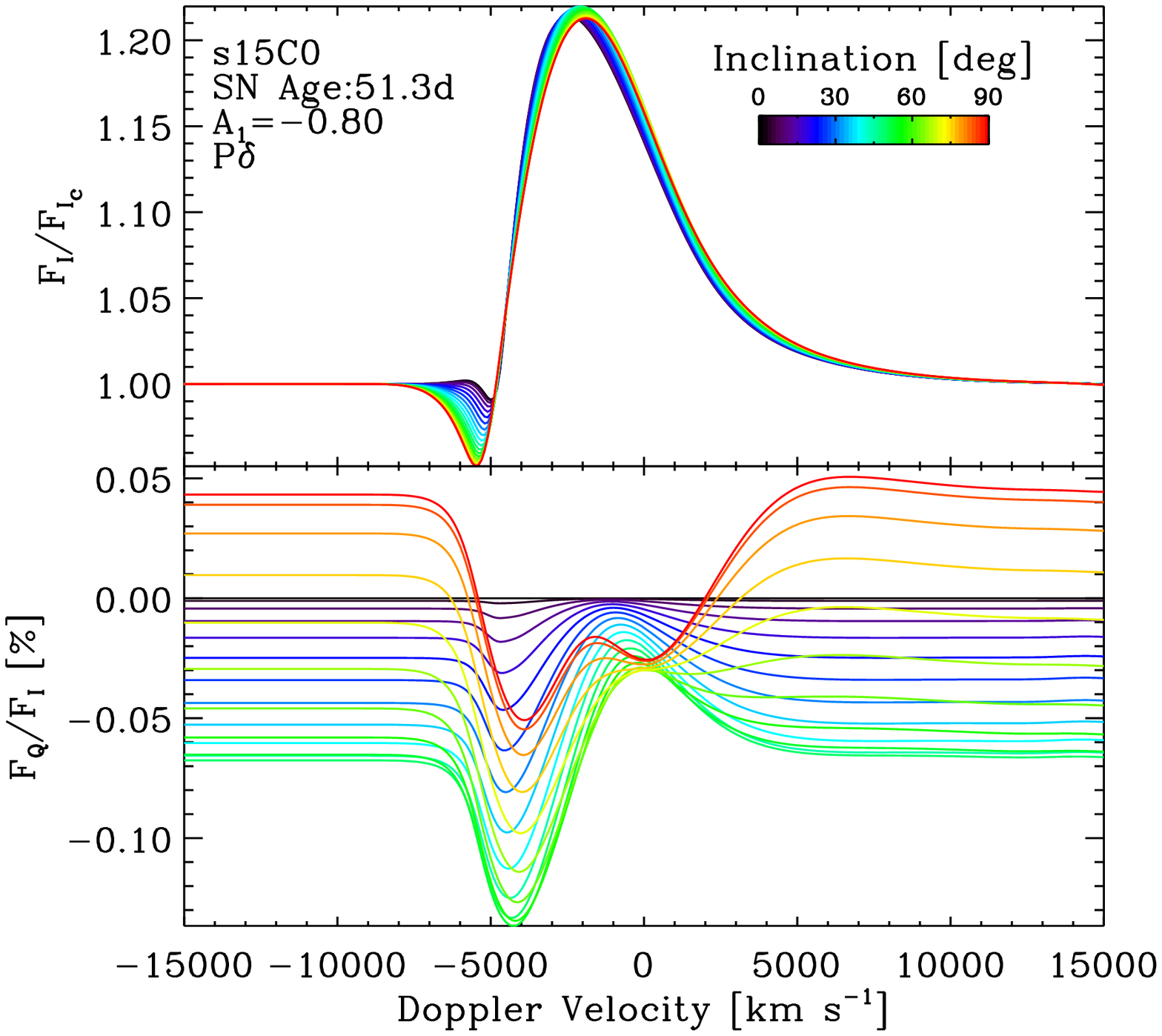,width=7.5cm}
\caption{Synthetic total and polarised flux for model s15CO at 51.3\,d after explosion ($A_1 = -0.8$)
and shown for Balmer (top; left to right shows H$\alpha$ and H$\delta$)
and Paschen (bottom; left to right shows P$\alpha$ and P$\delta$) lines.
\label{fig_var_hi}}
\end{figure*}

   We illustrate the inclination-dependent total and polarised flux for H$\alpha$, H$\delta$, P$\alpha$, and P$\delta$
   in Fig.~\ref{fig_var_hi}. The complicated behaviour of the continuum polarisation discussed in the previous section
   is again visible but now combined to the even more complex and diverse behaviour of the line polarisation -
   these signatures arise from the {\it same} ejecta!
   The properties of H$\alpha$ have been discussed before. H$\delta$ and P$\delta$ are so weak that they leave
   the overlying continuum polarisation unaffected, except in the P-Cygni trough where some absorption
   takes place and produces an enhanced relative polarisation.
   P$\alpha$ is a strong line like H$\alpha$ and produces a similar peak/red-wing polarisation.

\subsubsection{Polarisation for selected Fe\two, Na\one, and Ca\two\ lines}
\label{sect_other_lines}

\begin{figure*}
\epsfig{file=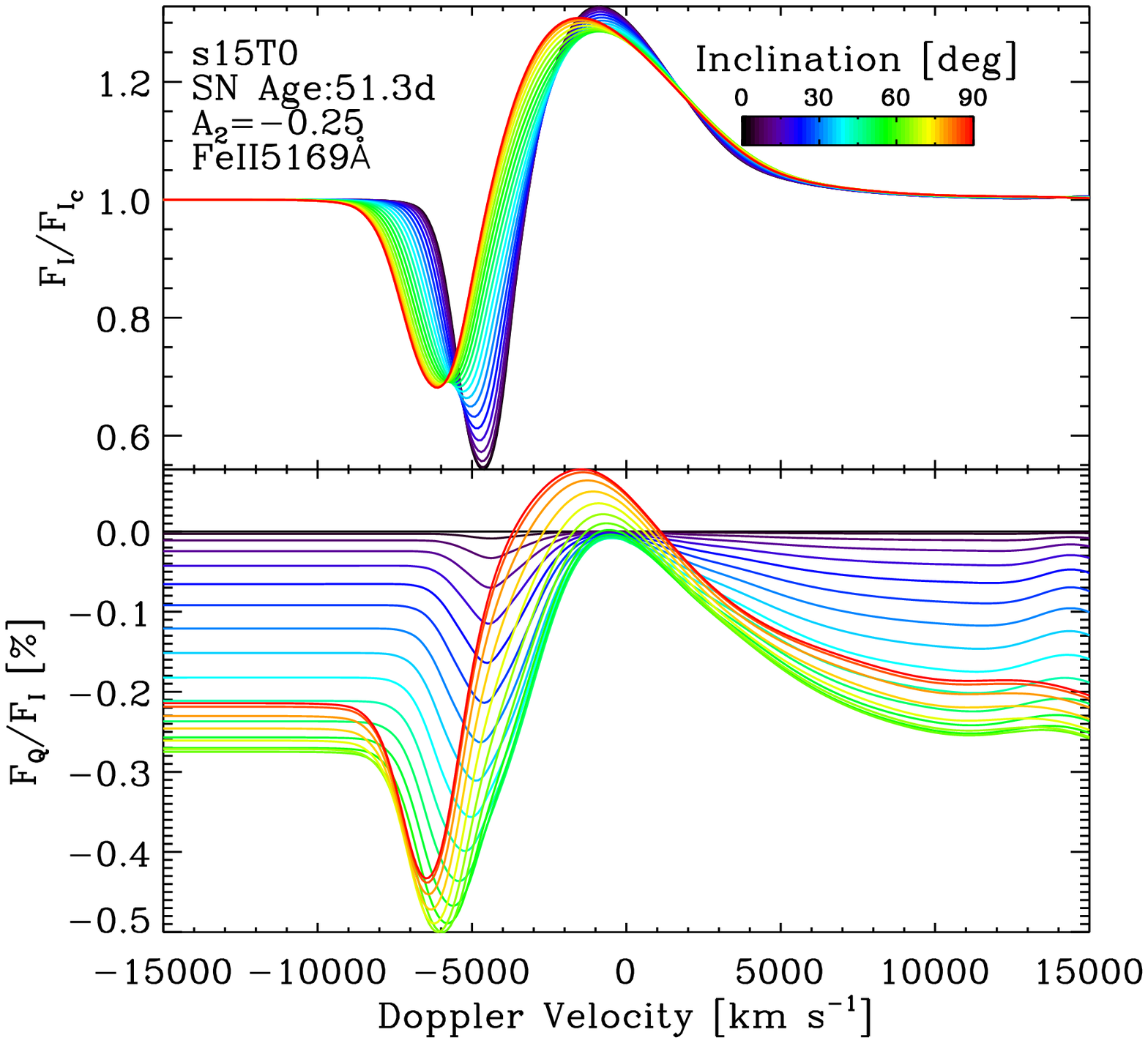,width=5.85cm}
\epsfig{file=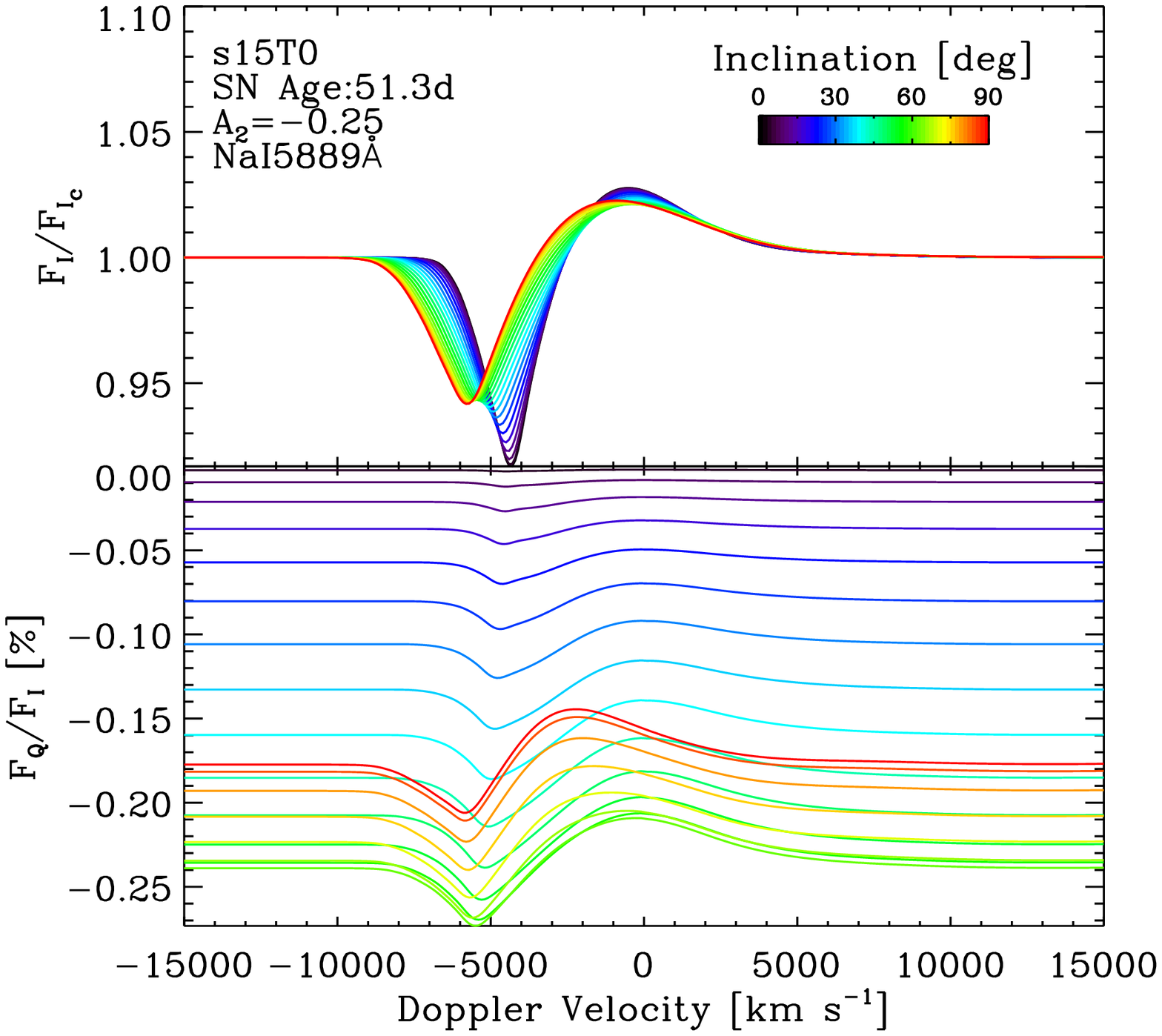,width=5.85cm}
\epsfig{file=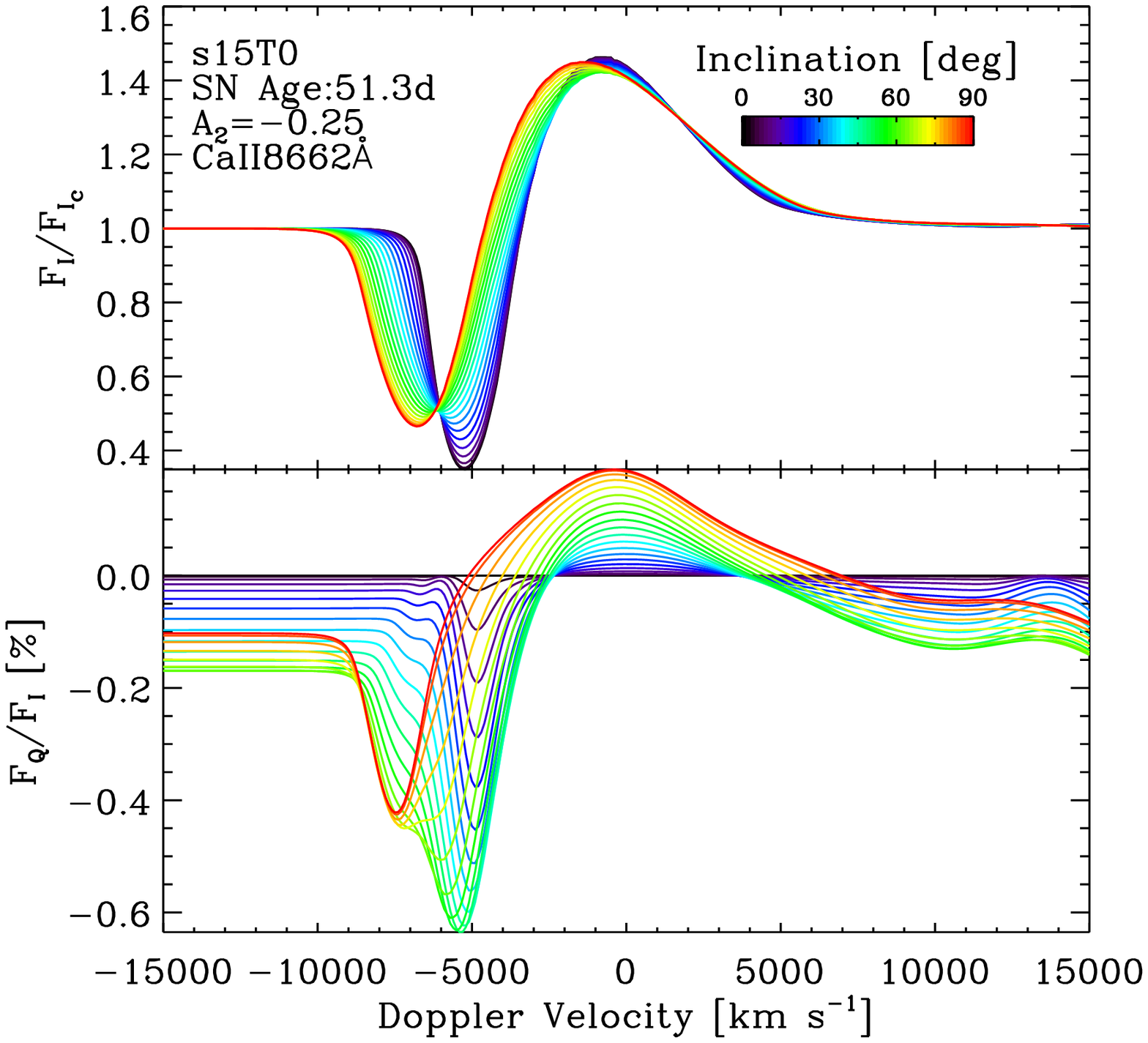,width=5.85cm}
\caption{{\it Left:} Variation of the normalised total (upper panel) and polarised (lower panel)
flux versus Doppler velocity and inclination for Fe\two\,5169\AA.
Ejecta properties are identical to those used in Fig.~\ref{fig_s15TO_incl}.
{\it Middle:} Same as left, but now for the Na\one\,5889\,\AA\ line.
{\it Right:} Same as left, but now for the Ca\two\,8662\,\AA\ line.
\label{fig_fenaca}}
\end{figure*}

    We have so far focused on H\one\ lines. As discussed in the previous section, although these
    lines all form through recombination, they show a considerable diversity of polarisation signatures
    owing to the range of strength and optical thickness they span.
    We now look at the polarisation signatures of lines from different species and formation processes.
    We include Fe\two\,5169\,\AA, a line that is a good indicator of the SN expansion rate of
    Type II SN ejecta at the recombination epoch \citep{DH05_epm,DH10a};
    Na\one\,5889\,\AA, a scattering line and blue component
    of the doublet Na\one\,D line; and Ca\two\,8662\,\AA\, which is a strong quasi-resonance line.
    For our present computations, we ignore resonant scattering, which may be questionable for the Ca\two\ line.
    The variation of the line optical depth for each of these is shown in Fig.~\ref{fig_tau_dist}.
    We show our results in Fig.~\ref{fig_fenaca}.

     The qualitative features of the line flux and polarisation are comparable for all three lines. Quantitatively,
     the results for the Ca\two\ line are the most analogous to those for H$\alpha$.
     It is a very optically-thick line which forms at and well beyond the photosphere, making it a
     key probe of asphericities in SN ejecta.
     The Fe\two\ and Na\one\ lines are rather thin and do not lead to a sizable change in polarisation
     in the peak/red-wing regions of the line profile at this post-explosion time - the changes in the total
     flux are however non trivial.

\subsection{Time evolution of polarisation}
\label{sect_time}

\begin{figure*}
 \epsfig{file= 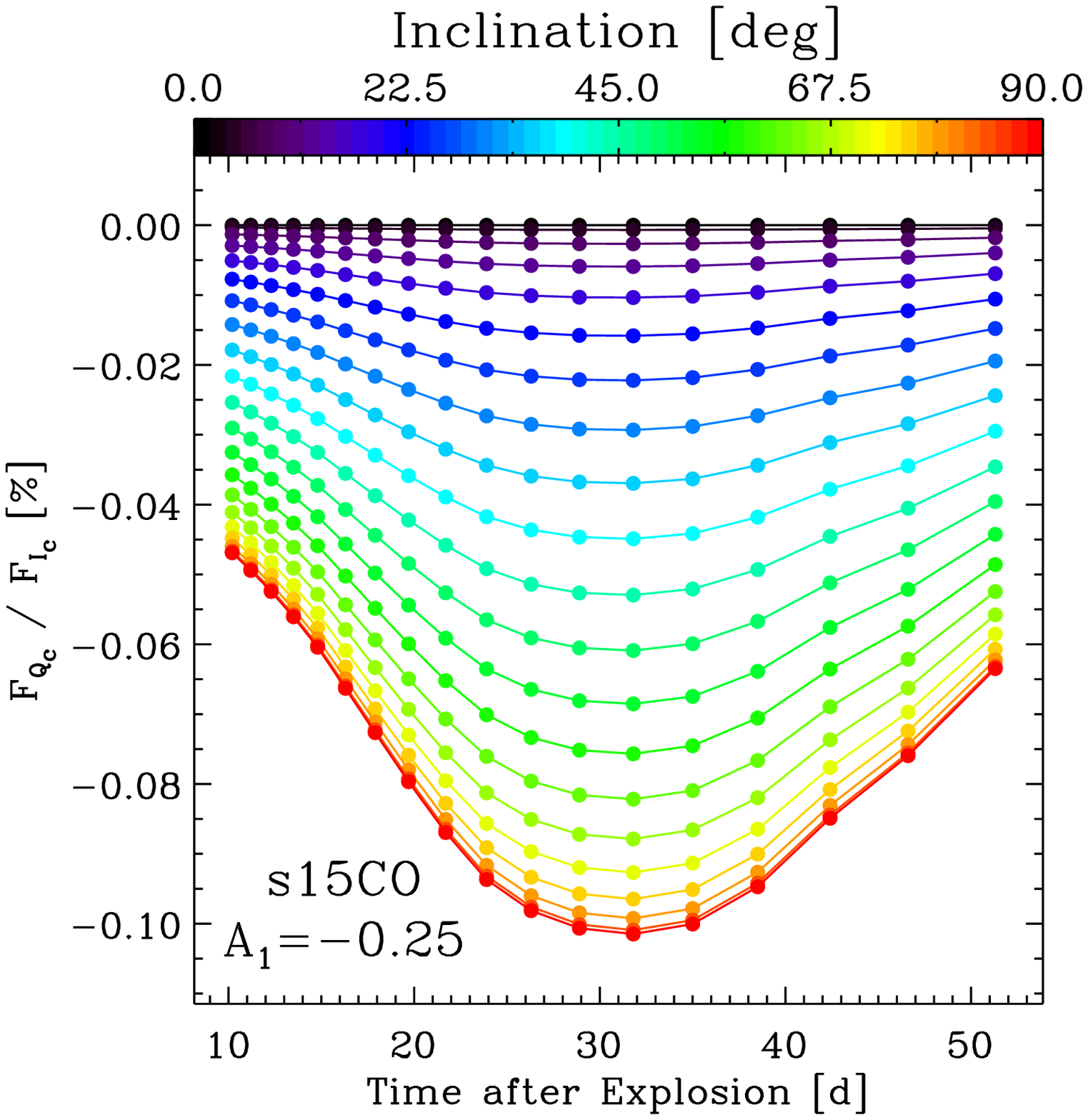,width=5.75cm}
 \epsfig{file= 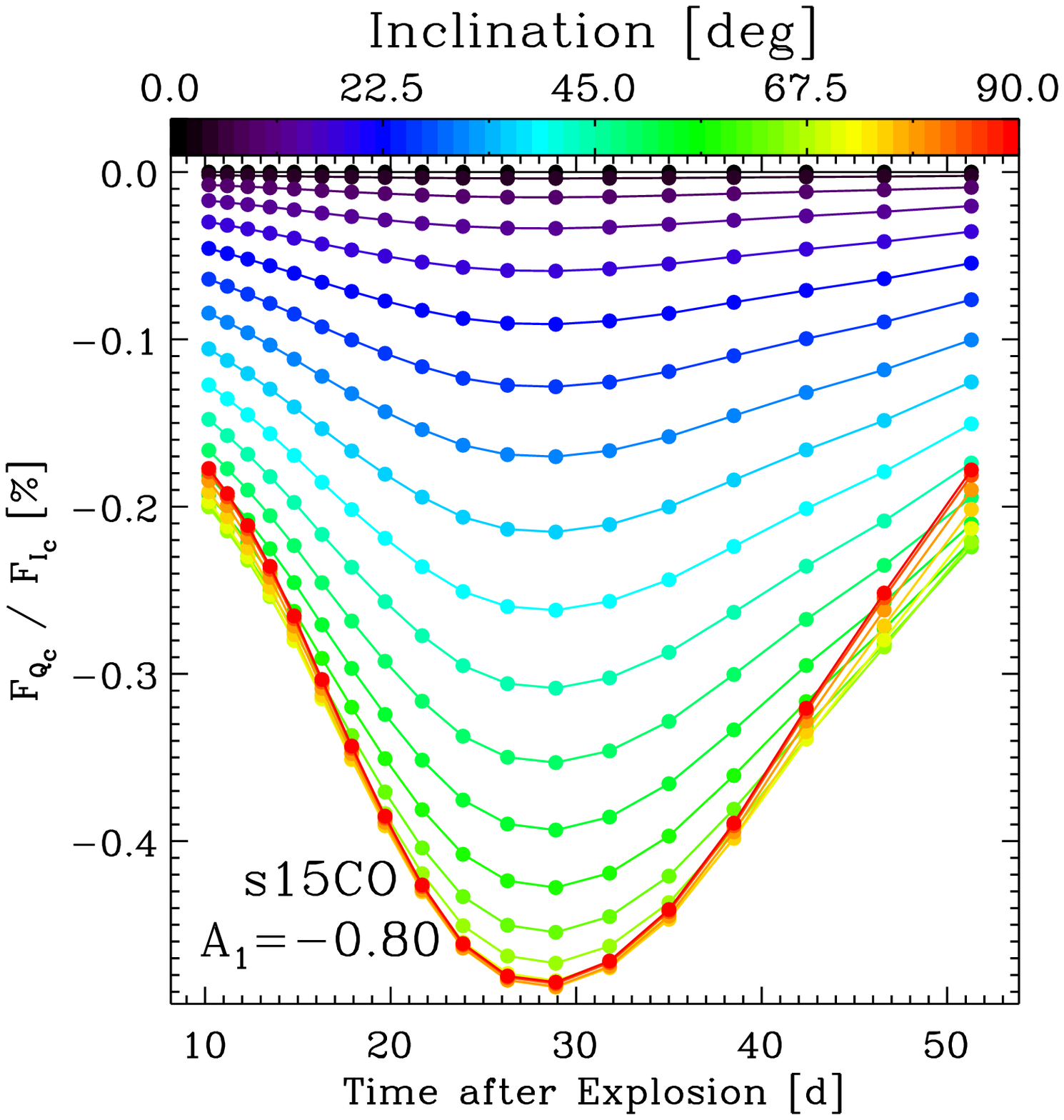,width=5.75cm}
 \epsfig{file= 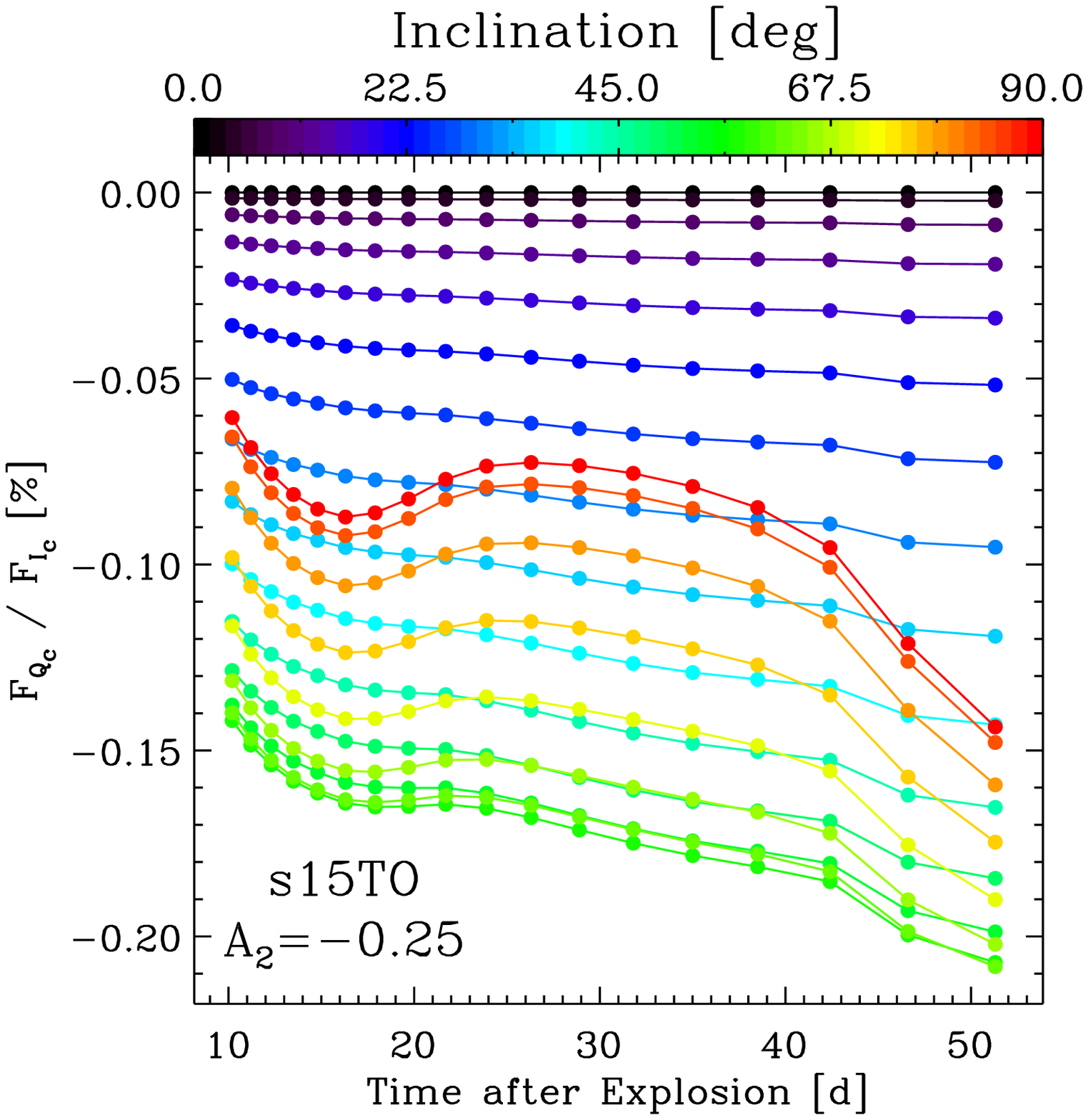,width=5.75cm}
 \epsfig{file= 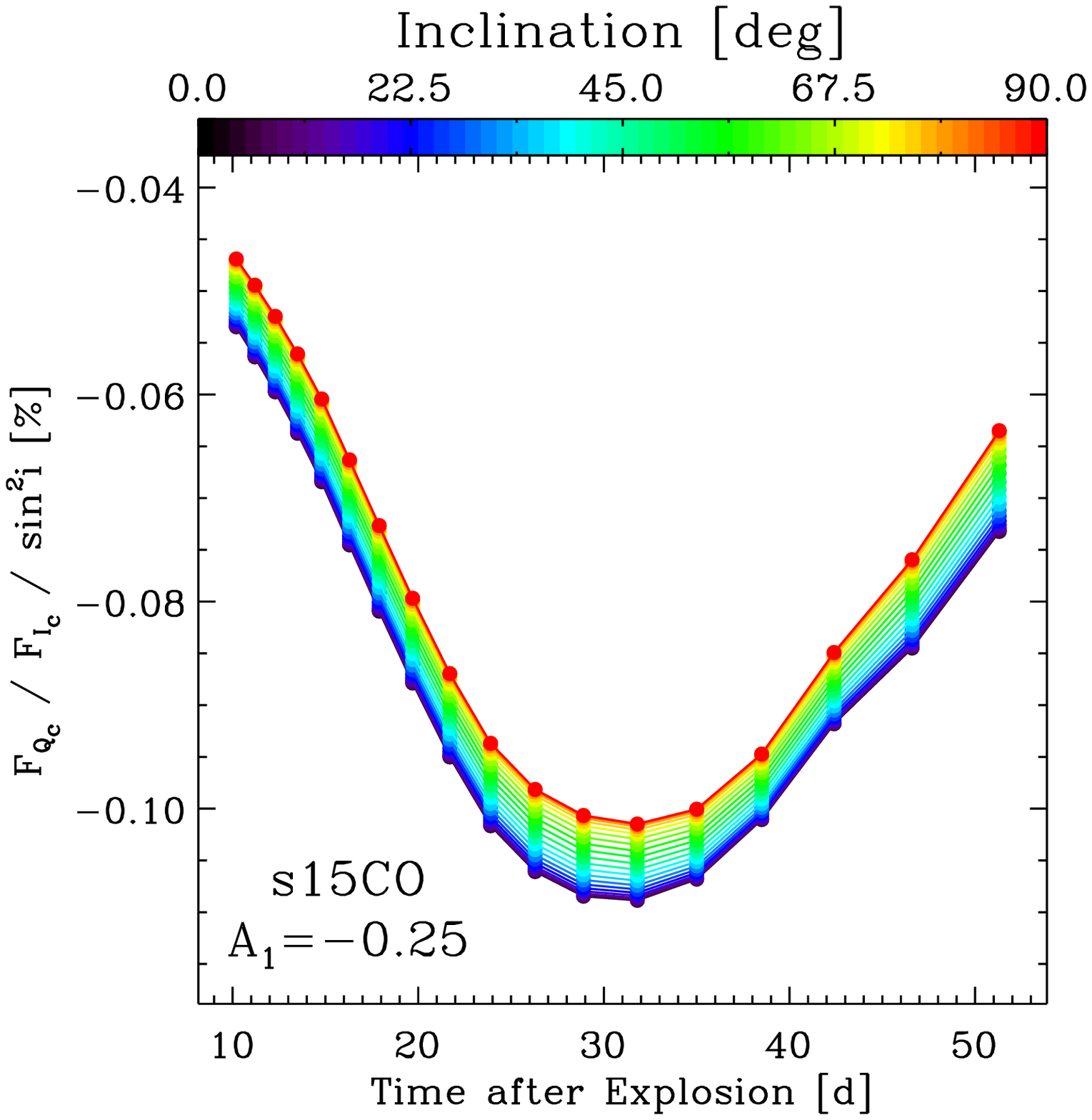,width=5.75cm}
 \epsfig{file= 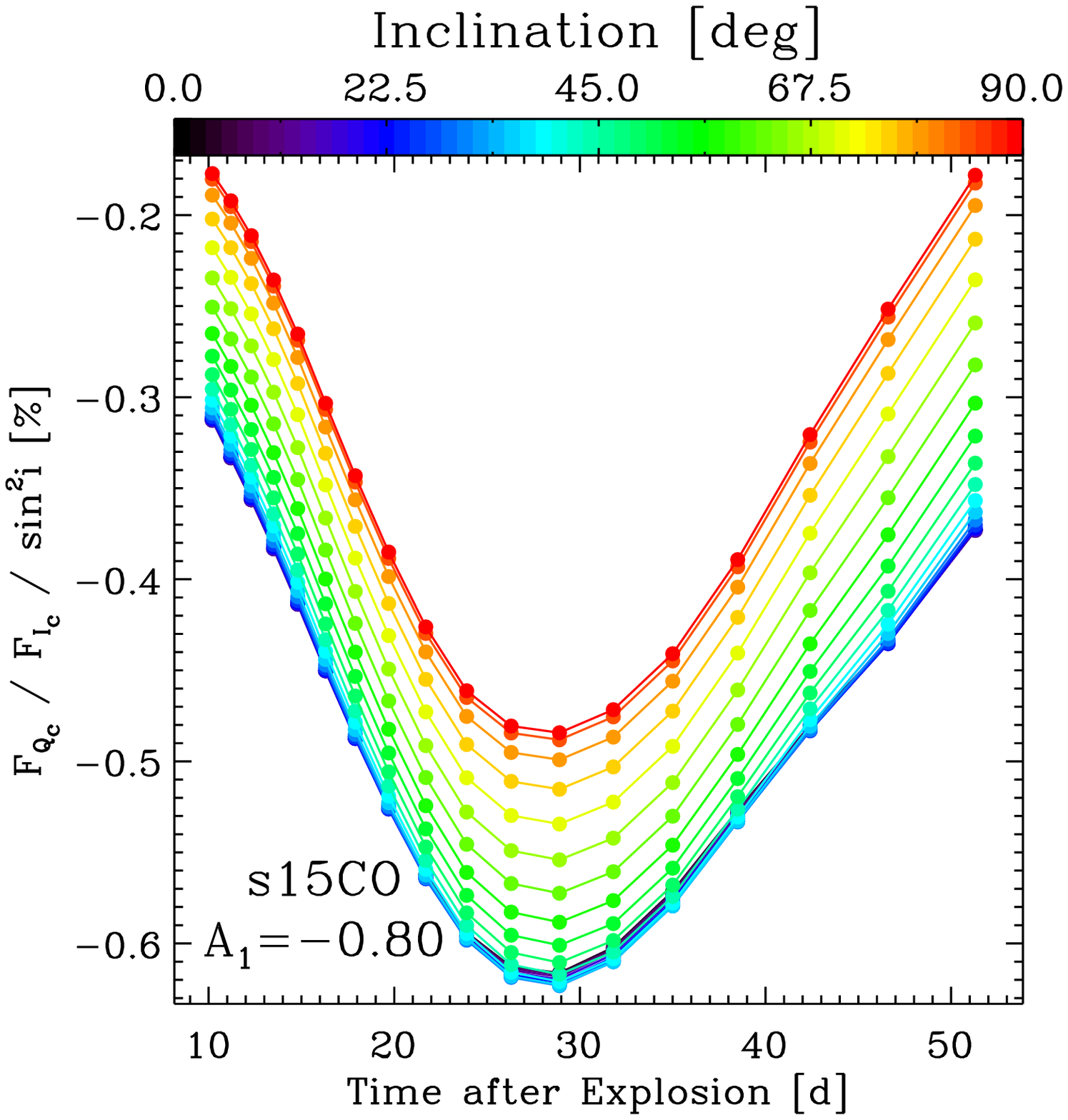,width=5.75cm}
 \epsfig{file= 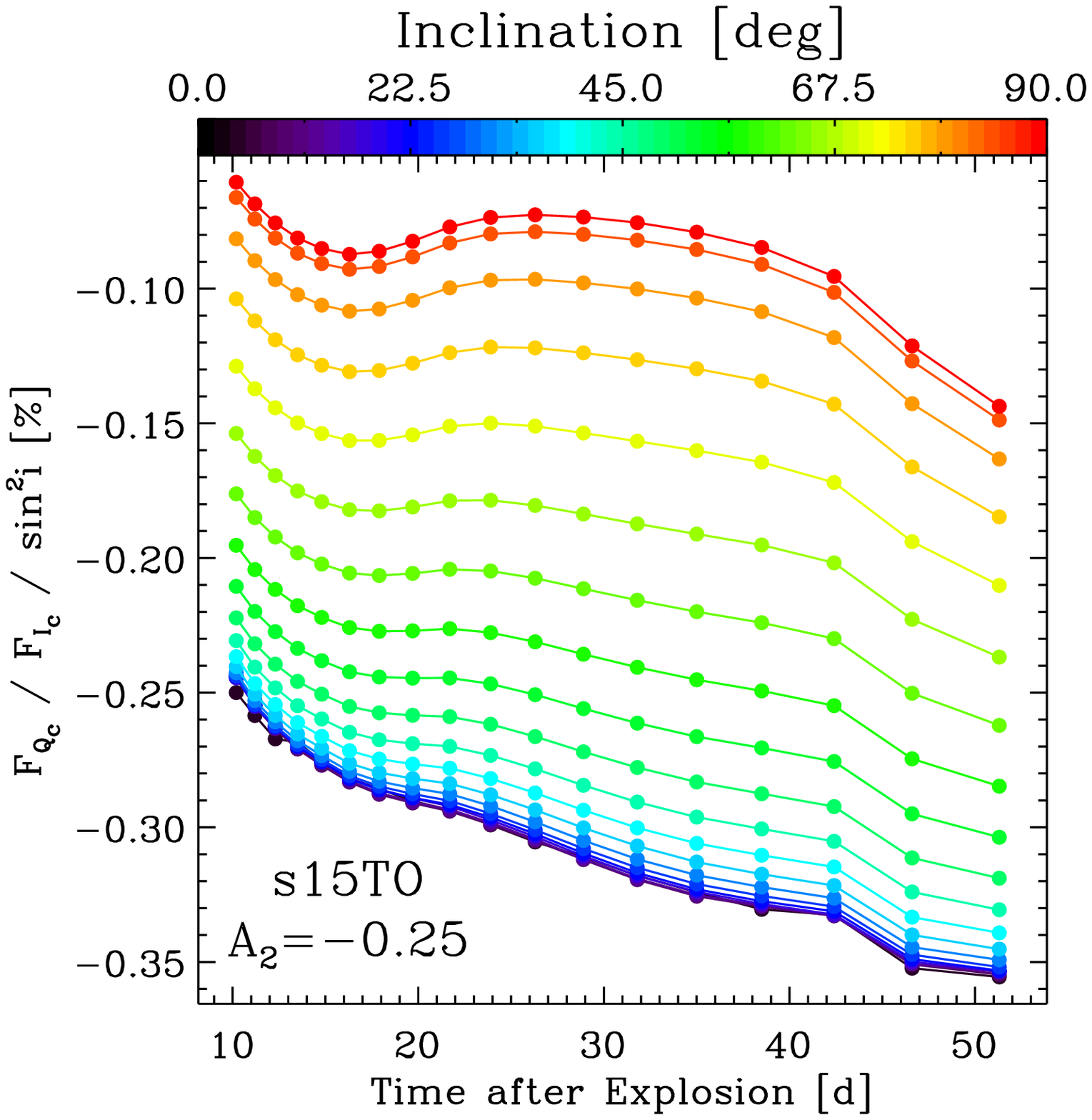,width=5.75cm}
 \caption{{\it Top row:}
 Evolution from 8.5 to 51.3\,d after explosion of the normalised continuum polarisation $F_{Qc}/F_{Ic}$
 in the H$\alpha$ region. The simulations are based on the s15e12 simulation, scaled (left two panels)
 or stretched (right panel) to yield an oblate spheroid.
 A colour coding gives the results for 21 different inclinations (0$^\circ$: Pole-on view;
 90$^\circ$: Equator-on view).
 {\it Bottom row:} Same as top row, but now scaling $F_{Qc}/F_{Ic}$ by $1/\sin^2i$.
   \label{fig_s15CO_ha_WR_cont} }
\end{figure*}

We first present results from the long-characteristic code at the photospheric phase,
before describing results from the Monte Carlo code at the end of the plateau
phase and during the nebular phase. These nebular-phase simulations are not
final since the effects of non-thermal excitation/ionisation were neglected \citep{DH11}.

\begin{figure*}
 \epsfig{file=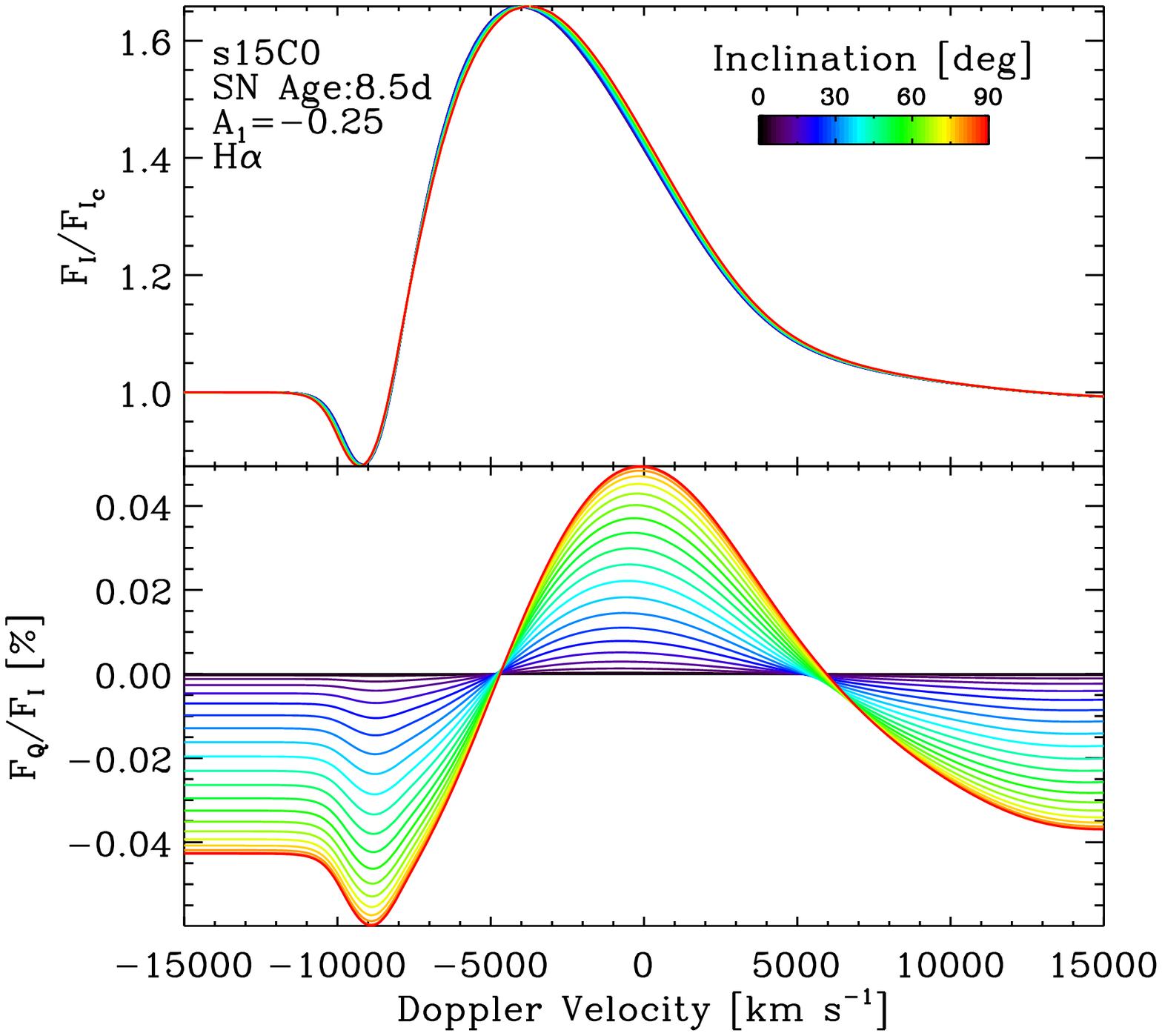,width=7.5cm}
 \epsfig{file=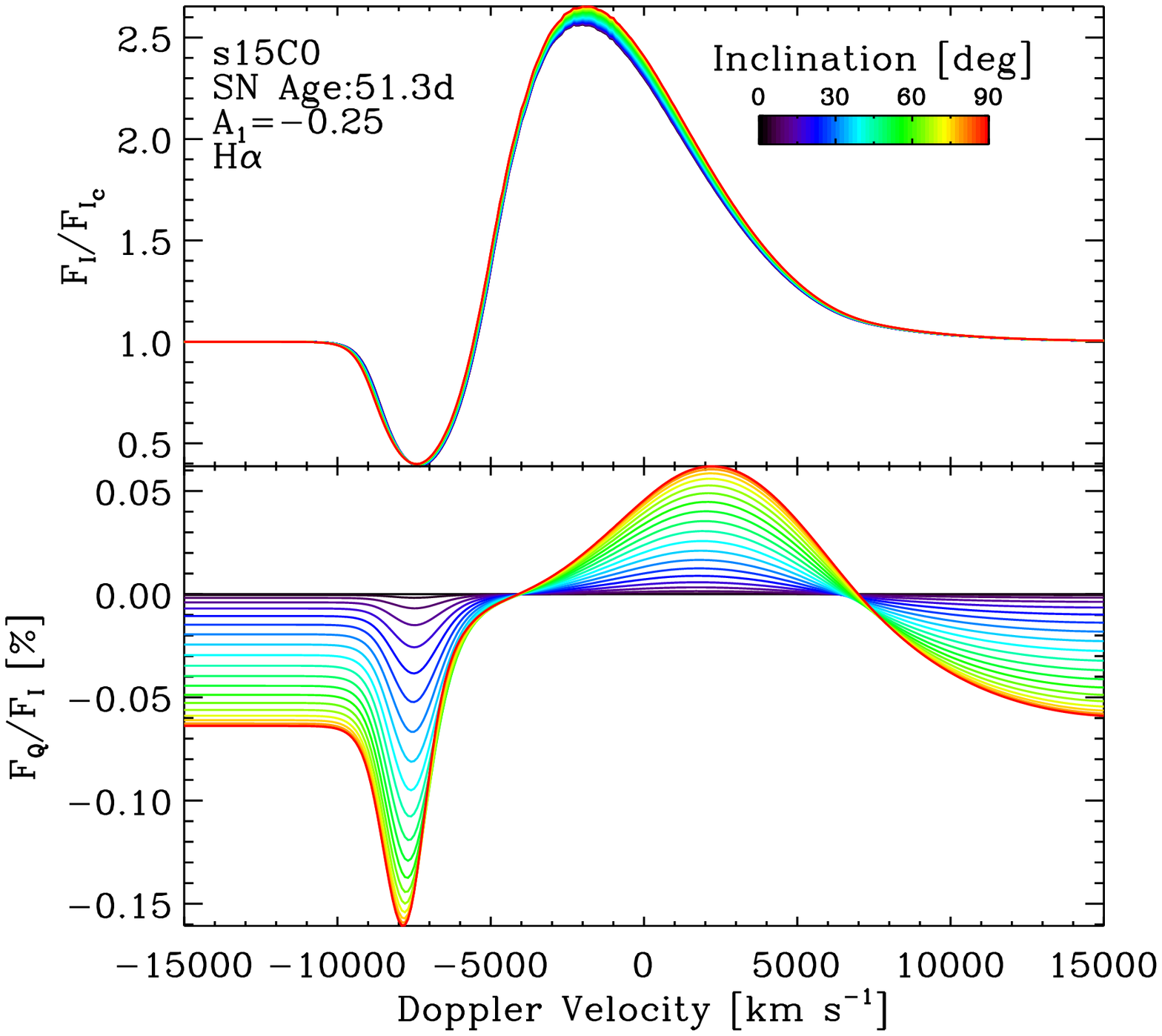,width=7.5cm}
 \epsfig{file=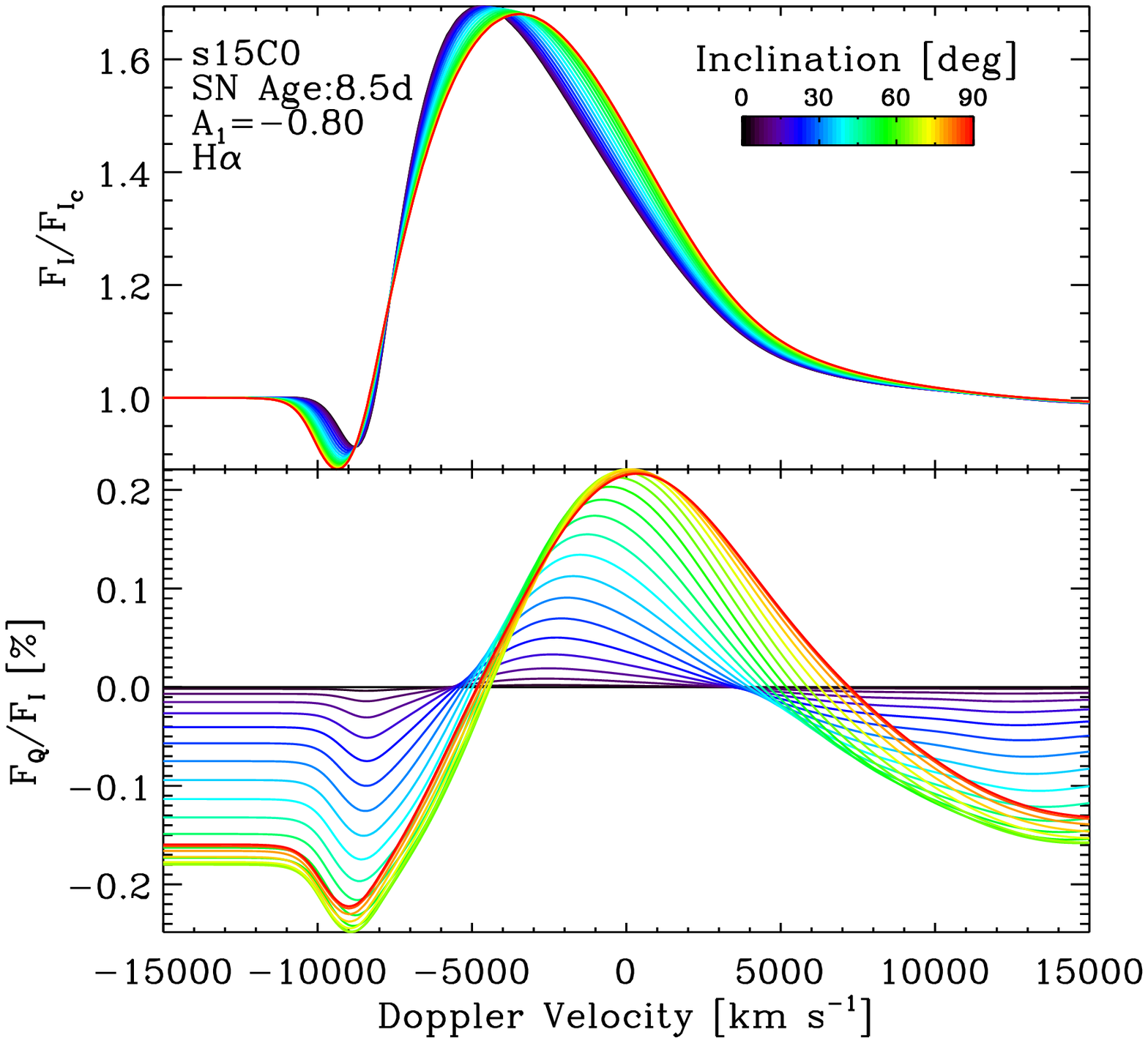,width=7.5cm}
 \epsfig{file=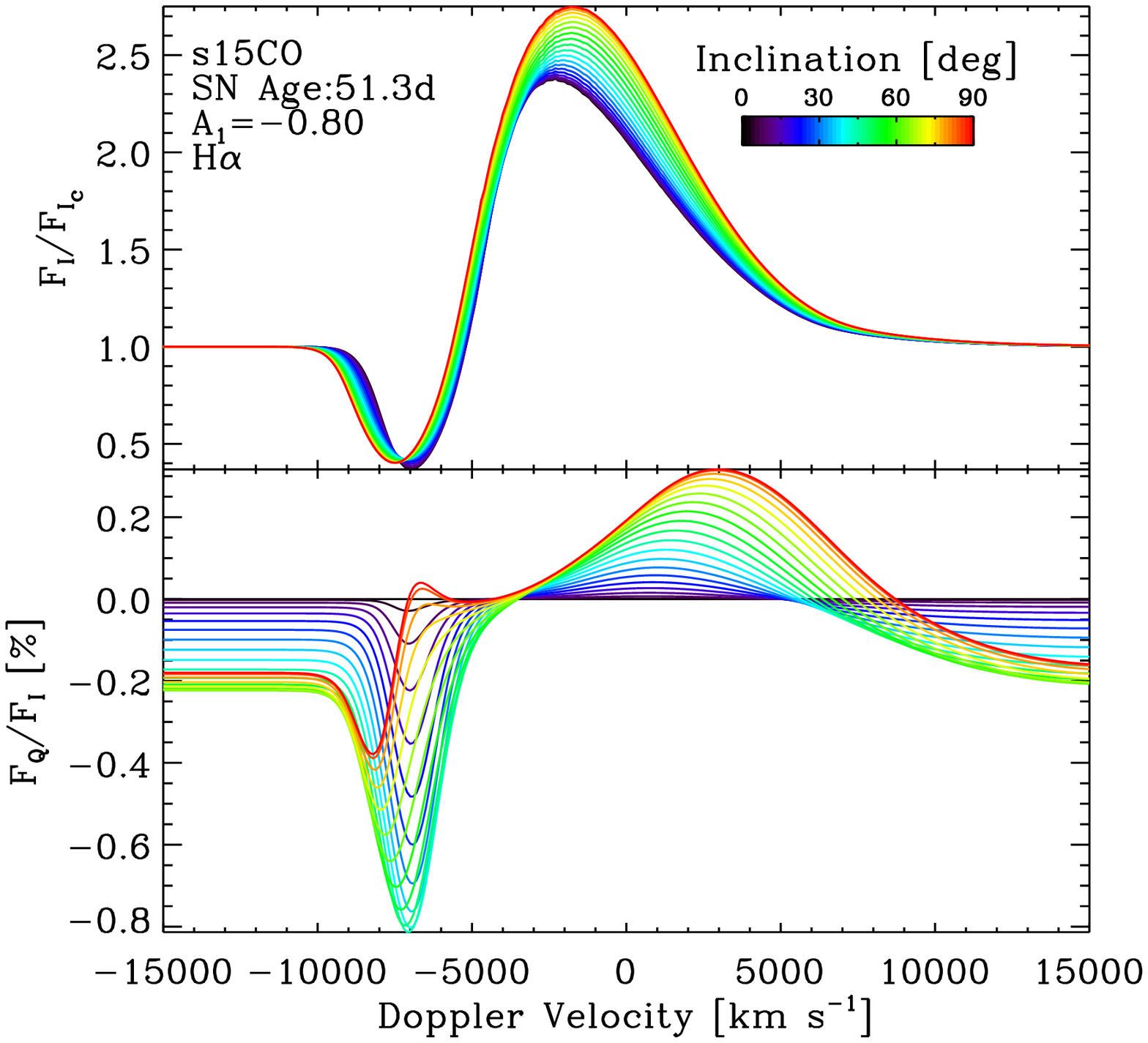,width=7.5cm}  
\caption{
{\it Top:} Synthetic H$\alpha$ normalised line flux (upper panel) and normalised linearly-polarised
 flux (lower panel) based on model s15e12 at 8.5\,d (left) and 51.3\,d (right) after explosion,
 and for an enforced oblateness with pole-to-equator density contrast of a factor of 0.75 ($A_1=-0.25$).
 {\it Bottom:} Same as top row, but now for an enforced oblateness with pole-to-equator density contrast of
a factor of 0.2 $(A_1=-0.8)$.
 \label{fig_s15CO_halpha_WR_pol} }
\end{figure*}

\subsubsection{Evolution during the photospheric phase}

  In Fig.~\ref{fig_s15CO_ha_WR_cont}, we show the evolution of the continuum polarisation
  for the first 50\,d after explosion in models s15CO ($A_1 = -0.25$ and $A_1 = -0.8$) and
  s15TO ($A_2 = -0.25$; ordered from left to right).
  Throughout this sequence, the photosphere remains in what used to be  the hydrogen-rich envelope
  of the progenitor star, so the composition remains essentially fixed and does not influence the radiative signatures
  we discuss \citep{DH11}.

  The continuum polarisation measures in model s15CO are consistently small, always negative,
  and show a non-monotonic temporal variation with a maximum magnitude around 30\,d.
  For the model s15TO, the evolution is more gradual and displays significant time variability.
  We find that $F_Q/F_{Ic}\sin^2 i$ is nearly inclination independent at all times considered if the asphericity
  is small, but this no longer holds as the asphericity is increased. Recovering the scaling of \citet{BL77} is in any case
  somewhat accidental since the conditions are strongly affected by optical-depth effects
  (Sections~\ref{sect_s15TO}--\ref{sect_incl}).

The polarisation evolution is conditioned here by two factors. First, the ionisation changes at and above the photosphere,
  causing a decrease in free-electron density past $\sim$30\,d and the associated appearance
  of a recombination front. Second, throughout this time span, the mass-density profile in the photosphere region flattens
  from a power-law density exponent of $N_d\sim$25 to $N_d\sim$10. This still corresponds to a very steep density fall-off.
  Then, at a given time, the variation with inclination stems from cancellation effects (Sections~\ref{sect_codes}
  and \ref{sect_incl}).

  For completeness, we also show the H$\alpha$ line polarisation at two epochs in Fig.~\ref{fig_s15CO_halpha_WR_pol}.
  For a small asphericity, the polarisation signature evolves weakly between 8.5 and 51.3\,d (top row),
  but for a large asphericity, the change in photospheric conditions strongly affects the polarisation measure
  in the continuum and in the P-Cygni trough. These again are related to subtle cancellation effects.
  In the peak/red-wing regions, the polarisation remains about the same for a fixed shape factor.

\begin{figure}
 \epsfig{file=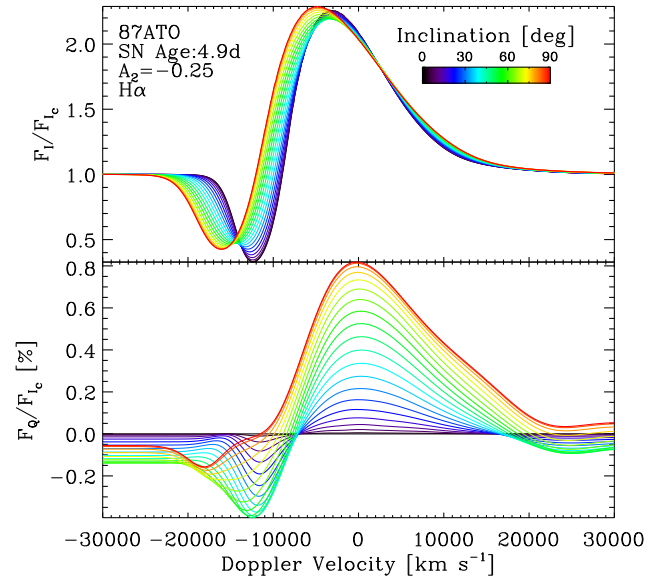,width=8.5cm}
\caption{Same as Fig.~\ref{fig_s15TO_incl}, which is based on model s15e12 at 51.3\,d after explosion,
but now for a calculation based on a SN 1987A ejecta model
at 4.9\,d after explosion \citep{DH10a}.
For both, the same 25\% radial stretching is used to yield an oblate spheroid.
Note the similarity between the two figures, despite the different
nature of the two ejecta resulting from the explosion of RSG and a BSG star.
In practice, at the epochs selected, the line and continuum formation processes are analogous
in those two ejecta, and thus produce essentially the same polarisation signatures for the same
imposed asphericity.
\label{fig_87A}
}
\end{figure}

  It is interesting to compare these results with those for SN 1987A.
  In this SN II-pec, the
  outer ejecta density distribution is well described by a power law with $N_d=$\,8 \citep{DH10a},
  which is comparable to that at the photosphere of model s15e12 at $\sim$50\,d after explosion.
  At an age of 4.9\,d, this SN 1987A ejecta model has a photospheric temperature of 6600\,K (compared to 5400\,K
  for s15e12 at 51.3\,d) and a photospheric velocity of 10,900\,\kms\ (compared to 5378\,\kms\ for
  s15e12 at 51.3\,d). And indeed, for the same asphericity, these two models of a SN II-pec and
  a SN II-P yield a similar polarisation signature (Fig.~\ref{fig_s15TO_incl}
  and Fig.~\ref{fig_87A}), although corresponding to very different post-explosion times.
  On similar grounds, we find that the proximity of ejecta properties for the s15e12 and s25e12 models described
  in \citet{DH11} for 15 and 25\,\msun\ progenitor stars causes their polarisation signatures to look alike
  for a given asymmetry.

\subsubsection{Evolution during the end of the plateau phase and nebular phase}

   For this set of calculations, performed at times halfway through the plateau up to the nebular phase,
   we use a density scaling since the density distribution in model s15e12 (and quite generally for the inner ejecta
   of Type II SNe) is rather flat over extended regions of space. Under such conditions,
   a radial stretching would not yield any asphericity.
   We adopt a latitudinal scaling to produce an oblate spheroid with a pole-to-equator density ratio of 0.8.

\begin{figure*}
\epsfig{file=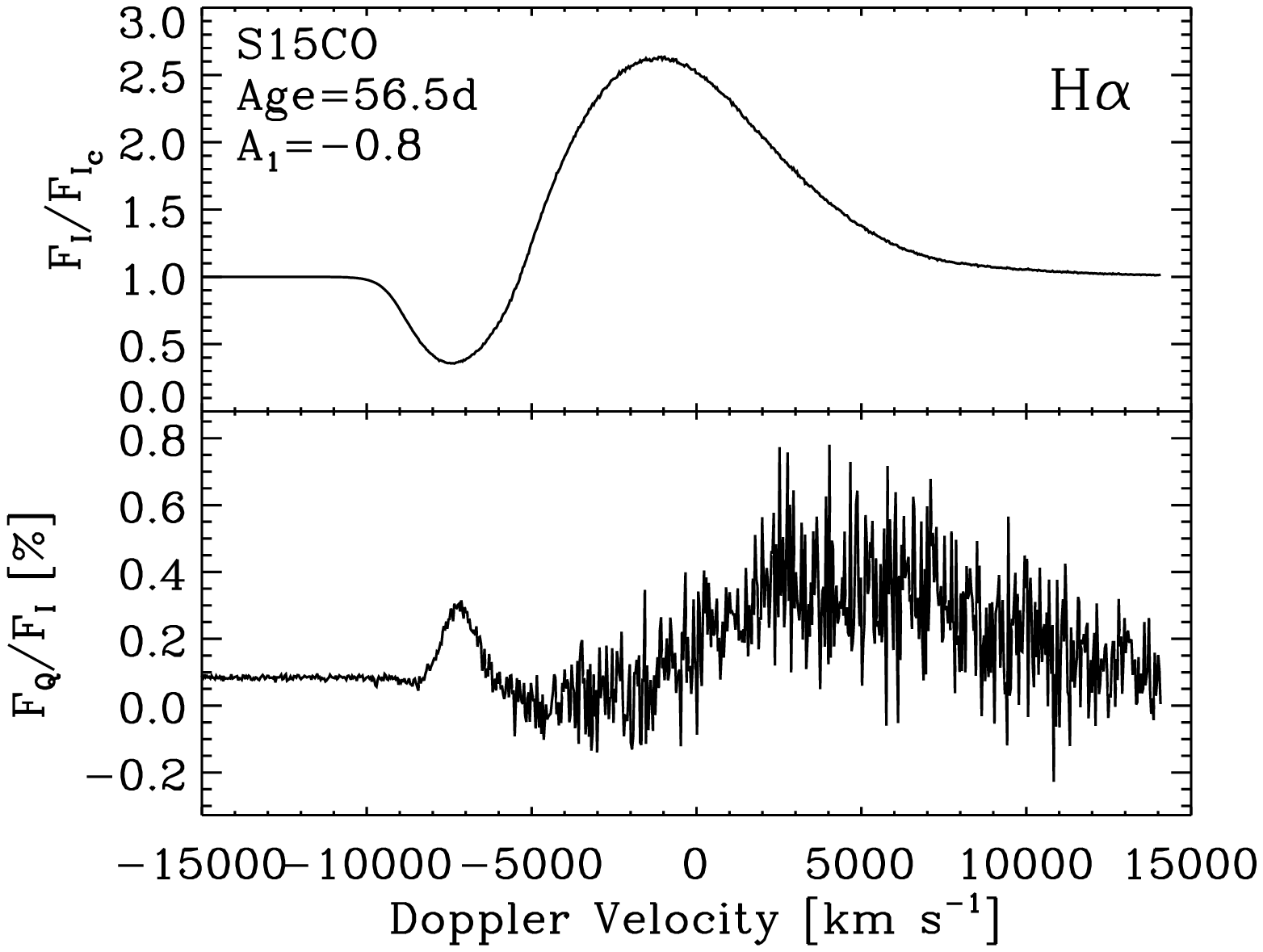,width=7.5cm}
\epsfig{file=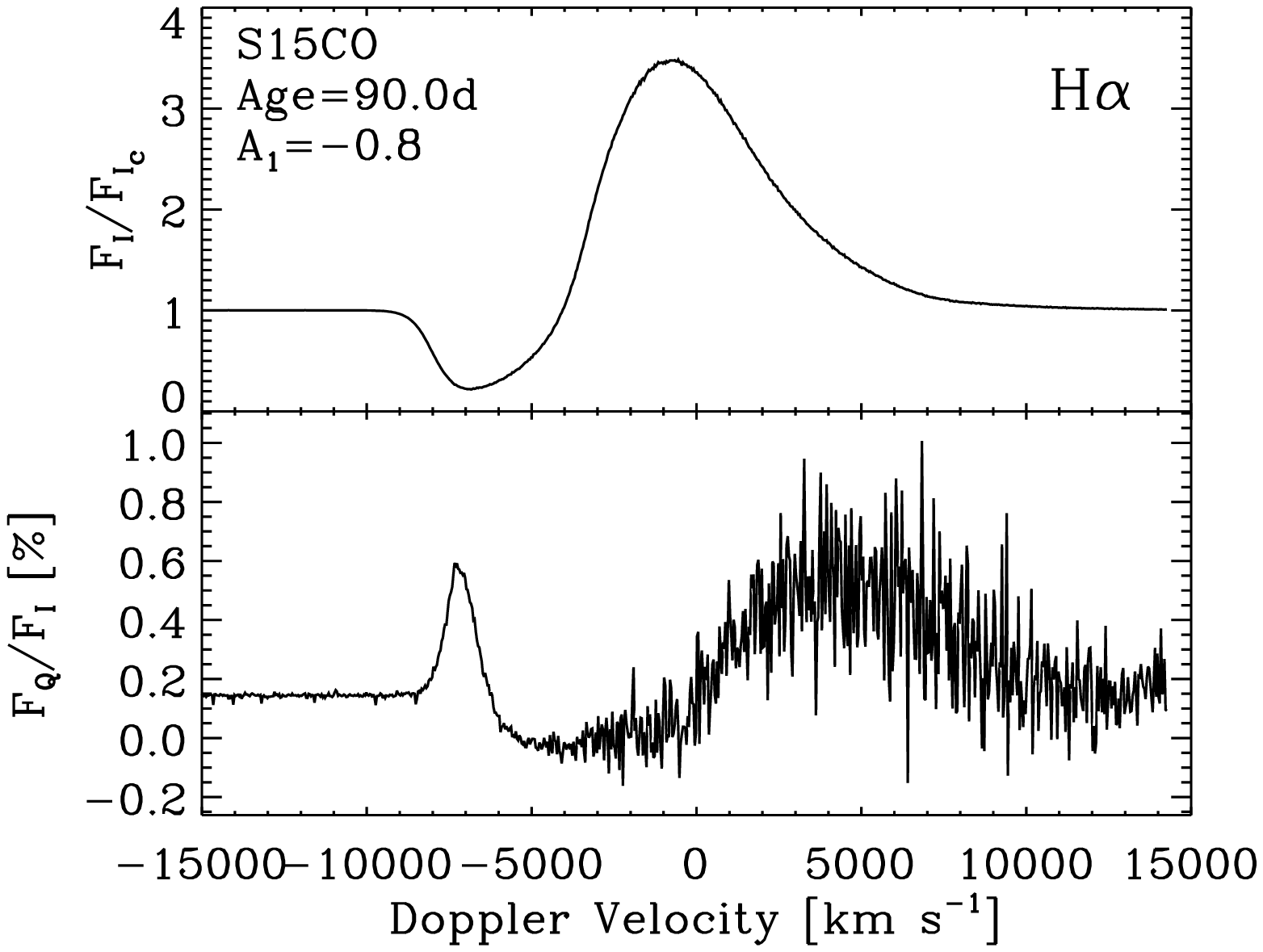,width=7.5cm}
\epsfig{file=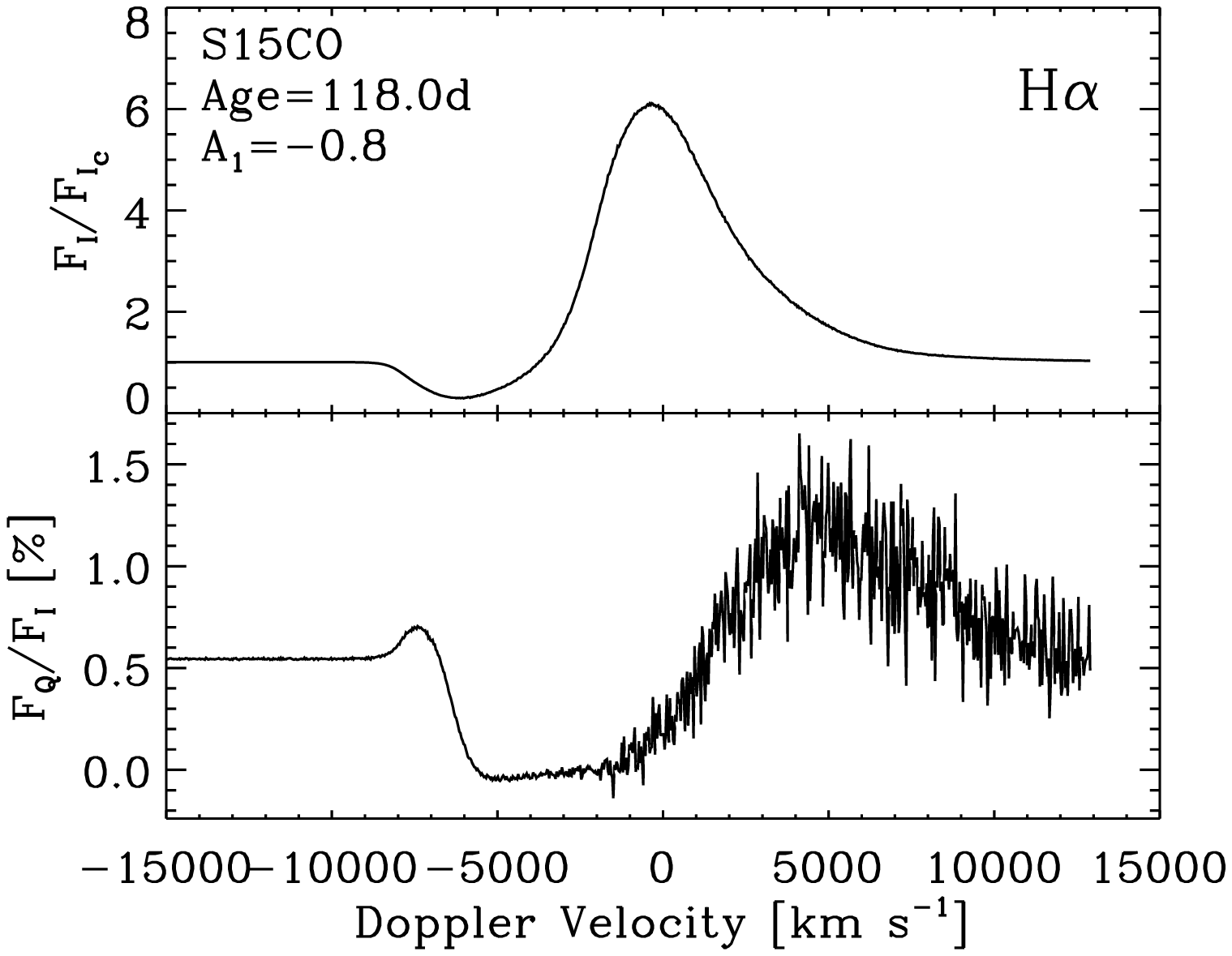,width=7.5cm}
\epsfig{file=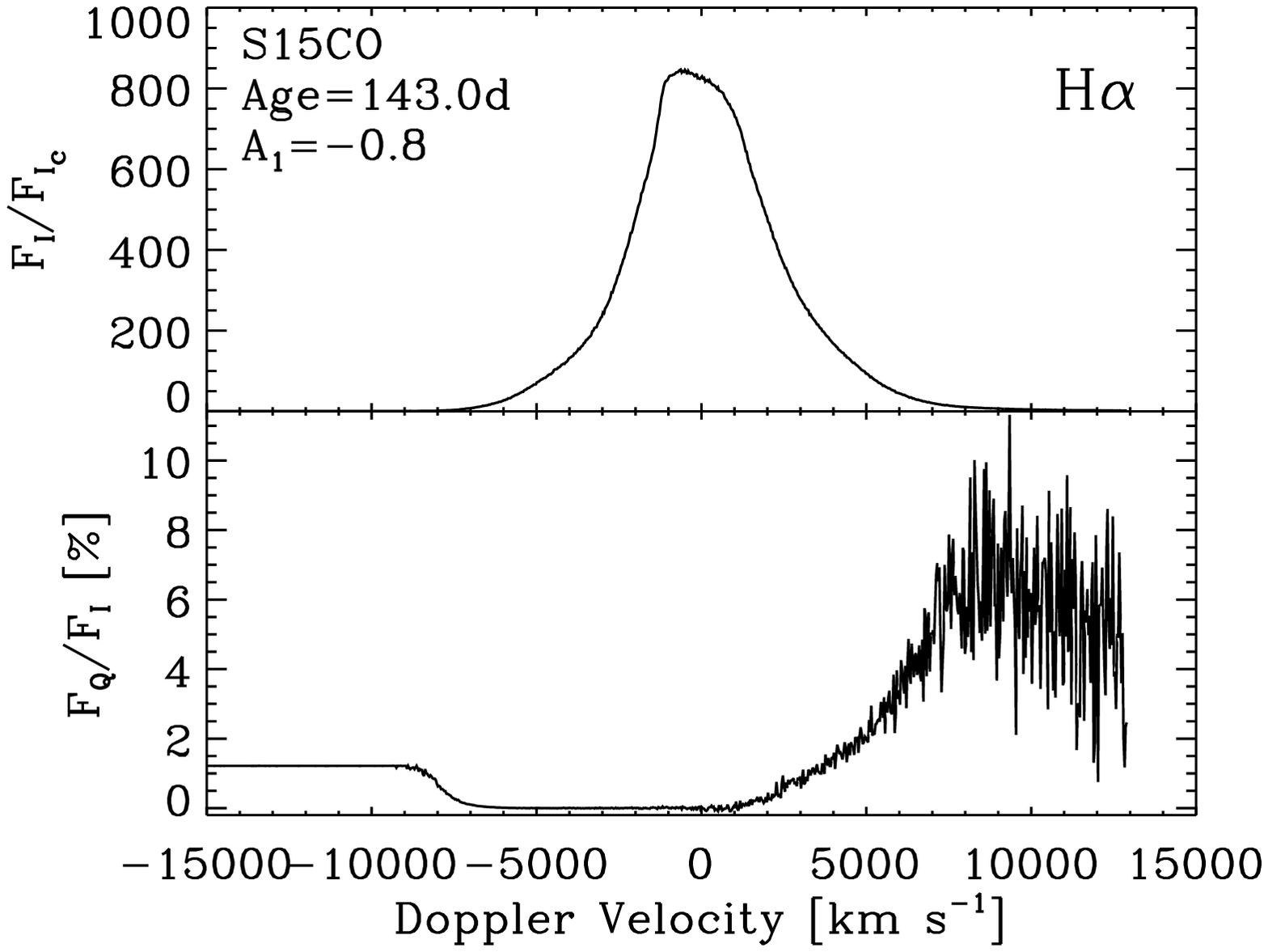,width=7.5cm}
\caption{Monte Carlo simulations of the H$\alpha$ line profile morphology and polarisation
for a viewing angle of 90$^\circ$ with respect to the axis of symmetry
based on model s15e12 and assuming an axisymmetric oblate spheroid, with a pole-to-equator
density ratio of 0.2.
Post-explosion times are 56.5, 90.0, 118.0, ad 143.0\,d, hence covering from optically-thick to
optically-thin conditions. \label{fig_MC_s15_ha}}
\end{figure*}

   We show the results from our Monte Carlo simulations of the H$\alpha$ region in Fig.~\ref{fig_MC_s15_ha} for post-explosion
   times of 56.5, 90.0, 118.0, and 143.0\,d.\footnote{These simulations use $N = 2 \times 10^7$ photons. We have performed
   tests with 10$^8$ photons and find that the results are unchanged, while errors go down with $1/\sqrt{N}$, as expected.}
   Throughout the sequence, the continuum polarisation is now
   positive, starting with small values but increasing to $\gtrsim$1\,\% at the last time.
   In the H$\alpha$ P-Cygni trough, the polarisation is enhanced compared to that in the continuum
   due to the relatively larger contribution of photons undergoing larger-angle scatterings at the edge
   of the photodisk.
   There is now an extended region of zero polarisation at the low-velocity edge of the trough,
   due to enhanced absorption at all impact parameters (see Section~\ref{sect_s15TO}).
   The peak/red-wing polarisation is now of the same sign as in the continuum and reaches large values
   up to a few percent (note that the continuum flux is strongly decreasing at such late times,
   in part because of our neglect of non-thermal excitation/ionisation effects).
   This behaviour across the profile suggests optical-depth effects are now weaker and that the polarising
   material acts more like a scattering nebula. This is also facilitated by the flattening density profile
   and the large volume over which the H$\alpha$ line now forms (as visible from its width in Doppler-velocity space).
   Because of composition stratification, hydrogen is absent from the inner regions where the bulk of the radiation
   originates, and so indeed, the hydrogen-rich layers of the SN ejecta appear as an external oblate ``nebula".
   Furthermore, we now find that varying the inclination angle leaves $F_Q/F_{Ic}\sin^2 i$ unchanged (not shown).

   Our simulations show a growing continuum polarisation with time, from 56.5\,d
   after explosion to 143\,d after explosion, even though the nature and magnitude of aspheriticy is kept fixed!
   Hence, an increasing polarisation may arise exclusively from the changes in line/continuum formation properties
   as a function of time, which reflect changes in ionisation and density distributions. At early post-explosion times,
   the SN polarisation is intrinsically small due to cancellation effects, no matter how asymmetric we made it
   (Section~\ref{sect_shape_factor}).

\begin{figure*}
 \epsfig{file=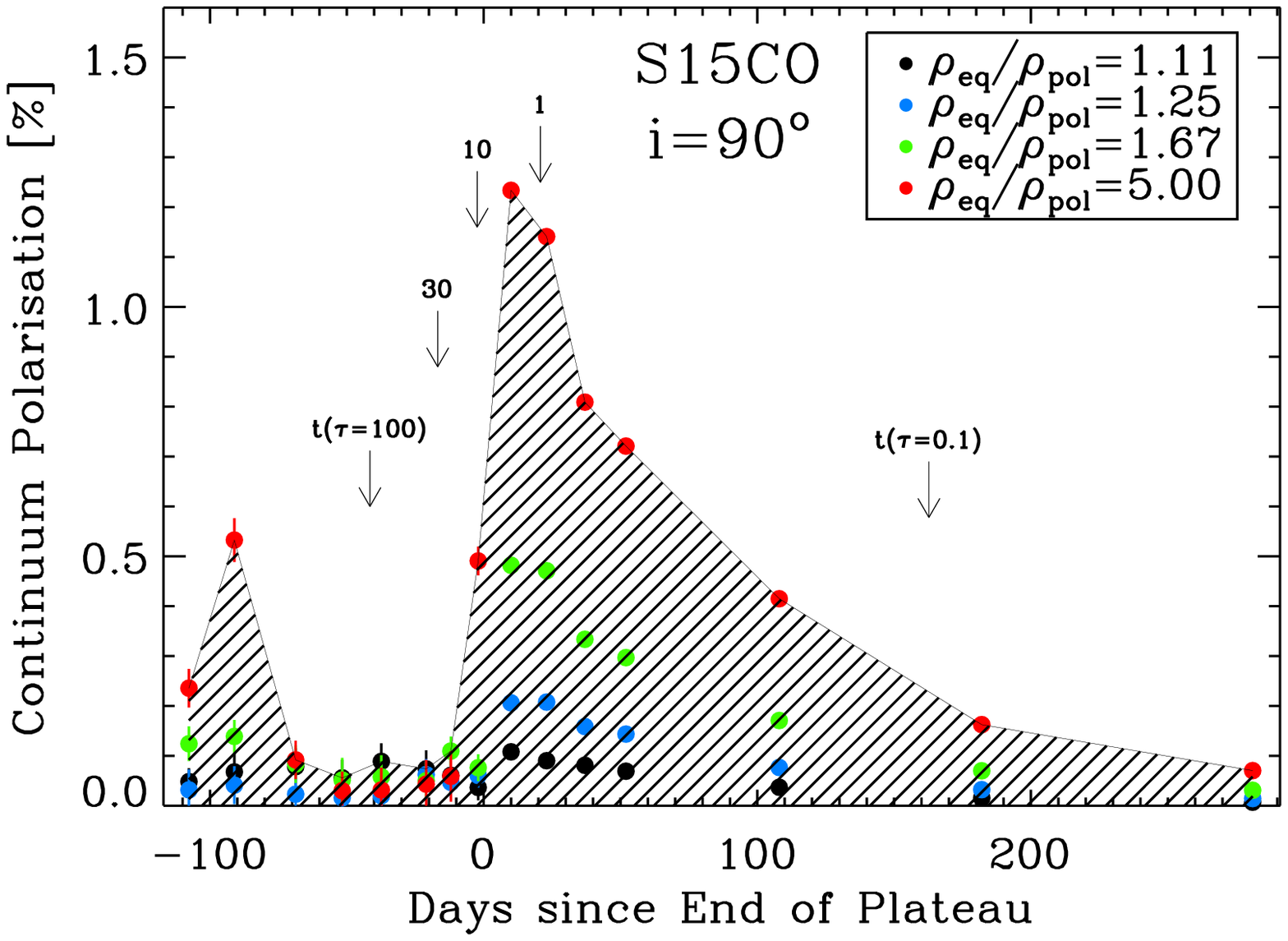,width=15cm}
\caption{Evolution of the continuum polarisation $F_P$ with respect to the end of the plateau
phase for model s15CO. Throughout each time sequence, we adopt a {\it fixed asphericity} 
corresponding to $\rho_{\rm eq} / \rho_{\rm pol}=$1.11 ($A_1=-0.1$, black), 
1.25 ($A_1=-0.2$, blue), 1.67 ($A_1=-0.4$, green), and 5.0 ($A_1=-0.8$, red). We show
results for an inclination of 90$^\circ$, which is the inclination of maximum polarisation at nebular
times but not necessarily at earlier times due to optical-depth effects (vertical arrows indicate the ejecta
electron-scattering optical depth for a selection of times).
Although conditioned by the asphericity of the ejecta, the rise in continuum polarisation that occurs 
suddenly and precisely at the  end of the plateau phase results primarily from the transition to an 
optically-thin ejecta. A similar occurrence has been observed in, e.g., SN2004dj \citep{leonard_etal_06}
and likely results from this sudden change in ejecta optical depth.
\label{fig_s15CO_time_seq}
}
\end{figure*}

   To synthesise our results for axisymmetric Type II SN ejecta and allow an easier confrontation to observations, 
   in particular of SNe II-P, we show in Fig.~\ref{fig_s15CO_time_seq} the evolution of the synthetic continuum 
   polarisation $F_P$ from early-on in the plateau phase and up to 300 days 
   into the nebular phase. These calculations are performed with the Monte Carlo code, adopt an oblate ejecta seen 
   edge on and a continuum wavelength of $\sim$7000\AA. We show four different cases for the equator-to-pole 
   density ratio, as indicated by colored labels.
   As discussed earlier, during the plateau phase, $F_P$ is generally low but its non-monotonic evolution may
   reach sizeable values prior to the recombination phase. Subsequently, the steep free-electron-density profile
   dwarfs the polarised flux. As the ejecta optical depth drops below $\sim$10 at the end of the plateau phase, $F_P$ 
   suddenly jumps to a value that is now commensurate with the magnitude of the asphericity, followed by a decline
   controlled by the $1/t^2$ dependence of the ejecta optical depth.
  Similar polarisation jumps of $\sim$0.5-1.5\% have been reported for SNe 2004dj \citep{leonard_etal_06},
   and 2006my, 2006ov, 2007aa \citep{chornock_etal_10}. Axisymmetric ejecta, perhaps of the oblate
   type just discussed, may explain these polarisation properties. This is at least suggestive and encouraging,
   but we leave to future work a more quantitative assessment of the ejecta morphology based on polarisation signatures.
   A crucial result of this work is that large changes in polarization over time do not require a large change in the asymmetry of the
   ejecta. Rather, such changes can simply arise from radiative transfer effects in an ejecta with a changing
   radial stratification.

\section{Asphericity effects on line-profile morphology}
\label{sect_line_profile}

   All SN spectra, apart from those arising from interaction, are characterised by two different
   line-profile morphologies. During the photospheric
phase, the ejecta is optically thick and all lines are intrinsically P-Cygni profiles. Line overlap
may alter this shape, as in blanketing regions of the blue part of the optical or in the Ca\two\
regions at $\sim$\,8500\,\AA. During the nebular phase, lines eventually become optically thin
and appear as pure emission, with a boxy profile if emitted from a constant-velocity shell \citep{castor_70}.
Departures from spherical symmetry can alter these morphologies. While the line and continuum
formation is essentially unchanged, the altered absorption/emission contribution on the plane of the sky
and in the receding/approaching regions of the ejecta can skew the profile to the red or the blue
or alter the absorption/emission strength.
In addition, small scale inhomogeneities may introduce localised, and likely time dependent, ripples
at a fixed Doppler velocity.
Such signatures suggestive  of departures from spherical symmetry have been observed, but primarily
at nebular times when line overlap and optical-depth effects are weaker or more easily identified
(see, e.g., \citealt{spyromilio_etal_90,spyromilio_94,matheson_etal_00,modjaz_etal_08,maeda_etal_08}).

 Such aspherical effects on line profile morphology can alter numerous inferences we make on SN ejecta.
 During the photospheric phase, oblate/prolate ejecta morphologies can cause a migration of the
 location of the P-Cygni absorption trough with inclination. This is visible in essentially all the simulations presented
 in this paper, and perhaps most vividly in Fig.~\ref{fig_s15TO_incl}.
 The line generally used for inferring Type II SN expansion rates during the photospheric phase at the recombination
 epoch is Fe\two\,5169\,\AA. In the case shown in Fig.~\ref{fig_fenaca} (middle panel), the velocity at
 maximum absorption varies by 30\% from 0 to 90$^{\circ}$ inclinations. This would change the
 inferred explosion energy by a factor of two, but more importantly would potentially compromise
 the desired accuracy of distance determinations based on this diagnostic, such as the Standard-Candle Method
 \citep{hamuy_pinto_02,poznanski_etal_09} - other methods using earlier time observations would
 be less concerned by this problem \citep{DH09_review}.
 Such an anomalous P-Cygni profile in otherwise standard SNe II-P properties could be inferred by
 analysis of the spectra. The steepness of the density distribution, which regulates the blueshift
 of peak emission and the line strength, would not provide a good fit to the observed peak
  \citep{DH05_qs_SN}.

\begin{figure*}
 \epsfig{file=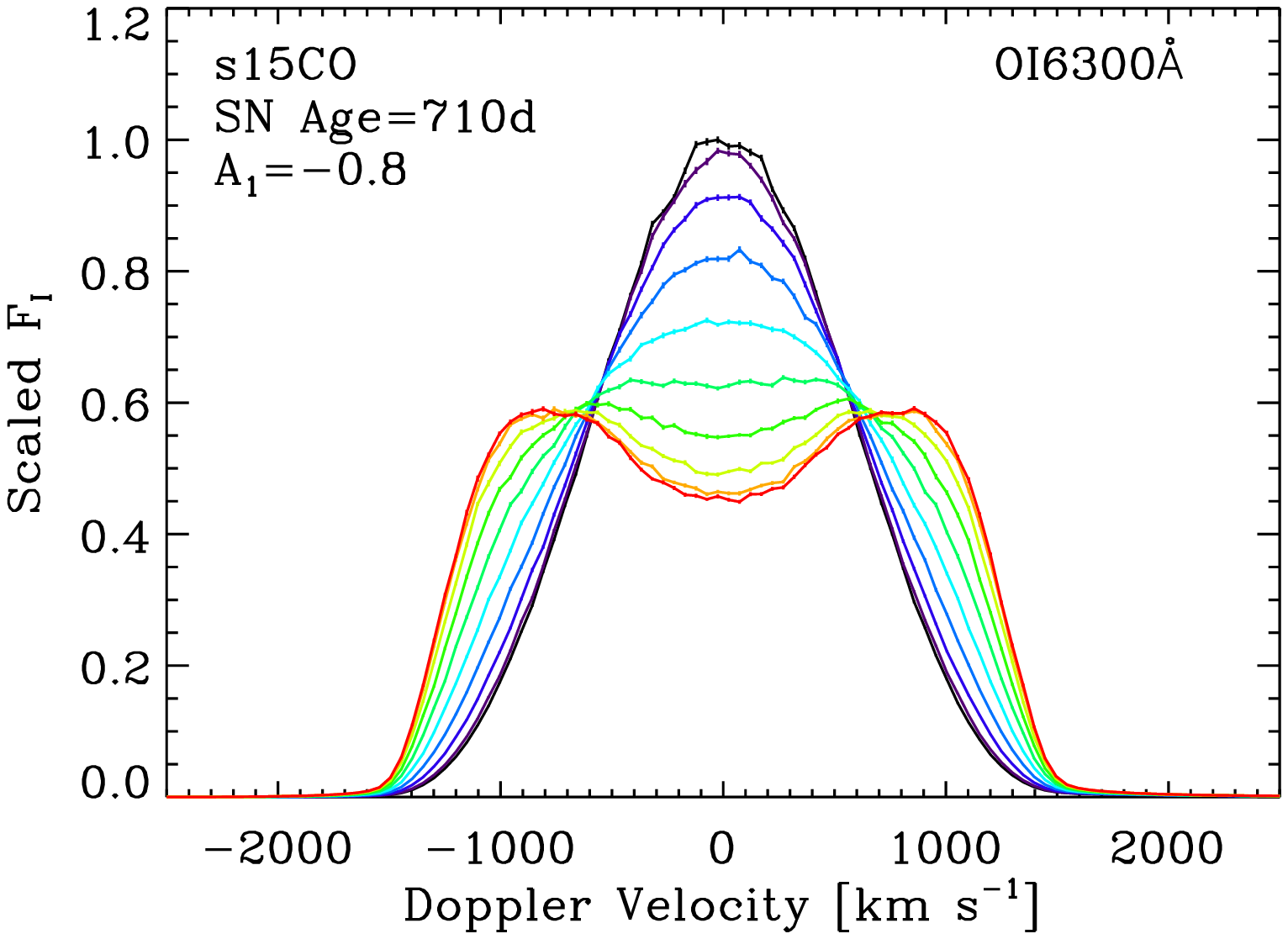,width=8.5cm}
 \epsfig{file=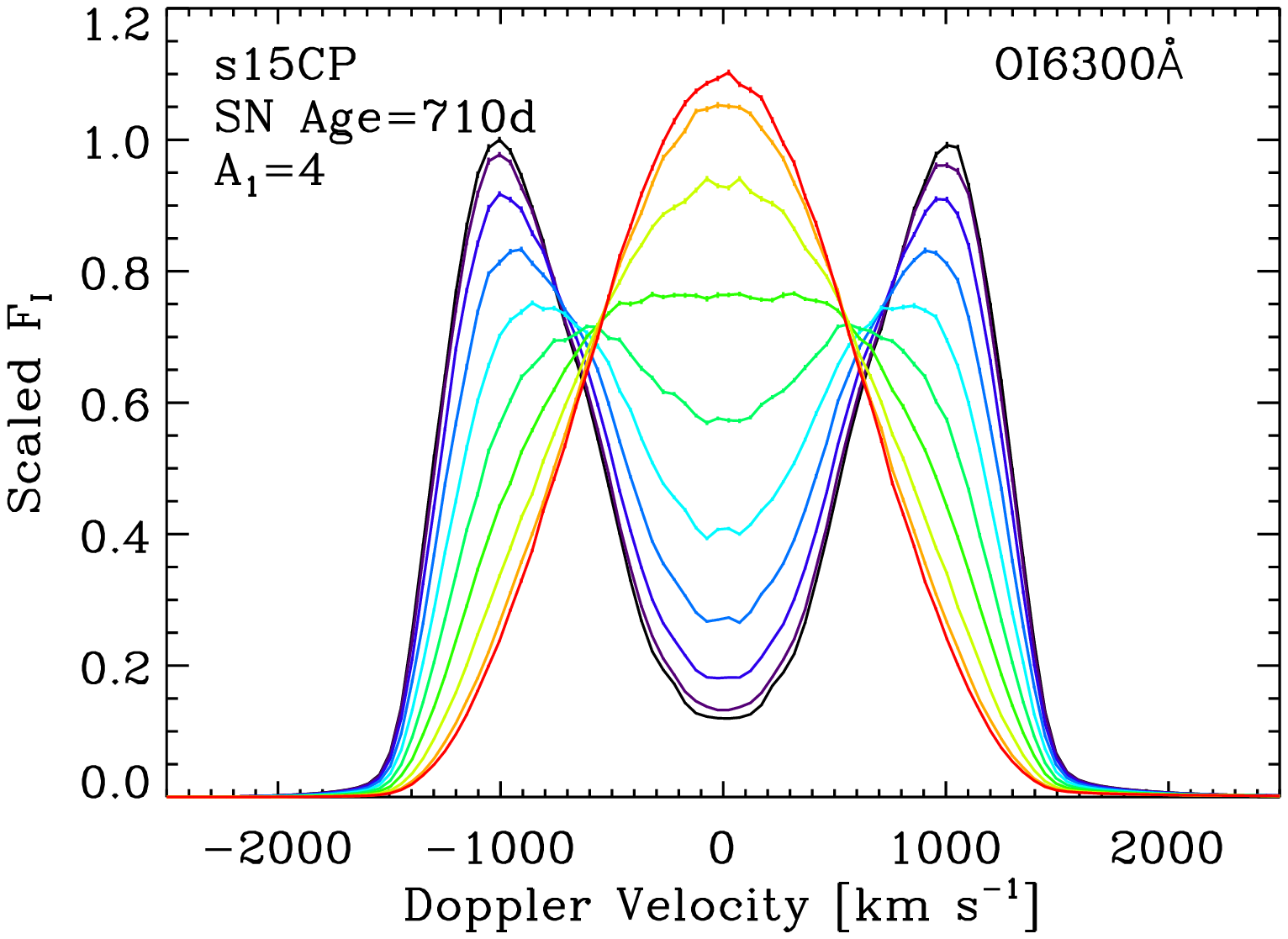,width=8.5cm}
\caption{{\it Left:}
Synthetic line-profile calculations for  O\one\,6300\,\AA\ (i.e., blue component of the corresponding
doublet line) obtained with the Monte Carlo programme
using model s15e12 at 710\,d after explosion and adopting an oblate deformation ($A_1 = -0.8$;
pole-to-equator density ratio is 0.2).
Using a colour coding, we show the profile variations as a function of inclination, from pole-on (black)
to equator-on (red) - we use a constant 10$^\circ$ increment.
{\it Right:} Same as left, but now for a prolate configuration.
\label{fig_s15CO_oi}}
\end{figure*}

During the nebular phase, Type II SNe show only a few strong lines, primarily of  O\one\ and Ca\two.
The O\one\ doublet line at 6300\,\AA\ has been widely used for speculating on the level and
nature of asphericity of core-collapse SN ejecta (see, e.g., \citealt{modjaz_etal_08,maeda_etal_08}).
We have carried out some Monte Carlo simulations of the blue component of this doublet
at 6300\,\AA\ for models s15CO and s15CP at 710\,d after explosion (oblate/prolate configuration with a
pole-to-equator density ratio of 0.8/1.2; Fig.~\ref{fig_s15CO_oi}).

For the oblate configuration, the profile broadens and develops a flat top as the inclination is increased.
A double-peak appears at large inclinations of 60$^{\circ}$ or more.
For arbitrary orientations, there is a probability $P(<\theta_0) = 1 - \cos\theta_0$ of seeing
such an axisymmetric SN with an inclination less that $\theta_0$, hence about a 50\% chance of
seeing a double-peak profile for such an oblate SN morphology.
For prolate configurations, the variation with inclination is reversed although the amplitudes
of the flux variations are much larger.
The oblate configuration seen pole-on and the prolate configuration seen edge-on correspond
to similar profile morphologies, i.e., a single broad emission.
Obviously, prolate and oblate configurations are difficult to disentangle from line-profile morphology.
However, in this last example, the former would yield no line or continuum polarisation by symmetry,
while the latter may.

\section{Asphericity effects on observed luminosity}
\label{sect_lbol}

Because of the variation of the area of the radiating layer with inclination, as well as the angular
dependence of the specific intensity, the flux and derived luminosity
of aspherical SN ejecta is viewing-angle dependent \citep{hoeflich_91,steinmetz_hoeflich_92}.
In the simulations presented in this paper, we find variations in flux/luminosity by up to 50\%.
This is relatively small if we are mostly concerned with understanding the general characteristics
of SN explosions, but this is problematic if the SN is to be used for distance determinations.
This uncertainty adds to that on the expansion rate. Here again, using early-time
observations, ideally prior  to the recombination phase when the photosphere resides in the outer
ejecta layers, is preferable.

\section{Conclusion}
\label{sect_concl}

In this paper, we have presented exploratory calculations of the linear polarisation
from aspherical but axially-symmetric SN ejecta, using 1D inputs from non-LTE time-dependent
radiative-transfer simulations of the SN II-pec 1987A and SNe II-P \citep{DH10a, DH11}.
Using a Monte-Carlo and a long-characteristic code for polarised radiative transfer,
we investigated the total and polarised flux signatures for both continuum and lines for a variety of 
oblate and prolate configurations, and as a function of inclination angle.
The aspherical axially-symmetric ejecta were produced by introducing a latitudinal stretching or
scaling of the 1D ejecta inputs.
Conditioned by our assumptions, particularly the assumption of axial symmetry,
our main results are the following:

\begin{enumerate}

\item Synthetic line and continuum polarisation during the early-plateau phase of Type II-P SNe
tends to be small, largely irrespective of the magnitude of the asphericity. In Type II SNe,
the steep density fall-off in the outer ejecta causes the formation regions of lines and continuum
to overlap and to be radially confined. Most of the SN flux comes from the photodisk, whose extent on
the plane of the sky is limited to the photospheric radius $R_{\rm phot}$. This inhibits large-angle scattering from
the side lobes and favors the contribution from forward-scattered, multiply-scattered, or unpolarised flux radiated
from the photodisk. The resulting polarisation suffers from near-complete cancellation.

\item We find a complex, even erratic, behaviour of line and continuum polarisation
with wavelength, inclination, or shape factors. We routinely obtain polarisation-sign reversals.
The naive visualisation of the polarising SN ejecta as a scattering nebulae surrounding
a radiating central source is unsupported at all times prior to the nebular phase.
The level of SN polarisation is affected by both the angular dependence of the flux and the
angular distribution of the scatterers.

\item We find that the polarisation at the center of strong lines like H$\alpha$ approaches zero only
at the nebular phase. At earlier times, such line photons scatter with
free electrons before escaping and thus cause a residual polarisation at the profile peak and red wing.
This occurs because line photons are in part emitted from regions that are
optically thick in the continuum.
Contrary to the continuum, the line polarisation in the peak/red-wing regions has the sign
expected for an oblate/prolate morphology at all times. In the continuum, the polarisation can
be any sign during the photospheric phase due to cancellation and optical-depth effects.

\item
We find that continuum polarisation can vary dramatically with wavelength, from the Balmer edge
to the near-IR. We interpret this as arising from an optical-depth effect, with the flux escaping primarily
from optically-thin regions and thereby biasing the observed polarisation. The effect is stronger at
shorter wavelengths, a consequence of the wavelength dependence of the albedo and
continuum source function.

\item
At the end of the plateau phase, the continuum polarisation generally increases. This is a consequence of
the lower ejecta optical depth and  the flattening density distribution.

\item
Ejecta asphericity can considerably alter line-profile morphology.
During the photospheric phase, we find that this affects P-Cygni line profiles, and in particular
the location of maximum absorption and the magnitude of the peak blueshift.
This compromises the inference of the SN expansion rate and explosion energy.
At nebular times, asphericity can cause double-peak emission profiles. Lacking the knowledge
of the inclination, oblate and prolate configurations cannot be easily disentangled.

\item
Finally, our axially-symmetric oblate/prolate configurations show an inclination-dependent
luminosity, with variations as high as 50\%. This, and the profile variations, will potentially affect the
accuracy in distances that can be achieved using SNe II-P.

\end{enumerate}

Future efforts are two-fold. On the modelling side, the polarised radiative-transfer codes need
to be generalised to include the treatment of resonant scattering as well as the handling
of a wide spectral range not limited to one line. This would allow the simulation of the full
optical range and the treatment of line overlap.
We also need to start from hydrodynamical inputs of axisymmetric explosions, accounting accurately for
the physical latitudinal variation of their gas properties.
Such modelling tools will then be used to re-analyse the spectropolarimetric observations of SN 1987A,
as well as the more recent ones obtained for SNe II-P and Ib/c \citep{dessart_etal_11}. 
By using all available information,
including line profiles, fluxes, continuum and line polarisation, and time variability, it should be possible
to place strong constraints on the asymmetries of core-collapse SNe, and hence on the explosion
mechanism.

\section*{Acknowledgments}

LD acknowledges financial support from the European Community through an
International Re-integration Grant, under grant number PIRG04-GA-2008-239184.
DJH acknowledges support from STScI theory grant HST-AR-11756.01.A and NASA theory grant NNX10AC80G.
Calculations presented in this work were performed in part at the French National Super-computing Centre  (CINES)
on the Altix ICE JADE machine.


\label{lastpage}

\end{document}